\documentclass[twocolumn,aps,rmp,showpacs,floatfix,amsmath,amssymb,eqsecnum]{revtex4}\def\tablefont{}
\usepackage[dvips]{graphics}

\newcommand{\jamode}[2]{#2}

\def\imo{i}
\def\Re#1{\mathrm{Re}(#1)}
\def\Im#1{\mathrm{Im}(#1)}
\let\pd\partial
\let\ds\displaystyle
\newcommand{\rem}[1]{}

\jamode{%
\def\arXivNum#1{}%
\newcommand{\extraref}[1]{.}%
}{%
\def\arXivNum#1{ [arXiv:#1]}%
\citestyle{nature}%
\newcommand{\extraref}[1]{; #1}
}%

\begin{document}
\title{Quasinormal modes of black holes: from astrophysics to string theory}
\author{R. A. Konoplya}\email{konoplya_roma@yahoo.com}
\affiliation{Department of Physics, Kyoto University, Kyoto 606-8501, Japan}
\affiliation{Theoretical Astrophysics (TAT), Eberhard-Karls University of T\"{u}bingen, T\"{u}bingen 72076, Germany}
\author{Alexander Zhidenko}\email{zhidenko@fma.if.usp.br}
\affiliation{Instituto de F\'{\i}sica, Universidade de S\~{a}o Paulo, 
C.P. 66318, 05315-970, S\~{a}o Paulo-SP, Brazil}
\affiliation{Centro de Matem\'atica, Computa\c{c}\~ao e Cogni\c{c}\~ao, Universidade Federal do ABC,\\ Rua Santa Ad\'elia, 166, 09210-170, Santo Andr\'e-SP, Brazil}
\begin{abstract}
Perturbations of black holes, initially considered in the context of possible observations of astrophysical effects, have been studied for the past ten years in string theory, brane-world models and quantum gravity. Through the famous gauge/gravity duality, proper oscillations of perturbed black holes, called quasinormal modes (QNMs), allow for the description of the hydrodynamic regime in the dual finite temperature field theory at strong coupling, which can be used to predict the behavior of quark-gluon plasmas in the nonperturbative regime. On the other hand, the brane-world scenarios assume the existence of extra dimensions in nature, so that multidimensional black holes can be formed in a laboratory experiment. All this stimulated active research in the field of perturbations of higher-dimensional black holes and branes during recent years. In this review recent achievements on various aspects of black hole perturbations are discussed such as decoupling of variables in the perturbation equations, quasinormal modes (with special emphasis on various numerical and analytical methods of calculations), late-time tails, gravitational stability, AdS/CFT interpretation of quasinormal modes, and holographic superconductors. We also touch on state-of-the-art observational possibilities for detecting quasinormal modes of black holes.
\end{abstract}
\maketitle
\tableofcontents

\renewcommand{\labelenumi}{\theenumi)}

\section{Introduction: New aspects of black hole perturbations}

Isolated black holes in equilibrium are intrinsically simple objects: They are much simpler than ordinary matter around us because they can be described by only a few parameters such as their mass, angular momentum, and charge. This simplicity, however, is apparent only because one never has an isolated black hole. Black holes in centers of galaxies or intermediate-mass black holes always have complex distributions of matter around them, such as galactic nuclei, accretion disks, strong magnetic fields, other stars, planets, etc., and are therefore actively interacting with their surroundings. However, even having been removed from all macroscopic objects and fields in space, a black hole will interact with the vacuum around it, creating pairs of particles and evaporating due to Hawking radiation.

Thus, \emph{a real black hole} can never be fully described by its basic parameters and is \emph{always in the perturbed state}. Whenever you want to know something about stability, gravitational waves, Hawking evaporation of black holes, or about the interaction of black holes with their astrophysical environment, you have to start from an analysis of their perturbations. Moreover, the first moments after the formation of a black hole due to gravitational collapse of matter, the black hole is in a perturbed state. A perturbed black hole can be described by the metric $g_{\mu \nu}$ that can be written as
\begin{equation}
g_{\mu \nu} = g_{\mu \nu}^{0} + \delta g_{\mu \nu}.
\end{equation}
The metric $g_{\mu \nu}^{0}$ is the space-time of the nonperturbed black hole when all perturbations have been damped. It is called the \emph{background}.
In the lowest, linear approximation, the \emph{perturbations} $\delta g_{\mu \nu}$ are supposed to be much less than the background
$\delta g_{\mu \nu} \ll g_{\mu \nu}^{0}$. For example, $g_{\mu \nu}^{0}$ can be the well-known Schwarzschild or Kerr solutions.

Once a black hole is perturbed it responds to perturbations by emitting gravitational waves whose evolution in time can be conditionally divided into three stages:
\begin{enumerate}
\item a relatively short period of initial outburst of radiation,
\item usually a long period of damping proper oscillations, dominated by the so-called quasinormal modes (``quasi'' here means that the system is open and loses energy through gravitational radiation),
\item at very large time the quasinormal modes are suppressed by power-law or exponential late-time tails.
\end{enumerate}

The major part of this review is devoted to the second stage of the evolution of perturbations represented by quasinormal modes. There are a number of reasons for such special interest in quasinormal modes. First, if a new generation of gravitational antennas such as LIGO, VIRGO, LISA, and others manage to detect a gravitational signal from black holes, the dominating contribution to such a signal will be the quasinormal mode with lowest frequency: the \emph{fundamental mode}.

Another motivation comes from the famous duality between supergravity in anti-de Sitter space-time (AdS) and conformal field theory (CFT), AdS/CFT
correspondence \cite{Maldacena:1997re,Aharony:1999ti}. The AdS/CFT correspondence says that quasinormal modes of a (D+1)-dimensional asymptotically AdS black hole or brane are poles of the retarded Green's function in the dual conformal field theory in $D$ dimensions at strong coupling. The latter allows us to describe various properties of strongly coupled quark-gluon plasmas which cannot be studied by traditional perturbative methods of quantum field theory. Recent achievement in this direction was the conjecture of the universal value $1/4\pi$ for the ratio of viscosity to the entropy density in quark-gluon plasma \cite{Kovtun:2004de}. The universal value $1/4\pi$, which follows from string theory, is the same for a variety of dual gravitational backgrounds and is close to the one observed at the Relativistic Heavy Ion Collider (RHIC) \cite{Shuryak:2003xe}.

Usually we can talk about the following types of black holes:
\begin{enumerate}
\item supermassive black holes in centers of galaxies with masses $M \sim (10^5 - 10^9) M_{\odot}$,
\item intermediate mass black holes  $M \sim 10^3 M_{\odot}$,
\item stellar mass black holes $M \sim 10 M_{\odot}$,
\item miniature black holes (intrinsically higher dimensional in string theory and brane-world scenarios) with masses of at least a few Planck masses or more.
\end{enumerate}
Perturbations of the first three types of black holes are important for astrophysical observations, while the fourth type gives its own motivation for study. It is worth remembering that in $D =4$ space-time, due to the uniqueness theorem, there is only one solution of the Einstein equations that describes a stationary rotating black hole -- the Kerr solution. Thus, we have only one candidate for a black hole metric in four dimensions, which is the Kerr metric.
In $D>4$ space-times the situation is different because there is no usual uniqueness theorem and, as a consequence, there are a number of ``black'' solutions with an event horizon such as black strings, branes, rings, saturns, etc \cite{Emparan:2008eg}.
These solutions have different topologies, for instance a black ring has the topology of a doughnut, and many of them are probably gravitationally unstable.
In order to learn which of these higher-dimensional solutions are stable, and thus could exist in nature, one needs to find their spectra of gravitational quasinormal modes. Stability is guaranteed when all quasinormal modes are damped. Stability of higher-dimensional AdS black holes is also important because the onset of instability of a black hole corresponds to the thermodynamic phase transition in the dual field theory \cite{Gubser:2000mm}. Thus, quasinormal modes of higher-dimensional black holes are connected with stability, thermodynamic phase transitions and the hydrodynamic regime of strongly coupled field theories.

We note that two reviews on quasinormal modes of astrophysical black holes and stars appeared in 1999 \cite{Nollert:1999ji,Kokkotas-review}, and one recent review on quasinormal modes of black holes and branes \cite{Berti:2009kk} appeared when we were writing this manuscript. When choosing the material for this review, we emphasized those questions which were not related in detail in these reviews. Thus, we made a review of methods for separation of variables in the perturbation equations (Sec.~\ref{sec:equations}), related presently used numerical and analytical methods for finding quasinormal modes (QNMs) (Sec.~\ref{sec:methods}), and discussed in detail the stability of black holes (Sec.~\ref{sec:stability}) and late-time tails (Sec.~\ref{sec:latetimetails}). In Sec.~\ref{sec:AdSCFT}, in addition to what has become the ``standard'' material on the hydrodynamic regime of field theories at strong coupling, we included a subsection that explains how ordinary WKB methods, usually used for QNMs of black holes, can be applied to finding the conductivity of holographic superconductors.

In most cases, by ``quasinormal modes'' we mean ``frequencies of quasinormal modes'' and not the corresponding amplitudes.
We will be mostly using the units $\hbar = c = G = 1$.

\renewcommand{\labelenumi}{\theenumi.}

\section{Master wave equations}\label{sec:equations}

\subsection{Equations of motion}

From a theoretical point of view, perturbations of a black hole space-time can be performed in two ways: by adding fields to the black hole space-time or by perturbing the black hole metric (the background) itself. In the linear approximation, i.~e., when a field does not backreact on the background, the first type of perturbation is reduced to the propagation of fields in the background of a black hole, which is, in many cases, a general covariant equation of motion of the corresponding field. The covariant form of the equation of motion is quite different for fields of different spin $s$ in curved backgrounds. Thus, for a scalar field $\Phi$ of mass $\mu$ in the background of the metric $g_{\mu \nu}$, the equation of motion is the general covariant Klein-Gordon equation \cite{Landau}
$$ (\nabla^\nu\nabla_\nu - \mu^{2}) \Psi  =0, $$
where $\nabla_\nu$ is the covariant derivative. The above equation can be written explicitly as follows:
\begin{equation}\label{scalar_covariant}
\frac{1}{\sqrt{-g}} \pd_\nu\left(g^{\mu \nu} \sqrt{-g} \pd_\mu\Psi\right) - \mu^{2} \Psi = 0 \quad (s=0).
\end{equation}

For massive Dirac fields in a curved background $g_{\mu \nu}$, the equation of motion reads \cite{Ivanenko}
\begin{equation}\label{em:Dirac}
(\gamma^ae_a^{~\mu}(\pd_\mu+\Gamma_\mu)+\mu)\Phi=0, \quad ( s =
\pm 1/2)
\end{equation}
where $\mu$ is the mass of the Dirac field, and $e_a^{~\mu}$ is the
tetrad field, defined by the metric $g_{\mu\nu}$:
$$ g_{\mu\nu}=\eta_{ab}e_\mu^{~a}e_\nu^{~b}, \quad
g^{\mu\nu}=\eta^{ab}e_a^{~\mu}e_b^{~\nu}, $$
\begin{equation}
\quad e_a^{~\mu}e_\mu^{~b}=\delta_a^b, \quad e_\mu^{~a}e_a^{~\nu}=\delta_\mu^\nu,
\end{equation}
where $\eta_{ab}$ is the Minkowskian metric, $\gamma^a$ are the Dirac
matrices: $$\{\gamma^a,\gamma^{b}\}=2\eta^{ab},$$ and $\Gamma_\mu$ is the spin
connection  \cite{Ivanenko},
\begin{equation}
\Gamma_\mu=\frac{1}{8}[\gamma^a,\gamma^b]~g_{\nu\lambda}~e_a^{~\nu}~\nabla_\mu e_{b}^{~\lambda}.
\end{equation}

For massive vector perturbations we have the general covariant
generalization of the Proca equations \cite{Proca}. For a vector
potential $A_{\mu}$, one has
\begin{equation}
\nabla^\nu F_{\mu \nu}- \mu^2 A_\mu=0, \qquad F_{\mu \nu} = \pd_\mu A_\nu - \pd_\nu A_\mu.
\end{equation}
In a curved space-time these equations read
\begin{equation}\label{em:Proca}
\frac{1} {\sqrt{-g}}\pd_\nu (F_{\rho\sigma} g^{\rho
  \mu} g^{\sigma \nu} \sqrt{-g}) - \mu^2 A^{\mu } =0. \qquad (s=1)
\end{equation}
When $\mu = 0$ in the above form of the Proca equation, we obtain
the Maxwell equation
\begin{equation}\label{em:Maxwell}
\pd_\nu((\pd_\alpha A_\sigma - \pd_\sigma A_\alpha) g^{\alpha \mu}
g^{\sigma \nu}
\sqrt{-g}) = 0.
\end{equation}

There may be various generalizations of the massive scalar, spinor, and vector fields considered above. Thus, if we study perturbations of massive charged particles in scalar electrodynamics in a curved charged background, we have to deal with a complex scalar field
$$ (D^\nu D_\nu - \mu^{2}) \Psi  =0, $$
where $D_\nu=\nabla_\nu-\imo e A_\nu$ is an ``extended'' covariant derivative, and $e$ is the charge of the particle. Finally, we find that the equation of motion of the charged scalar field in a curved space-time reads \cite{Hawking}
\begin{eqnarray}
\frac{1}{\sqrt{-g}} \pd_\nu\left(g^{\mu \nu} \sqrt{-g} (\pd_\mu\Psi-\imo e A_\mu\Psi)\right) - \imo e A^\nu\pd_\nu\Psi\nonumber\\
- (\mu^{2}+e^2A^\nu A_\nu)\Psi  = 0.
\label{em:scalar}
\end{eqnarray}

In a similar fashion, the massive charged Dirac particle is described by the equation of motion with an extended derivative $\pd_\mu\rightarrow\pd_\mu-\imo e A_\mu$,
\begin{equation}
(\gamma^ae_a^{~\nu}(\pd_\nu+\Gamma_\nu-\imo eA_\nu)+\mu)\Phi=0.
\end{equation}

Another type of perturbation, metric perturbations, can be written in the linear approximation in the form
\begin{equation}
g_{\mu \nu} =  g_{\mu \nu}^{0} + \delta g_{\mu \nu},
\end{equation}
\begin{equation}\label{Einsteinperturb}
\delta R_{\mu \nu} = \kappa~\delta\left( T_{\mu \nu} -\frac{1}{D-2}T g_{\mu\nu}\right)+\frac{2\Lambda}{D-2}\delta g_{\mu\nu}.
\end{equation}
Linear approximation means that in Eq.~(\ref{Einsteinperturb}) the terms of order $\sim \delta g_{\mu \nu}^{2}$ and higher are neglected. The unperturbed space-time given by the metric $g_{\mu\nu}^{0}$ is called the background.

\subsection{Separation of variables and radial equations of wavelike form}

The first step toward the analysis of the black hole perturbation equations is their reduction to the two-dimensional wavelike form with decoupled angular variables. Once the variables are decoupled, a wave like equation for radial and time variables usually has the Schr\"odinger-like form for stationary backgrounds,
\begin{equation}\label{wave1}
-\frac{d^2R}{dr_*^2}+ V(r, \omega) R=\omega^2R, 
\end{equation}
and can be treated by a number of sophisticated and well-developed numerical, analytical, and semianalytical methods.

As a simple exercise, one can see that for the massless scalar field on the Schwarzschild background ($g_{\mu \nu}^{0}:$ $g_{tt} = -g_{rr}^{-1} = 1-2 M/r$, $g_{\theta \theta} = g_{\phi \phi} \sin^{-2} \theta = r^2$), Eq.~(\ref{scalar_covariant}), after using a new variable
$$d r_{*} = \frac{d r}{1-2 M/r}.$$
The coordinate $r_*$ maps the semi-infinite region from the horizon to infinity into the $(-\infty, +\infty)$ region and is, therefore, called the \emph{tortoise coordinate}.

The wave function
$$\Psi(t,r,\theta,\phi) = e^{-\imo\omega t}Y_{\ell}(\theta,\phi)R(r)/r,$$
produces (\ref{wave1}) with a potential
\begin{equation}\label{SBH-scalar}
V(r) = \left(1 - \frac{2 M}{r}\right)\left(\frac{\ell (\ell +1)}{r^2} + \frac{2 M(1-s^2)}{r^3}\right),
\end{equation}
where $s=0$, and $\ell$ is the multipole quantum number, which arises from the separation of angular variables by expansion into spherical harmonics
$$\Delta_{\theta,\phi}Y_{\ell} (\theta, \phi)=-\ell(\ell+1)Y_{\ell} (\theta, \phi),$$
exactly in the same way as it happens for the hydrogen atom problem in quantum mechanics when dealing with the Schr\"odinger equation.

When $s=1$ the effective potential (\ref{SBH-scalar}) corresponds to the Maxwell field. When $s=2$ we obtain the effective potential of the gravitational perturbations of the axial type, which was derived by Regge and Wheeler in \cite{Regge:1957td}.

The separation of variables, however, is not always so easy. The variables in perturbation equations cannot be decoupled for perturbations of an arbitrary metric. For this to happen, the metric must possess sufficient symmetry, expressed in existence of the Killing vectors, Killing tensors, and Killing-Yano tensors \cite{Carter:1968rr,Frolov:2008jr}. The choice of appropriate coordinates is crucial for separation of variables \cite{separatists2}. Having left readers with literature on this subject \cite{separatists3}, in the next two subsections we considered some more generic and practically useful examples of the separation of variables for fields of various spin $s=0, 1/2, 1, 3/2, 2$ in various black hole backgrounds. The two most efficient and general approaches to black hole perturbations will be related here: the general formalism of the spin $s$ perturbations with the help of the Newman-Penrose tetrads \cite{Newman} and the gauge-invariant method for gravitational perturbations \cite{Ishibashi-Kodama-gauge-invarinat}.

{
\subsection{Separation of variables procedure for the Kerr-Newman-de Sitter geometry}

We considered the perturbation equations for fields of various spin in the the Kerr-Newman-(A)dS background, describing the charged, rotating black hole in asymptotically de Sitter (or anti-de Sitter) universe. This solution described quite a wide class of black holes which possesses a number of parameters: the black hole mass, its charge, momentum, and the cosmological constant. Since the electromagnetic and  gravitational fields couple through the electric charge of the black hole, there is no pure electromagnetic or pure gravitational perturbations unless the charge $Q = 0$, but one type of perturbation induces the other. With the help of the Newman-Penrose formalism shown below, Teukolsky in \cite{Teukolsky} separated variables in the perturbation equations for the Kerr geometry for massless fields of various spin.

In Boyer-Lindquist coordinates the Kerr-Newman-(A)dS metric has the form,
$$ ds^2 = -\rho^2
\left(\frac{dr^2}{\Delta_r}+\frac{d\theta^2}{\Delta_\theta}\right)
-\frac{\Delta_\theta \sin^2\theta}{(1+\alpha)^2 \rho^2} [a dt-(r^2+a^2)d\varphi]^2  $$
\begin{eqnarray}
+\frac{\Delta_r}{(1+\alpha)^2 \rho^2}(dt-a\sin^2\theta d\varphi)^2,
\end{eqnarray}
where
\begin{eqnarray}
\Delta_r&=&(r^2+a^2)\left(1-\frac{\alpha}{a^2}r^2\right)-2Mr+Q^2\\\nonumber
 &=&-\frac{\alpha}{a^2}(r-r_+)(r-r_-)(r-r'_+)(r-r'_-),
\end{eqnarray}
\begin{eqnarray}
 \Delta_\theta=1+\alpha\cos^2\theta, &\quad&
 \alpha=\frac{\Lambda a^2}{3},  \nonumber\\
\bar{\rho}=r+ia\cos\theta, &\quad&
\rho^2=\bar{\rho}\bar{\rho}^*.\nonumber
\end{eqnarray}
Here $\Lambda$ is the cosmological constant, $M$ is the mass of the black hole, $Q$ is its charge, and $a$ is the rotation parameter. The electromagnetic field due to the charge of the black hole is given by
\begin{equation}
A_{\mu}dx^{\mu}
   =-\frac{Qr}{(1+\alpha)^2\rho^2}(dt-a\sin^2\theta d\varphi).
\end{equation}
In particular, we adopt the following vectors as the null tetrad:
\begin{subequations}\label{npformalism}
\begin{eqnarray}
l^\mu &=& \left(\frac{(1+\alpha)(r^2+a^2)}{\Delta_r}, 1, 0,
\frac{a(1+\alpha)}{\Delta_r}\right), \nonumber \\
n^\mu &=& \frac{1}{2\rho^2}\left((1+\alpha)(r^2+a^2), -\Delta_r, 0,
a(1+\alpha)\right), \nonumber \\
m^\mu &=& \frac{1}{\bar{\rho}\sqrt{2\Delta_\theta}}
\left(ia(1+\alpha)\sin\theta, 0, \Delta_\theta,
\frac{i(1+\alpha)}{\sin\theta}\right), \nonumber \\
\bar{m}^\mu &=& {m^*}^\mu.
\end{eqnarray}
Because of the existence of the Killing vectors $\partial_t$ and $\partial_\phi$ for the unperturbed background, the time and azimuthal dependence of the fields has the form $e^{-i(\omega t-m\varphi)}$, the tetrad components of the derivative and the electromagnetic field are
\begin{equation}
\begin{array}{cc}
l^\mu \partial_\mu={\cal D}_0, &
n^\mu \partial_\mu=\displaystyle{
-\frac{\Delta_r}{2\rho^2}{\cal D}_0^\dag}, \\
m^\mu \partial_\mu=\displaystyle{
\frac{\sqrt{\Delta_\theta}}{\sqrt{2}\bar{\rho}}
{\cal L}_0^\dag}, &
\bar{m}^\mu \partial_\mu=\displaystyle{
\frac{\sqrt{\Delta_\theta}}{\sqrt{2}\bar{\rho}^*}{\cal L}_0},\\
l^\mu A_\mu=\displaystyle{-\frac{Qr}{\Delta_r}}, &
n^\mu A_\mu=\displaystyle{-\frac{Qr}{2\rho^2}}, \\
m^\mu A_\mu=\bar{m}^\mu A_\mu=0, &
\end{array}
\end{equation}
where
\begin{eqnarray}
{\cal D}_n&=&\partial_r-\frac{i(1+\alpha)K}{\Delta_r}
+n\frac{\partial_r \Delta_r}{\Delta_r}, \nonumber \\
{\cal D}_n^\dag&=&\partial_r+\frac{i(1+\alpha)K}{\Delta_r}
+n\frac{\partial_r \Delta_r}{\Delta_r},  \\
{\cal L}_n&=&\partial_\theta+\frac{(1+\alpha)H}{\Delta_\theta}
+n\frac{\partial_\theta(\sqrt{\Delta_\theta} \sin\theta)}
{\sqrt{\Delta_\theta} \sin\theta}, \nonumber \\
{\cal L}_n^\dag&=&\partial_\theta-\frac{(1+\alpha)H}{\Delta_\theta}
+n\frac{\partial_\theta(\sqrt{\Delta_\theta} \sin\theta)}
{\sqrt{\Delta_\theta} \sin\theta}, \nonumber
\end{eqnarray}
\end{subequations}
with $K=\omega(r^2+a^2)-am$ and
$\displaystyle{H=-a\omega\sin\theta+\frac{m}{\sin\theta}}$.

Making use of the above equations (\ref{npformalism}) and after some algebra the covariant equations of motion can be cast into the form with separated variables
\begin{eqnarray}
&& \left[
\sqrt{\Delta_\theta}{\cal L}_{1-s}^{\dag}\sqrt{\Delta_\theta}
{\cal L}_s
- 2(1+\alpha)(2s-1)a\omega\cos\theta \right.\nonumber \\
&&  \left.
- 2\alpha(s-1)(2s-1)\cos^2\theta+\lambda \right]S_s(\theta) = 0,
\label{eqn:Ss}\\
&&  \left[
\Delta_r {\cal D}_1 {\cal D}_s^\dag +2(1+\alpha)(2s-1)i\omega r
-\frac{2\alpha}{a^2}(s-1)(2s-1)  \right.  \nonumber \\
&&  \left.
+\frac{-2(1+\alpha)eQKr+iseQr\partial_r \Delta_r+e^2Q^2r^2}
{\Delta_r}  \right.  \nonumber \\
&&  \left.
 -2iseQ-\lambda \right]R_s(r) = 0, \label{eqn:Rs}
\end{eqnarray}
}%
where $s$ is the spin of the field, and $e$ is its charge. In the general case of the Kerr-Newman-de Sitter (and anti-de Sitter) space-times, spin 0 and $1/2$ fields can be  separated for the above equations \cite{Suzuki:1998vy}. For Kerr-de Sitter black holes ($Q=0$), the separation is possible for fields of spin $0$, $1/2$, $1$, $3/2$, and $2$.

{
\def\A{{\cal A}}
\def\B{{\cal B}}
\def\C{{\cal C}}
\def\D{{\cal D}}
\def\E{{\cal E}}
\def\F{{\cal F}}
\def\G{{\cal G}}
\def\H{{\cal H}}
\def\I{{\cal I}}
\def\J{{\cal J}}
\def\K{{\cal K}}
\def\LL{{\cal L}}
\def\M{{\cal M}}
\def\N{{\cal N}}
\def\OO{{\cal O}}
\def\P{{\cal P}}
\def\Q{{\cal Q}}
\def\R{{\cal R}}
\def\SS{{\cal S}}
\def\T{{\cal T}}
\def\U{{\cal U}}
\def\V{{\cal V}}
\def\W{{\cal W}}
\def\X{{\cal X}}
\def\Y{{\cal Y}}
\def\Z{{\cal Z}}
\def\GAMMA{{\mit \Gamma}}
\def\DELTA{{\mit \Delta}}
\def\THETA{{\mit \Theta}}
\def\LAMBDA{{\mit \Lambda}}
\def\XI{{\mit \Xi}}
\def\PI{{\mit \Pi}}
\def\SIGMA{{\mit \Sigma}}
\def\UPSILON{{\mit \Upsilon}}
\def\PHI{{\mit \Phi}}
\def\PSI{{\mit \Psi}}
\def\OMEGA{{\mit \Omega}}
%

\def\vol#1{{\bf #1}}
\def\linebreak{\hfill\break}
\def\Exists{$\exists$}
\def\Forall{$\forall$}
\def\Then{$\then$ }
\def\Equivalent{ $\equivalent$ }
\def\newterm#1{{\it #1}}
\def\Ref#1{(\ref{#1})}
\def\Eqref#1{Eq.(\ref{#1})}
\def\Eqsref#1#2{Eqs.(\ref{#1})-(\ref{#2})}
\def\Prop#1{Proposition \ref{#1}}
\def\Cor#1{Corollary \ref{#1}}
\def\Lemma#1{Lemma \ref{#1}}
\def\Theorem#1{Theorem \ref{#1}}

\def\Nucl#1#2#3{\hbox{${}_{#2}^{#3}${#1}}}
\def\el{\r{e}}
\def\p{\r{p}}
\def\n{\r{n}}
\def\Mpl{M_\r{pl}}
\def\Lpl{L_\r{pl}}
\def\tpl{t_\r{pl}}
\def\Epl{E_\r{pl}}
\def\sun{\odot}
\def\Msun{M_\sun}
\def\bra<#1|{\langle #1\rvert}
\def\ket|#1>{\lvert#1 \rangle}
\def\braket<#1|#2>{\langle #1|#2 \rangle}


\def\mod{{\rm mod}}
\def\sign{{\rm sign}}
\def\index{{\rm ind}}
\def\pararell{\mathrel{/\!/}}
\def\para{{\scriptscriptstyle /\!/}}
\def\orth{\perp}
\def\pfrac#1#2{\left(\frac{#1}{#2}\right)}

\def\tend{\rightarrow}
\def\then{\Rightarrow\quad}
\def\equivalent{\quad\Leftrightarrow\quad}
\def\therefore{\mbox{\setbox0=\hbox{X}\hbox{$\ldotp$}\raise0.7\ht0\hbox{$\ldotp$}\hbox{$\ldotp$}} \quad }
\def\because{\mbox{\setbox0=\hbox{X}\raise0.7\ht0\hbox{$\ldotp$}\hbox{$\ldotp$}\raise0.7\ht0\hbox{$\ldotp$}}\kern0pt }

\def\r#1{{\rm #1}}
\def\b#1{{\bf #1}}
\def\e#1{{10^{#1}}}
\def\bm#1{\boldsymbol{#1}}
\let\bg=\bm
\def\Slash#1{\hbox{\rlap{$\:/$}$#1$}}

\def\NN{{{\mathbb N}}}
\def\ZR{{{\mathbb Z}}}
\def\QF{{{\mathbb Q}}}
\def\RF{{{\mathbb R}}}
\def\CF{{{\mathbb C}}}
\def\HF{{{\mathbb H}}}
\def\FF{{{\mathbb F}}}
\def\sZR{{\bm{\scriptstyle Z}}}
\def\sQF{{\bm{\scriptstyle Q}}}
\def\sRF{{\bm{\scriptstyle R}}}
\def\sCF{{\bm{\scriptstyle C}}}
\def\sHF{{\bm{\scriptstyle H}}}
\def\sFF{{\bm{\scriptstyle F}}}
\def\GL{{\rm GL}}
\def\gl{{\rm gl}}
\def\IGL{{\rm IGL}}
\def\SL{{\rm SL}}
\def\sl{{\rm sl}}
\def\SO{{\rm SO}}
\def\so{{\rm so}}
\def\ISO{{\rm ISO}}
\def\SU{{\rm SU}}
\def\su{{\rm su}}
\def\Sp{{\rm Sp}}
\def\sp{{\rm sp}}
\def\PSL{{\rm PSL}}
\def\OG{{\rm O}}
\def\IO{{\rm IO}}
\def\UG{{\rm U}}
\def\Pin{{\rm Pin}}
\def\Spin{{\rm Spin}}
\def\Cliff{{{\rm C}\!\ell }}
\def\CCliff{ {\CF\!\ell} }

\def\id{{\rm id}}
\def\range#1{{\rm ran}\; #1}
\def\domain#1{{\rm dom}\; #1}
\def\image#1{{\rm Im}\; #1}
\def\closure#1{{\rm cl}(#1)}
\def\interior#1{{\overset{\circ}{#1}}}
\def\support#1{{\rm supp}\;#1}
\def\maps{\rightarrow}
\def\mapsnamed#1{\stackrel{#1}{\longrightarrow}}
\def\Set#1{\left\{#1\right\}}
\def\SetDef#1#2{\left\{#1 \mid #2\right\}}

\def\Frac(#1/#2){\left(\frac{#1}{#2}\right)}
\def\lsim{\stackrel{<}{\sim}}
\def\gsim{\stackrel{>}{\sim}}
\def\defas{\stackrel{\rm def}{=}}
\def\bop{\mathchoice{{\scriptstyle\circ}}{{\scriptstyle\circ}}{{\scriptscriptstyle\circ}}{\circ}}

\def\ad{{\rm ad}}
\def\Ad{{\rm Ad}}
\def\Tr{{\rm Tr}}
\def\Tp#1{\,{}^t\! #1}
\def\TS#1#2{\mbox{\bf T}^{#1}_{#2}}
\def\rank{{\rm rank}}
\def\Hom{{\rm Hom}}
\def\End{{\rm End}}
\def\Aut{{\rm Aut}}
\def\kernel#1{{\rm Ker}\; #1}
\def\Ker{{\rm Ker}}
\def\Im{{\rm Im}}
\def\cokernel#1{{\rm Coker}\; #1}
\def\sdsum{\mathrel{\hbox{$\tilde{+}$}}}
\def\dsum{\mathrel{\hbox{$\dot{+}$}}}
\def\sdprod{\mathrel{\hbox{$\stackrel{\sim}{\times}$}}}
\def\rsdp{\rtimes}
\def\lsdp{\ltimes}
\def\Otimes{\mathop{\otimes}\limits}
\def\Wedge{\mathop{\wedge}\limits}
\def\lact{\rhd}
\def\ract{\lhd}

\def\BG#1{\stackrel{0}{#1}{}\!\!}
\def\LPart#1{\stackrel{1}{#1}{}\!\!}
\def\Order#1{\r{O}\!\left(#1\right)}
\def\Tdot#1{{{#1}^{\hbox{.}}}}
\def\Tddot#1{{{#1}^{\hbox{..}}}}
\def\Tdddot#1{{{#1}^{\hbox{...}}}}
\def\dddot#1{\stackrel{...}{#1}{}\!\!}

\def\VNA{W^*}
\def\WOT{{\rm WOT}}
\def\SOT{{\rm SOT}}
\def\spectrum#1{{\rm spec}(#1)}

\def\Lie{\hbox{\rlap{$\cal L$}$-$}}
\def\vfield{\hbox{\rlap{$\cal X$}$-$}}
\def\DiffG#1{\r{Diff}_0(#1)}
\def\Hodge{{}*\!}
\def\dual{{}\,*\,\! }
\def\chiral{{}^+\! }
\def\Ricci{{\r Ricci}}
\def\Isom{{\rm Isom}}
\def\RP{{\RF P}}
\def\CP{{\CF P}}
\def\dS{{\rm dS}}
\def\AdS{{\rm AdS}}

\def\Eq#1{\begin{equation} #1 \end{equation}}
\def\Eqn#1{\Eq{#1 \nonumber}}
\def\Eqr#1{\begin{eqnarray} #1 \end{eqnarray}}
\def\Eqrn#1{\begin{eqnarray*} #1 \end{eqnarray*}}
\def\Eqrsub#1{\begin{subequations}\Eqr{#1}\end{subequations}}
\def\Eqrsubl#1#2{\begin{subequations}\label{#1}\Eqr{#2}\end{subequations}}
\def\leq#1{\begin{flalign} #1 \end{flalign} }
\def\Bitm{\begin{itemize}}
\def\Eitm{\end{itemize}}
\def\Blist#1#2{\begin{list}{#1}{\parsep=0pt \itemsep=0pt%
  \listparindent=0pt #2}}
\def\Elist{\end{list}}
\def\ignore#1#2{\def\ignoreflag{#1}\long\def\tmptext{#2}
  \ifnum\ignoreflag>1 #2 \fi}

\def\THB{{\mathbb T}}
\def\VHB{{\mathbb V}}
\def\SHB{{\mathbb S}}

\subsection{Gravitational perturbations of D-dimensional black holes}


The Newman-Penrose tetrad approach to the separation of variables given in the previous subsection was efficient for $D=4$ black holes. The higher-dimensional cases were not decoupled until the recent papers of Ishibashi and Kodama \cite{Kodama:2003jz,Ishibashi:2003ap,Kodama:2003kk}, although the $D$-dimensional generalizations of the Schwarzschild and Kerr solutions, called the Tangherlini \cite{Tangherlini:1963bw} and Myers-Perry \cite{Myers-Perry} solutions, respectively, have been known for a long time.

For the purposes of string theory and higher-dimensional gravity mentioned in the introduction, we considered a general class of the higher-dimensional black holes, whose space-time can be represented as the direct sum of two-dimensional space-time $(r, t)$ and $(D-2)$-dimensional space $ \M^D\approx \N^2 \times \K^{D-2}$. The corresponding metric has the form
\Eq{
ds^2=g_{ab}(y)dy^a dy^b + r^2(y)d\sigma_{(D-2)}^2,
\label{BG:metric}}
where $d\sigma_{D-2}^2=\gamma_{ij}(z)dz^i dz^j$ is a metric of the $(D-2)$-dimensional complete Einstein space $\K^{D-2}$,
$$\hat R_{ij}=(D-3)K \gamma_{ij}\qquad (K=0,\pm1),$$ and $g_{ab}$ is a static
metric of the two-dimensional space-time $\N^2$:
\Eq{
g_{ab}dy^ady^b= -f(r)dt^2+\frac{dr^2}{f(r)}.
\label{24-sec2}
}
%
In this subsection we assumed that the metric (\ref{BG:metric}) is a solution of the vacuum Einstein equations with the cosmological constant $\Lambda$. Therefore $f(r)$ can be written in the form \cite{Tangherlini:1963bw}
\Eq{
f(r)=K - \frac{2M}{r^{D-3}} -\frac{2 \Lambda  r^2}{(D-2)(D-1)}.
\label{f:neutral}
}

For $K=1$, we have a regular black hole if
$$\frac{2 \Lambda}{(D-2)(D-1)} M^{2/(D-3)}<(D-3)/(D-1)^{(D-1)/(D-3)}.$$
For $K=0$ or $-1$, $\K^{D-2}$ may not be compact, and the space-time contains a regular black hole only for $\Lambda<0$.

We now start the analysis of linear perturbations of the above static D-dimensional space-times (\ref{BG:metric}). The convenient perturbation variables are
$$\psi_{\mu\nu}=h_{\mu\nu}-h g_{\mu\nu}/2, \quad h_{\mu\nu}=\delta g_{\mu\nu}.$$
Then, the perturbed Einstein equations can be written in the form
\Eq{
-\nabla^2 \psi_{\mu\nu} -2R_{\mu\alpha\nu\beta}\psi^{\alpha\beta}
  +2\nabla_{(\mu}\nabla^\alpha \psi_{\nu)\alpha}
  -\nabla^\alpha\nabla^\beta\psi_{\alpha\beta} g_{\mu\nu}=0.
\label{PerturbationEq:general}}
These equations are invariant under the gauge transformations
\Eq{
x^\mu \maps x^\mu + \xi^\mu\quad \quad
\delta h_{\mu\nu}= -\nabla_\mu\xi_\nu -\nabla_\nu\xi_\mu.
}
In order to extract the dynamics of the physical degrees of freedom from the above perturbation equations (\ref{PerturbationEq:general}) the gauge freedom must be eliminated.

The main technical challenge of the separation of variables problem for gravitational perturbations is that Eq.~\eqref{PerturbationEq:general} is a set of coupled equations with $D(D+1)/2$ unknown functions.
For any metric that can be written in the form \eqref{BG:metric}, the separation of variables is possible with the help of the following tensorial decomposition of $h_{\mu\nu}$. According to the tensorial behavior on $\K^{D-2}$, $h_{\mu\nu}$ can be divided into the following components $h_{ab}$, $h_{ai}$ and $h_{ij}$. Here letters $a$, $b$, and $c$ are used for the radial and time coordinates, while coordinates on $\K^{D-2}$ are denoted by $i,j\ldots$ Then, the vector and tensor components can be decomposed as follows \cite{Ishibashi-Kodama-gauge-invarinat}
\Eqr{
&& h_{ai}=\hat D_i \Psi_a + \Psi_{ai};\quad \hat D^i\Psi_{ai}=0,\\
&& h_{ij}=\Psi_L\gamma_{ij}+\Psi_{Tij}; \quad \Psi^j_{Tj}=0,\\
&& h_{Tij}=\left(  \hat D_i\hat D_j
      -\frac{1}{n}\gamma_{ij}\hat\triangle\right)\Psi_{T}
   +2\hat D_{(i}\Psi_{T j)}+\Psi_{T ij};\notag\\
&&   \hat D^j \Psi_{T j}=0,\
   \hat D^j \Psi_{T ij}=0,
}
where $\hat D_i$ is the covariant derivative with respect to the metric $\gamma_{ij}$ on $\K^{D-2}$, $\hat\triangle=\hat D\cdot\hat D$. In this way one can obtain the following three groups of variables, which are classified by their transformation law with respect to the coordinate transformations on $\K^{D-2}$:
\Blist{$\bullet$}{}
\item { scalar-type} variables       $\Psi_{ab}$, $\Psi_a$, $\Psi_L$, and $\Psi_T$;
\item { vector-type} variables       $\Psi_{ai}$ and $\Psi_{Ti}$;
\item and a{ tensor-type} variable   $\Psi_{Tij}$.
\Elist
The Einstein equations written in this form split into three subsets each of which contains only variables belonging to one of the above three sets of variables. Then, the gauge-invariant variables can be constructed from the harmonic expansion coefficients. Through this procedure the Einstein equations can be transformed to a set of gauge-invariant equations with a small number of unknown functions \cite{Kodama:2003jz}. In particular, for tensor perturbations, this procedure gives a single wave equation for a single unknown function (\ref{wave1}).

When $D=4$ we have a particular case of the general approach. The scalar type of gravitational perturbation was called ``polar'' by Chandrasekhar  \cite{Chandrabook}. The polar (scalar) perturbations are responsible for deformations of the black hole horizon. The vector type is known as ``axial'' in four dimensions. This type of perturbation is connected with small rotations of the black hole. The tensor perturbations are not dynamical and can be gauged off in four dimensions. Note that it is important not to confuse the scalar type of gravitational perturbations (sometimes simply called ``scalar perturbations'') with perturbations of the test scalar field (\ref{scalar_covariant}).

}

\section{Methods for quasinormal modes calculations}\label{sec:methods}

\subsection{Definitions and main properties of the quasinormal modes}\label{sec:defQNMs}

Quasinormal modes are solutions of the wave equation (\ref{wave1}), satisfying specific boundary conditions at the black hole horizon and far from the black hole. At the event horizon this boundary condition is a requirement of the pure ingoing waves
\begin{equation}\label{BC1}
\Psi \sim pure~ingoing~wave, \quad r_* \rightarrow -\infty.
\end{equation}
Another boundary condition, imposed at spatial infinity or at the de Sitter horizon for asymptotically de Sitter black holes, is different in the astrophysical and string theory contexts. For astrophysical purposes one has
\begin{equation}\label{BC2}
\Psi \sim pure~outgoing~wave, \quad r_* \rightarrow +\infty.
\end{equation}
In string theory the boundary condition at infinity depends on the perturbed field under consideration. For the simplest scalar field the wave function $\Psi$ must vanish at infinity, while for higher spin fields some gauge-invariant combination of field components, dictated by the AdS/CFT correspondence \cite{Son:2007vk}, must vanish. Thus in string theory the boundary condition at infinity is the Dirichlet one
\begin{eqnarray}\label{BC3}
\Psi \rightarrow 0, &\quad& r \rightarrow \infty \quad (s=0),\\\nonumber
gauge~inv.~comb. \rightarrow 0, &\quad& r \rightarrow \infty \quad (higher~s).
\end{eqnarray}

Remembering that $\Psi \sim e^{-i \omega t}$, we write the QNM frequencies in the following form:
\begin{equation}\label{choise-omega}
\omega = \omega_{Re} - i \omega_{Im}.
\end{equation}
Here $\omega_{Re}$ is the real oscillation frequency of the mode and $\omega_{Im}$ is proportional to its damping rate. Positive $\omega_{Im}$ means that $\Psi$ is damped, negative $\omega_{Im}$ means an instability.

We now review the main properties of the quasinormal modes of black holes.
\begin{enumerate}
\item Quasinormal modes of non-AdS black holes do not form a complete set. Therefore, the signal cannot be represented as a sum over the quasinormal modes all of the time. Thus, at asymptotically late time, $t \rightarrow \infty$, the power law or exponential tails usually dominate in a signal (see Fig.~\ref{time-domain-profile}). An exception to this rule is the asymptotically AdS black holes, for which, due to their special boundary conditions, quasinormal modes dominate at all time \cite{Horowitz:1999jd}.

\item The quasinormal frequencies do not depend on a way by which the black hole or a field around it was perturbed. Thus, quasinormal modes are completely determined by a black hole's parameters and were called, therefore, ``fingerprints'' or ``footprints'' of black holes.

\item For Kerr black holes as well as for other astrophysical or string theory motivated cases, quasinormal modes form a countable set of discrete frequencies

\item Quasinormal modes calculated in the linear approximation are in good agreement with those obtained by the fully nonlinear integration of the Einstein equations, at least at sufficiently late time \cite{Barreto:2004fn,Yoshino:2006kc}.
\end{enumerate}

Practically, it is important to calculate QNMs with a {\it high accuracy} because considerable changing of black hole parameters frequently changes quasinormal frequency just by a few percent. Therefore, numerical and semianalytical methods for solving the quasinormal eigenvalue problem with high accuracy have gained considerable attention during the past years.

\subsection{The Mashhoon method: Approximation with the P\"oschl-Teller potential}\label{sec:PoschTeller}

Here we relate possibly the easiest method for calculation of the quasinormal modes, which is also illustrative as to the physical essence of the problem. This method was suggested by B. Mashhoon \cite{Mashhoon}.

We start from the usual wavelike equation, with an effective potential, which depends on some parameter $\alpha$:
\begin{equation}
\frac{d^2\Psi}{dr_*^2}+ (\omega^2- V(r_{*}, \alpha))) \Psi= 0.
\end{equation}
According to (\ref{BC1}, \ref{BC2}) the astrophysically relevant quasinormal boundary conditions for asymptotically flat black holes are
\begin{equation}\label{BC4}
\Psi  \sim  e^{ \pm i \omega r_{*}},  \quad  r_{*} \quad \rightarrow \pm \infty.
\end{equation}

Because of ``symmetric'' boundary conditions for the QNM problem  at both infinities $r_{*} \rightarrow \pm \infty$, it is reasonable to consider transformations $r_{*} \rightarrow - i r_{*}$ and $p \rightarrow p^{\prime}$, such that the potential is invariant under these transformations
\begin{equation}
V(r_{*}, \alpha ) = V(- i r_{*}, \alpha^{\prime} ).
\end{equation}
The wave function $\Psi$ and the quasinormal frequency  $\omega$ transform as
\begin{equation}
\Psi(r_{*}, \alpha) = \Phi(- i r_{*}, \alpha^{\prime}), \quad \omega(\alpha) = \Omega(\alpha^{\prime})
\end{equation}

Then, the wave equation for $\Phi$ and the boundary conditions will read
\begin{equation}
\frac{d^2\Phi}{dr_*^2}+ (-\Omega^2 + V) \Phi= 0, \quad \Phi \sim e^{\mp \Omega r_{*}}, \quad  r_{*} \rightarrow \mp \infty.
\end{equation}
These boundary conditions correspond to a vanishing wave function at the boundaries, so that the QNM problem is now reduced to the bound states problem for an inverse potential $ V \rightarrow - V$, which is smooth potential gap, approaching some constant values at the infinite boundaries. This gap can be approximated by a potential for which we know the analytic solution of the wave equation: It is the P\"oschl-Teller potential \cite{Mashhoon},
\begin{equation}
 V_{PT} = \frac{V_{0}}{\cosh^{2} \alpha (r_{*} - r_{*}^{0})}.
\end{equation}
Here $V_{0}$ is the height of the effective potential and $- 2V_{0} \alpha^2$ is the curvature of the potential at its maximum. The bound states of the P\"oschl-Teller potential are well known
\cite{Quanti}
\begin{equation}
\Omega = \alpha^{\prime}\left(-\left(n+\frac{1}{2}\right) + \left(\frac{1}{4} + \frac{V_{0}}{(\alpha^{\prime})^2} \right)^{1/2} \right).
\end{equation}
The quasinormal modes $\omega$ can be obtained from the inverse transformation $\alpha^{\prime} = i \alpha$,
\begin{equation}\label{PTformula}
\omega = \pm \sqrt{V_{0} - \frac{1}{4} \alpha^2 } - \imo \alpha \left(n + \frac{1}{2} \right), \quad n=0, 1, 2, \ldots
\end{equation}

Technically one has to fit a given black hole potential to the inverted P\"oschl-Teller potential. In the forthcoming sections we will see that for the low-lying QNMs, in the majority of cases the behavior of the effective potential is essential only in some region near the black hole, so that the fit of the height of the effective potential and of its curvature is indeed sufficient for estimation of quasinormal frequencies.

This method gives quite accurate results for the regime of high multipole numbers $\ell$, i.e. for the eikonal (geometrical optics) approximation. In particular, for gravitational perturbations of the $D=4$ Schwarzschild black holes, the fundamental quasinormal modes obtained by the Mashhoon method have a relative error of not more than 2\% for the lowest multipole $\ell =2$, and of about fractions of a percent for higher multipoles.

There are cases when the effective potential of a black hole is \emph{exactly} the P\"oschl-Teller potential. These are Schwarzschild-de Sitter and Reissner-Nordstr\"om-de Sitter black holes with extremal value of the $\Lambda$-term \cite{Cardoso:2003sw,Molina:2003ff}. After some modifications the Mashhoon method can be used for rotating black holes \cite{Mashhoon2}. Other exactly solvable potentials and their applications to the analysis of the QN spectra were reviewed in \cite{Boonserm-Visser}.

\subsection{Chandrasekhar-Detweiler and shooting methods}

By the substitution
\begin{equation}
\Psi = exp \left(i \int^{r_{*}} \Phi d r_{*}  \right),
\end{equation}
the wave equation (\ref{wave1}) can be reduced to the Riccati equation
\begin{equation}
i d\Phi/dr_{*} + \omega^{2} - \Phi^{2} - V(r_{*}) = 0.
\end{equation}

The quasinormal boundary conditions (\ref{BC4}) give
\begin{eqnarray}
\Phi &\rightarrow& + \omega, \quad r_{*} \rightarrow + \infty,\\
\Phi &\rightarrow& - \omega, \quad r_{*} \rightarrow - \infty.
\end{eqnarray}

Then in order to obtain QNMs one needs to integrate the Riccati equation numerically \cite{Chandrasekhar:1975zz}. The easiest way is to ``shoot'' the two asymptotic solutions at the horizon and in the far region at some common intermediate point, which is usually the peak of the potential barrier. The
shooting procedure is the following. We take some fixed (complex) $\omega$ with $\Psi \propto e^{i \omega r_{*}}$ and integrate from infinity backward to some intermediate value of $r_{*} = r_{*}^{0}$. Then, for the same $\omega$ we integrate the wave equation with  $\Psi \propto e^{-i \omega r_{*}}$ from the horizon until the chosen intermediate value of $r_{*}^{0}$. In the common point, the Wronskian of the two solutions must vanish, which gives the equation for the quasinormal modes \cite{Chandrasekhar:1975zz}. This shooting scheme, when applied to the Riccati equation, works quite well for lower modes, yet, there
is a technical difficulty in this procedure when one uses it for the usual second-order wave equation (\ref{wave1}). For large positive $r_{*}$, the solution $\Psi$ with asymptotic $\Psi \propto e^{i \omega r_{*}}$ gets contaminated by the admixture of waves $\propto e^{- i \omega r_{*}}$ and vice versa; for large negative $r_{*}$ the solution with asymptotic $\Psi \propto e^{- i \omega r_{*}}$ gets admixture with waves $\propto e^{+ i \omega r_{*}}$.

Since for the AdS case the asymptotical behavior of the function is not exponential it is possible to adopt the shooting method for the second-order differential equation. Namely, we integrate Eq.~(\ref{wave1}), starting from the event horizon, where the quasinormal boundary condition is applied for a fixed value of $\omega$. At large value of $r$ we find a fit of the function $\Psi$ by the two linearly independent solutions, in which the series expansions can be found analytically up to any order at spatial infinity. One of these solutions $\Psi_D$ vanishes at infinity, satisfying the boundary condition inspired by string theory, while the other solution $\Psi_N$ grows. The fitting procedure allows us to find the contributions for each of these solutions for the particular value of $\omega$:
$$\Psi(r)\approx C_D(\omega)\Psi_D(r)+C_N(\omega)\Psi_N(r), \quad r\gg r_+.$$

The requirement of vanishing of the function $\Psi$ at spatial infinity implies that $C_N(\omega)=0$, if the frequency $\omega$ is a quasinormal mode. Because of growing of the function $\Psi_N(r)$, the fitting procedure provides good accuracy for the corresponding coefficient and allows us to find accurate values of quasinormal modes by minimizing $C_N(\omega)$.

For the asymptotically flat and de Sitter cases the exponentially growing solution satisfies the quasinormal boundary conditions, while the other solution decays exponentially for large $r$. The exponentially decaying contribution is difficult to calculate. That is why the shooting method for the second-order equation does not allow us to find quasinormal modes for stable non-AdS backgrounds.

Yet, if the quasinormal spectrum contains a growing mode ($\omega_{Im}<0$), the corresponding eigenfunction $\Psi$ must be zero at the spatial infinity. Therefore, we are able to use the shooting method to check stability of black holes. In order to do this one can check for all possible values of $\omega$, for which the corresponding $\Psi$ does not vanish at spatial infinity (or at the cosmological horizon).

In practice, we consider only the parametric region of $\omega$, where we were unable to prove analytically that the growing mode is not consistent with the equation and the quasinormal boundary conditions. In this region we calculate the prefactor for the growing solution. If the prefactor is zero for some value of $\omega$, this is a growing quasinormal mode.

\subsection{WKB method}\label{sec:WKBmethod}

\begin{figure}
\resizebox{\linewidth}{!}{\includegraphics*{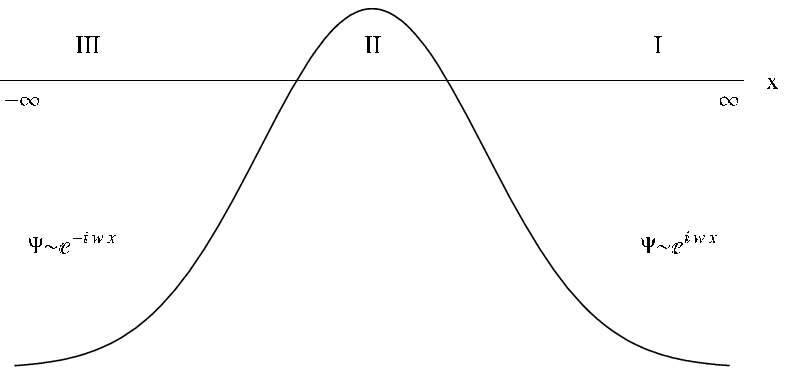}}
\caption{The three regions separated by the two turning points $Q(x)=0$.}\label{WKBMethod}
\end{figure}

Here we describe the semianalytic technique for finding low-lying quasinormal modes, based on the JWKB approximation. The method was first applied by Schutz and Will to the problem of scattering around black holes \cite{Will_Schutz}. This method can be used for effective potentials which have the form of a potential barrier and take constant values at the event horizon and spatial infinity (see Fig.~\ref{WKBMethod}). The method is based on matching of the asymptotic WKB solutions at spatial infinity and the event horizon with the Taylor expansion near the top of the potential barrier through the two turning points.

We rewrite the wave equation (\ref{wave1}) in a slightly different form
\begin{equation}\label{WKBwave}
\frac{d^2 \Psi}{d x^2} + Q(x) \Psi(x) = 0.
\end{equation}
Here we used the new designations $x = r_{*}$ and $Q(x) = \omega^2 - V$. The asymptotic WKB expansion at both infinities has the following general form:
\begin{equation}\label{WKBexpansion}
\Psi \sim exp\left(\sum_{n=0}^{\infty} \frac{S_{n}(x) \epsilon^n}{\epsilon} \right).
\end{equation}
We introduced the WKB parameter $\epsilon$ in order to track orders of the WKB expansion.
Substituting the expansion (\ref{WKBexpansion}) into the wave equation (\ref{WKBwave}) and equating the same powers of $\epsilon$, we find
\begin{equation}\label{s0}
S_{0}(x) = \pm i \int^{x} Q(\eta)^{1/2} d \eta,
\end{equation}
\begin{equation}\label{s1}
S_{1}(x) = -\frac{1}{4} \ln Q(x).
\end{equation}
The two choices of sign in (\ref{s0}) correspond to either incoming or outgoing waves at either of the infinities $x = \pm \infty$. Thus, when $x \rightarrow + \infty$ (region I of Fig.~\ref{WKBMethod}), $Q(x) \rightarrow \omega^2$ in the dominant order, so that $S_{0} \rightarrow +i \omega x$ for the outgoing to the infinity wave and $S_{0} \rightarrow -i \omega x$ for the incoming from infinity wave. In a similar fashion, at the event horizon $x \rightarrow - \infty$ (region III), $S_{0} \rightarrow +i \omega x$ is for wave incoming from the event horizon, while $S_{0} \rightarrow -i \omega x$ is for a wave outgoing to the event horizon. We designate these four solutions as $\Psi^{I}_{+}$, $\Psi^{I}_{-}$, $\Psi^{III}_{+}$ and $\Psi^{III}_{-}$ respectively for plus and minus signs in $S_{0}$ in regions I and III. Thus,
\begin{equation}
\Psi^{I}_{+} \sim e^{+ i \omega x}, \quad  \Psi^{I}_{-} \sim e^{- i \omega x}, \quad x \rightarrow + \infty,
\end{equation}
\begin{equation}
\Psi^{III}_{+} \sim e^{+ i \omega x}, \quad  \Psi^{III}_{-} \sim e^{- i \omega x}, \quad x \rightarrow - \infty.
\end{equation}
The general solutions in regions I and III are
\begin{equation}\label{solution_I}
\Psi \sim Z_{in}^{I} \Psi^{I}_{-} + Z_{out}^{I} \Psi^{I}_{+}, \quad region \quad I,
\end{equation}
\begin{equation}\label{solution_III}
\Psi \sim Z_{in}^{III} \Psi^{III}_{+} + Z_{out}^{III} \Psi^{III}_{-}, \quad region \quad III.
\end{equation}
The amplitudes at $+\infty$ are connected with the amplitudes at $-\infty$ through the linear matrix relation
\begin{equation}\label{Mmatrix}
\left(\begin{array}{c}
Z_{out}^{III}\\
Z_{in}^{III}
\end{array}\right)= \left(\begin{array}{cc}S_{11} & S_{12}\\ S_{21} & S_{22}\end{array}\right) \left(\begin{array}{c}Z_{out}^{I}\\  Z_{in}^{I}\end{array}\right).
\end{equation}
Now we need to match both WKB solutions of the form (\ref{WKBexpansion}) in regions I and III with a solution in region II through the two turning points $Q(x) = 0$.

If the turning points are closely spaced, i.e. if $-Q(x)_{max} \ll Q(\pm \infty)$, then the solution in region II can be well approximated by the Taylor series
\begin{equation}\label{Taylor}
Q(x) = Q_{0} + \frac{1}{2} Q_{0}^{\prime \prime}(x- x_{0})^{2} + O((x- x_{0})^{3}),
\end{equation}
where $x_{0}$ is the point of maximum of the function $Q(x)$, $Q_{0} = Q(x_{0})$,  and $Q_{0}^{\prime \prime}$ is the second derivative with respect to $x$ at the point $x = x_{0}$. Region II corresponds to
\begin{equation}\label{regionII}
| x - x_{0} |  < \sqrt{\frac{-2 Q_{0}}{Q_{0}^{\prime \prime}}} \approx \epsilon^{1/2}, \quad region \quad II.
\end{equation}
The latter relation gives also the region of validity of the WKB approximation: $\epsilon$ must be small.

We introduce new functions
\begin{equation}\label{newfunctions1}
k = \frac{1}{2} Q_{0}^{\prime \prime}, \quad  t = (4 k)^{1/4} e^{i \pi/4} (x - x_{0}),
\end{equation}
\begin{equation}\label{newfunctions2}
\nu + \frac{1}{2} = - i Q_{0}/(2 Q_{0}^{\prime \prime})^{1/2}.
\end{equation}
Then, the wave equation (\ref{WKBwave}) takes the form
\begin{equation}\label{parabolic}
\frac{d^{2}\Psi}{d t^{2}} + \left(\nu + \frac{1}{2} - \frac{1}{4} t^2 \right)\Psi = 0.
\end{equation}
The general solution of this equation can be expressed in terms of parabolic cylinder functions $D_{\nu}(t)$,
\begin{equation}\label{solution_parabolic}
\Psi = A D_{\nu}(t) + B D_{- \nu -1}(i t).
\end{equation}
Large $| t |$ asymptotics of this solution are
$$ \Psi \approx B e^{\frac{- 3 i \pi (\nu + 1)}{4}} (4 k)^{-\frac{\nu + 1}{4}} (x-x_{0})^{-(\nu +1)} e^{\frac{i k^{1/2} (x-x_0)^{2}}{2}}$$
$$+ (A + B (2 \pi)^{1/2 e^{- i \nu \pi/2}}/\Gamma(\nu + 1)) \times$$
\begin{equation}\label{large_t_sol1}
 e^{\frac{i \pi \nu}{4}} (4 k)^{\frac{\nu}{4}} (x-x_{0})^{\nu}
e^{\frac{-i k^{\frac{1}{2}} (x-x_0)^{2}}{2}}, \quad x \gg x_2,
\end{equation}
$$ \Psi \approx A e^{- 3 i \pi \nu /4} (4 k)^{\nu/4} (x-x_{0})^{\nu} e^{-i k^{1/2} (x-x_0)^{2}/2} + $$
$$ \left(\frac{B - i A(2 \pi)^{1/2} e^{- i \nu \pi/2}}{\Gamma(-\nu)}\right)
e^{\frac{i \pi (\nu + 1)}{4}} (4 k)^{\frac{-(\nu + 1)}{4}} \times$$
\begin{equation}\label{large_t_sol2}
(x-x_{0})^{-(\nu + 1)} e^{i k^{1/2} (x-x_0)^{2}/2}, \quad x \ll x_1
\end{equation}

Equating the corresponding coefficients in (\ref{large_t_sol1}),
(\ref{large_t_sol2})  and eliminating $A$ and $B$, we obtain the
elements of $S$ matrix,
\begin{equation}\label{Mmatrix2}
\left(\begin{array}{c}
Z_{out}^{III}\\
Z_{in}^{III}
\end{array}\right)= \left(\begin{array}{cc} e^{i \pi \nu} & \frac{i R^2  e^{i \pi \nu} (2 \pi)^{1/2}}{\Gamma(\nu +1)} \\
\frac{R^{-2} (2 \pi)^{1/2}}{\Gamma(-\nu)} & - e^{i \pi \nu}  \end{array}\right) \left(\begin{array}{c}Z_{out}^{I}\\  Z_{in}^{I}\end{array}\right),
\end{equation}
where
\begin{equation}
R = (\nu +1)^{(\nu + 1/2)/2} e^{-(\nu + 1/2)/2}.
\end{equation}
When expanding to higher WKB orders, $S$ matrix has the same general form (\ref{Mmatrix2}), though with modified expression for $R$, which still depends only on $\nu$. We note that for a black hole there are no waves ``reflected by the horizon'', so that $Z_{in}^{III} = 0$, and due to quasinormal mode boundary conditions, there are no waves coming from infinity, i.~e., $Z_{in}^{I} = 0$. Both these conditions are satisfied by (\ref{Mmatrix2}), only if
\begin{equation}
\Gamma(-\nu) = \infty,
\end{equation}
and, consequently, $\nu$ must be an integer. Then, from the relation  (\ref{newfunctions2}) we find
\begin{equation}\label{QNM_WKB1}
n + \frac{1}{2} = - i Q_{0}/(2 Q_{0}^{\prime \prime})^{1/2}, \quad n = 0, 1, 2,...
\end{equation}
The latter relation gives us the complex quasinormal modes labeled by an overtone number $n$ at the first WKB order \cite{Will_Schutz}. Later this approach was extended to the third WKB order beyond the eikonal approximation by Iyer and Will \cite{WKBorder1} and to the sixth order by Konoplya \cite{Konoplya:2003dd,Konoplya:2004ip}. In order to extend the WKB formula to higher orders, it is sufficient to take higher orders in the $\epsilon$ WKB series (\ref{WKBexpansion}) and to take an appropriate number of consequent terms in the Taylor expansion (\ref{Taylor}). Since the $S$-matrix (\ref{Mmatrix}) depends only on $\nu$, its elements $S_{i j}$  can be found simply by solving the interior problem in region II at higher orders in $\epsilon$ \cite{WKBorder1}, and  without explicit matching of the interior solution with WKB solutions in regions I and III at all the same orders.

Going over from $Q$ to the effective potential $V$, the sixth order WKB formula has the form \cite{Konoplya:2003dd}
\begin{equation}\label{QNM_WKB6}
\frac{i (\omega^2 - V_{0})}{\sqrt{- 2 V_{0}^{\prime \prime}}} - \sum_{i=2}^{6}\Lambda_i - = n + \frac{1}{2}, \quad n = 0, 1, 2,\ldots
\end{equation}
where the correction terms $\Lambda_i$ depend on the value of the effective potential and its derivatives (up to the $i-$th order) in the maximum. The explicit form of the WKB corrections can be found in \cite{WKBorder1} ($\Lambda_2$, $\Lambda_3$) and in \cite{Konoplya:2003ii} ($\Lambda_4$, $\Lambda_5$, $\Lambda_6$).

In addition to solving the quasinormal mode problem, the $S$-matrix allows us to solve the standard scattering problem, which describes tunneling of waves and particles through the potential barrier of a black hole. One can easily check from Eq.~(\ref{Mmatrix2}) that for real $Q(x)$ (for real energy of the incident wave and spherically symmetric backgrounds)
\begin{equation}
S_{11}^{*} = S_{22}, \quad S_{12} = S_{21}^{*}, \quad |S_{21}|^{2} - |S_{11}|^{2} =1.
\end{equation}
The transmission coefficient is
\begin{equation}
T = \frac{|Z_{out}^{III}|^{2}}{|Z_{in}^{I}|^{2}} = S_{21}^{-1},
\end{equation}
and the reflection coefficient is $R = 1 - T$.

It was shown in \cite{Konoplya:2003dd,Konoplya:2004ip} that the WKB formula, extended to the sixth order, gives the relative error about 2 orders less than that of the third WKB order. Strictly speaking the WKB series converge only asymptotically, so that the consequent decreasing of the relative error in each WKB order is not guaranteed. Therefore it is reasonable to develop a modified WKB technique in the so-called optimal order \cite{Froman}. The latter gives better results for moderately higher overtones $n$ and especially when $n > \ell$. Yet, in many cases when $n \leq \ell$ the usual sixth order WKB formula gives much better results than the optimal order approach. For example, for the fundamental (odd parity) gravitational mode of the Schwarzschild black hole, the sixth WKB QNM $\omega M = 0.3736 - 0.0890 \imo$ is much closer to the accurate numerical value $\omega M = 0.3737 - 0.0890 \imo$ than the optimal (third) order value $\omega M = 0.3734 - 0.0891 \imo$ given by the phase-integral approach \cite{Froman}. This is in agreement with the general experience of various applications  of the WKB method: Extension to higher orders frequently gives better accuracy than expected \cite{WKB_book}.

In some cases the WKB approach related here needs modifications. For instance, when considering a massive scalar field in a black hole background, the effective potential has a local minimum far from a black hole. This local minimum induces two changes in the WKB procedure: First, there are three turning points which separate all space into four regions, so that three matchings are required \cite{Matukkin-Galtsov}. Second, an influent subdominant term in the asymptotic WKB expansion at spatial infinity (\ref{solution_III}) appears \cite{Ohashi:2004wr,Konoplya:2004wg}.

The higher-order WKB approach proved to be useful for finding lower overtones of the quasinormal spectrum \cite{Zhang:2003vb,Cornell:2005ux,Liu:2008mj,Konoplya:2009ig} and is in good agreement with accurate numerical data as was shown in \cite{Berti:2005eb,Yoshino:2005ps}. The higher-order WKB corrections lead also to better accuracy for the reflection and transmission coefficients \cite{Konoplya:2010kv}.

\subsection{Integration of the wavelike equations}\label{sec:time-domain-method}

We rewrite the wavelike equation (\ref{wave1}) \textit{without} implying the stationary ansatz ($\Psi \sim e^{-i \omega t}$)

\begin{equation}\label{wavelike}
\frac{\partial^2\Phi}{\partial t^2}-\frac{\partial^2\Phi}{\partial x^2}+V(t,x)\Phi=0,
\end{equation}
where $x$ is the tortoise coordinate. The technique of integration of the above wave equation in the time domain was developed by Gundlach, Price and Pullin \cite{Gundlach:1993tp}. We shall rewrite the wavelike equation (\ref{wavelike}) in terms of the so called light-cone coordinates $du = dt - dx$ and $dv = dt + dx$;
\begin{equation}\label{light-cone}
\left(4\frac{\partial^2}{\partial u\partial v}+V(u,v)\right)\Phi(u,v)=0.
\end{equation}
In these coordinates the operator of the time evolution is
\begin{eqnarray}\nonumber
&&\exp\left(h\frac{\partial}{\partial t}\right)=\exp\left(h\frac{\partial}{\partial u}+h\frac{\partial}{\partial v}\right)= \\ \nonumber
&&=\exp\left(h\frac{\partial}{\partial u}\right)+\exp\left(h\frac{\partial}{\partial v}\right) - 1 +
\\\nonumber &&+ \frac{h^2}{2}\left(\exp\left(h\frac{\partial}{\partial u}\right)+\exp\left(h\frac{\partial}{\partial v}\right)\right)\frac{\partial^2}{\partial u\partial v} + \mathcal{O}(h^4).
\end{eqnarray}
Acting by this operator on $\Phi$ and taking account of (\ref{light-cone}), one finds
\begin{eqnarray}
\Phi(N)= \Phi(W)+\Phi(E)-\Phi(S) -\nonumber\\
\frac{h^2}{8}V(S)\left(\Phi(W)+\Phi(E)\right) + \mathcal{O}(h^4),\label{integration-scheme}
\end{eqnarray}
where we introduced letters to mark the points as follows: $S=(u,v)$, $W=(u+h,v)$, $E=(u,v+h)$, and $N=(u+h,v+h)$.
\begin{figure}
\begin{center}
\resizebox{.7\linewidth}{!}{\includegraphics*{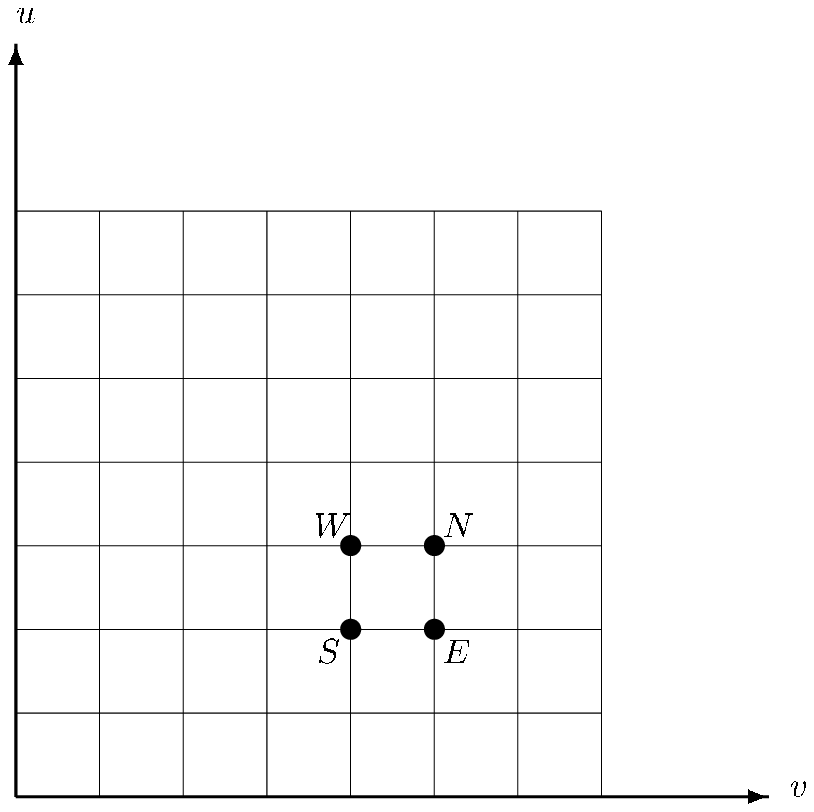}}
\rem{
\setlength{\unitlength}{.85cm}
\begin{picture}(10,10)
  \thicklines
  \put(0,0){\vector(0,1){9}}
  \put(0,0){\vector(1,0){9}}
  \put(0,9.25){$u$}
  \put(9.25,0){$v$}
  \thicklines
  \linethickness{0.08mm}
  \multiput(0,0)(1,0){8}{\line(0,1){7}}
  \multiput(0,0)(0,1){8}{\line(1,0){7}}
  \put(3.6,1.6){$S$}
  \put(3.56,3.1){$W$}
  \put(5.1,1.6){$E$}
  \put(5.1,3.1){$N$}
  \put(4,2){\circle*{0.25}}
  \put(4,3){\circle*{0.25}}
  \put(5,2){\circle*{0.25}}
  \put(5,3){\circle*{0.25}}
\end{picture}
}%
\end{center}
\caption{The integration grid. Each cell of the grid represents an integration step. The thick points illustrate the choice of ($S$, $W$, $E$, and $N$) for the particular step of the integration. The initial data are specified on the left and bottom sides of the rhombus.}\label{integration-square}
\end{figure}
Equation (\ref{integration-scheme}) allows us to calculate the values of $\Phi$ inside the rhombus, which is built on the two null-surfaces $u=u_0$ and $v=v_0$ (see Fig.~\ref{integration-square}), starting from the initial data specified on them. As a result we can find the time profile data $\{\Phi(t=t_0),\Phi(t=t_0+h),\Phi(t=t_0+2h),\ldots\}$ in each point of the rhombus. These values of the function can be used for calculations of quasinormal modes.

\begin{figure}
\resizebox{\linewidth}{!}{\includegraphics*{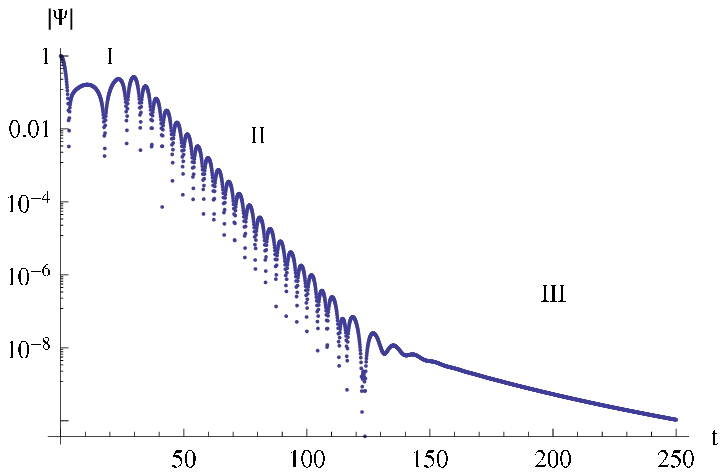}}
\caption{An example of a time-domain profile for the Schwarzschild black hole gravitational perturbations ($l=2$ vector type, in the point $r=11r_+$).}\label{time-domain-profile}
\end{figure}

In this way, one can obtain a time-domain profile of the perturbation, an example of which is shown in Fig.~\ref{time-domain-profile}. When looking at Fig.~\ref{time-domain-profile}, a natural question is to which value of the quasinormal frequency the above profile corresponds. The answer can be found with the help of the Prony method of fitting of the profile data by superposition of damped exponents (see, e.~g., \cite{Berti:2007dg})
\begin{equation}\label{damping-exponents}
\Phi(t)\simeq\sum_{i=1}^pC_ie^{-\imo\omega_i t}.
\end{equation}
We suppose that the quasinormal ringing epoch starts at $t_0=0$ and ends at $t=Nh$, where $N$ is an integer and $N\geq2p-1$. Then the formula (\ref{damping-exponents}) is valid for each value from the profile data:
\begin{equation}
x_n\equiv\Phi(nh)=\sum_{j=1}^pC_je^{-\imo\omega_j nh}=\sum_{j=1}^pC_jz_j^n.
\end{equation}
The Prony method allowed us to find $z_i$ in terms of known $x_n$ and, since $h$ was also known, to calculate the quasinormal frequencies $\omega_i$. In order to do this, we defined a polynomial function $A(z)$ as
\begin{equation}
A(z)=\prod_{j=1}^p(z-z_j)=\sum_{m=0}^{p}\alpha_m z^{p-m}, \qquad \alpha_0=1.
\end{equation}
Consider the following sum:
$$\sum_{m=0}^p\alpha_mx_{n-m}=\sum_{m=0}^p\alpha_m\sum_{j=1}^pC_jz_j^{n-m}=$$
$$\sum_{j=1}^pC_jz_j^{n-p}\sum_{m=0}^p\alpha_mz_j^{p-m}=\sum_{j=1}^pC_jz_j^{n-p}A(z_j)=0.$$
Since $\alpha_0=1$ we find
\begin{equation}\label{Prony-equation}
\sum_{m=1}^p\alpha_mx_{n-m}=-x_{n}.
\end{equation}
Substituting $n=p,...,N$ into Eq.~(\ref{Prony-equation}) we obtain $N-p+1\geq p$ linear equations for $p$ unknown coefficients $\alpha_m$.

We rewrite these equations in the matrix form
$$\left(\begin{array}{llll}
    x_{p-1} & x_{p-2} & \ldots & x_0 \\
    x_p & x_{p-1} & \ldots & x_1 \\
    \vdots & \vdots & \ddots & \vdots \\
    x_{N-1} & x_{N-2} & \ldots & x_{N-p} \\
  \end{array}\right)
\left(\begin{array}{c}
    \alpha_1 \\
    \alpha_2 \\
    \vdots \\
    \alpha_p\\
\end{array}\right)=-
\left(\begin{array}{c}
    x_p \\
    x_{p+1} \\
    \vdots \\
    x_N\\
\end{array}\right).
$$
Such a matrix equation
$$X\alpha=-x$$
can be solved in the least-squares sense
\begin{equation}
\alpha=-(X^+X)^{-1}X^+x,
\end{equation}
where $X^+$ denotes the Hermitian transposition of the matrix $X$.

Since the coefficients $\alpha_m$ of the polynomial function $A(z)$ are found, we can calculate numerically the roots $z_j$ of the polynomial and the quasinormal frequencies $\omega_j$:
$$\omega_j=\frac{\imo}{h}\ln(z_j).$$

While the above described integration scheme (\ref{integration-scheme}) was efficient for asymptotically flat or de-Sitter black holes, for asymptotically AdS black holes its convergence is too slow. For the latter case, an alternative integration scheme \cite{Wang:2004bv},
\begin{gather}
\left[ 1 + \frac{h^{2}}{16}V(S)\right] \Psi(N) =
\Psi(E)+ \Psi(W) - \Psi(S)\nonumber \\
-\frac{h^{2}}{16}
       \left[V(S)\Psi(S)+V(E)\Psi(E)+V(W)\Psi(W)\right], \,\,
\label{discrete2}
\end{gather}
is more stable.

\subsection{Fit and interpolation approaches}

For asymptotically flat (or de Sitter) black holes, the wavelike equation usually has the form (\ref{wave1}). The general solution of Eq.~(\ref{wave1}) at infinity can be written in the form
\begin{equation}
\Psi = A_{in} \psi_{in} + A_{out} \psi_{out}, \quad r_{*} \rightarrow \infty.
\end{equation}
The quasinormal modes, by definition, are the poles of the reflection coefficient $A_{out}/A_{in}$.  In astrophysical context one does not work with the true \textit{asymptotic} spatial infinity $r_{*} = \infty$, but a more practical notion of the ``far zone'' can be used instead. The far zone, or \textit{``astrophysical infinity''}, is separated from a black hole by a distance that is much larger than the black hole radius $r\gg r_+$. The basic intuitive physical idea of what we call here the ``fit and interpolation approaches'' is that \textit{scattering properties as well as quasinormal modes of black holes} (in an astrophysical context) \textit{depend only on the behavior of the effective potential in some region near the black hole and do not depend on behavior of the potential at infinity} \cite{Zhidenko:2005mv,fit-interpolation,Zhidenko:2007sj}. The immediate confirmation of this idea would is the WKB formula (\ref{QNM_WKB6}) or the Mashhoon formula for $\omega$ that depends only on derivatives of the effective potential in its maximum, i.~e., where the main process of scattering is localized. However, such a confirmation is almost trivial as it is based on the Taylor expansion of the effective potential near its maximum, which takes account of the behavior of the potential \emph{only} near its peak.

\begin{figure}
\resizebox{\linewidth}{!}{\includegraphics*{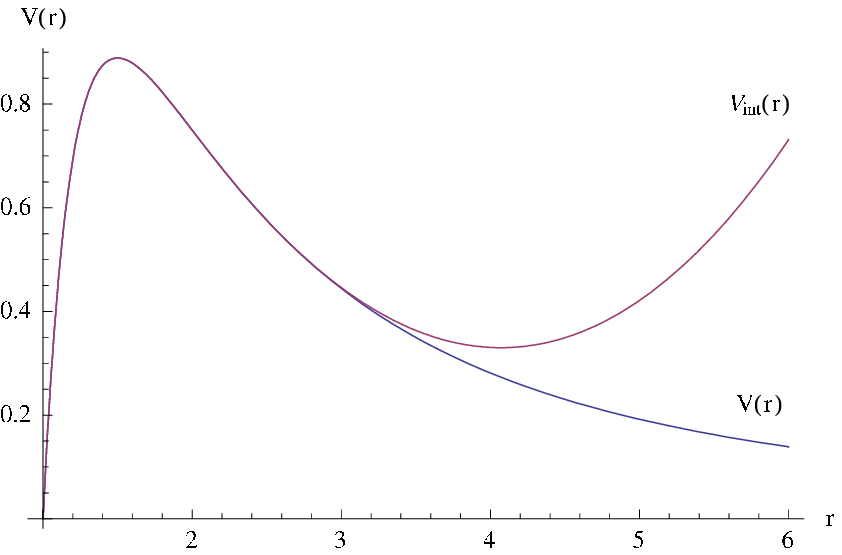}}
\caption{Potential  for electromagnetic perturbations near the Schwarzschild black hole ($r_+=1$, $\ell=2$) and the same potential interpolated numerically near its maximum. Despite the behavior of the two potentials being different in the full region of $r$, except for a small region near the black hole, low-lying quasinormal modes for both potentials are close.}\label{spot1}
\end{figure}

Nevertheless, one can check that the above statement about the dominance of the ``near zone'' in scattering is true by considering the well-known potential for the Schwarzschild black hole $V(r)$ and also two other potentials which lay perfectly close to the Schwarzschild potential near its maximum, but have strikingly different behavior far from the black hole. Then it can be shown that all three potentials, quite surprisingly, produce the same quasinormal modes.

The two alternatives to the Schwarzschild potential are chosen in the following way. We plotted the function $V(r)$ of the well-known analytical Schwarzschild potential (\ref{SBH-scalar}) and found numerical values of plot points of $V(r)$ near its peak, which served as a numerical data for our first alternative potential. We call it $V_{int}$, which is an {\it interpolation} of these points near the potential maximum by cubic splines (see Fig.~\ref{spot1}). The second potential, called $V_{fit}$, is a {\it fit} of the above plot's points near the maximum by a ratio of polynomial functions. From Fig.~\ref{spot1} one can see that far from the black hole, $V_{int}$ is indeed different from the Schwarzschild potential because it diverges at large $r$.

The method that is not ``trivial'' for our purpose is the time-domain integration described in Sec.~\ref{sec:time-domain-method}. Integrating the wave equation with the above two effective potentials in time domain and comparing the obtained QNMs with their WKB values one can show that indeed the quasinormal modes of all three potentials are the same up to the small numerical error \cite{fit-interpolation}.

The advantage of this approach is that one does not need to know the analytic form of the effective potential (or even of the background metric). Thus, it can be applied to a number of problems of astrophysical interest, where a black hole is surrounded by some distributions of matter or, for instance, for some alternative theories of gravity \cite{Jacobson}, where an exact solution for a black hole cannot be found. What one needs is only a numerical solution for the background metric {\it near} a black hole. Strictly speaking, the notion ``near'' depends on the distance at which the black hole effective potential reaches its maximum. Usually it is sufficient to know a numerical solution for the black hole metric in the region that starts from the event horizon and finishes at about $4-5$ black hole radii \cite{fit-interpolation}.

\subsection{Frobenius method}\label{sec:Frobenius}

Eq.~(\ref{wave1}) can be written in a slightly different form
\begin{equation}\label{radial}
\left(\frac{d^2}{dr^2}+p(r)\frac{d}{dr}+q(r)\right)R(r)=0,
\end{equation}
where the functions $p(r)$ and $q(r)$ depend on the eigenfrequency $\omega$.

We started from the analysis of singularities of Eq.~(\ref{radial}). There are two points which are always singular: the event horizon $r=r_+$ and the cosmological horizon (or spatial infinity) $r=r_\infty$. Usually, there are also other singular points, which depend on $p(r)$ and $q(r)$. By definition, quasinormal modes are eigenvalues $\omega$ which satisfy the boundary conditions corresponding to the outgoing wave at spatial infinity and the ingoing wave at the horizon. Thus, we are able to determine the function $\Psi(r)$ as a multiplication of a divergent (at these points) function by a series, which is
convergent in the region $r_+\leq r\leq r_\infty$. If $p(r)$ and $q(r)$ are rational functions of $r$, we can construct such a series in terms of the rational functions:
\begin{widetext}
\begin{equation}\label{Frobenius}
R(r)=\left\{
  \begin{array}{ll}
    \displaystyle\left(\frac{r-r_\infty}{r-r_0}\right)^{\imo\Omega}\left(\frac{r-r_+}{r-r_0}\right)^{-\imo a} \sum_{k=0}^\infty b_k\left(\frac{r-r_+}{r-r_0}\frac{r_\infty-r_0}{r_\infty-r_+}\right)^k, & r_\infty<\infty, \\
    \displaystyle e^{\imo\Omega r}(r-r_0)^{\sigma}\left(\frac{r-r_+}{r-r_0}\right)^{-\imo a} \sum_{k=0}^\infty b_k\left(\frac{r-r_+}{r-r_0}\right)^k,  & r_\infty=\infty.
  \end{array}
\right.
\end{equation}
\end{widetext}
$\Omega$, $\sigma$, and $a$ are defined in order to satisfy Eq.~(\ref{radial}) in the singular points $r=r_+$ and $r=r_\infty$. The quasinormal boundary conditions fix $\Re{\Omega}$ and $\Re{a}$, which must be chosen to be the same sign as $\Re{\omega}$.

The Frobenius series is
\begin{equation}\label{Frobenius_series}
u(z)=\sum_{k=0}^\infty b_k z^k, \qquad z=\frac{r-r_+}{r-r_0}\frac{r_\infty-r_0}{r_\infty-r_+}\,.
\end{equation}
If all the singular points of Eq.~(\ref{radial}) satisfy $|z|>1$, the series (\ref{Frobenius_series}) is convergent at $z=1$ ($r=r_\infty$) if and only if the value of $\omega$ is the eigenfrequency of Eq. (\ref{radial}). If there is at least one singular point inside the unit circle, one has to continue the Frobenius series (\ref{Frobenius_series}) through some midpoints (see Sec.~\ref{sec.midpoints}), in order to test the convergence at spatial infinity or the cosmological horizon.

Note that the definition of $z$ contains an arbitrary parameter $r_0<r_+$. In most cases, it can be chosen in order to move all of the singularities outside the unit circle.

\subsection{Method of continued fractions}

Substituting (\ref{Frobenius}) into (\ref{radial}), one can obtain an $N$-term recurrence relation for the coefficients $b_i$:
\begin{equation}\label{rrelation}
\sum_{j=0}^{min(N-1,i)} c_{j,i}^{(N)}(\omega)\,b_{i-j}=0,\quad
{\rm for}\,\,i>0\,,
\end{equation}
where the coefficients $c_{j,i}^{(N)}(\omega)$ ($0\leq j\leq min(N-1,i)$) depend on $\omega$.

We now decrease the number of terms in the recurrence relation
\begin{equation}\label{srcRE}
\sum_{j=0}^{min(k,i)}c_{j,i}^{(k+1)}(\omega)\,b_{i-j}=0
\end{equation}
by one, i.~e., we find $c_{j,i}^{(k)}(\omega)$ which satisfy the following equation
\begin{equation}\label{finRE}
\sum_{j=0}^{min(k-1,i)}c_{j,i}^{(k)}(\omega)\,b_{i-j}=0\,.
\end{equation}
For $i\geq k$ we can rewrite the above expression as
\begin{equation}\label{subsRE}
\frac{c_{k,i}^{(k+1)}(\omega)}{c_{k-1,i-1}^{(k)}(\omega)}
\sum_{j=1}^{k}c_{j-1,i-1}^{(k)}(\omega)\,b_{i-j}=0.
\end{equation}
Subtracting (\ref{subsRE}) from (\ref{srcRE}) we find the relation (\ref{finRE}) explicitly. Thus, we obtain
\begin{eqnarray}
&&c_{j,i}^{(k)}(\omega) = c_{j,i}^{(k+1)}(\omega),\qquad
{\rm for}\,\,j=0,\,\,\mbox{or}\,\,i<k,\nonumber \\[2mm]
&&c_{j,i}^{(k)}(\omega) = c_{j,i}^{(k+1)}(\omega)-\frac{c_{k,i}^{(k+1)}(\omega)\,
c_{j-1,i-1}^{(k)}(\omega)}{c_{k-1,i-1}^{(k)}(\omega}\,.
\nonumber
\end{eqnarray}
This procedure is called the \emph{Gaussian eliminations} and allows us to determine the coefficients in the three-term recurrence relation
\begin{subequations}\label{recurrence-relation}
\begin{eqnarray}\label{three-terms}
&&c_{0,i}^{(3)}\,b_i+c_{1,i}^{(3)}\,b_{i-1}+c_{2,i}^{(3)}\,b_{i-2}=0, 
\,\,\,i>1,\\\label{two-terms}
&&c_{0,1}^{(3)}\,b_1+c_{1,1}^{(3)}\,b_0=0,
\end{eqnarray}
\end{subequations}
numerically for a given $\omega$ up to any finite $i$. The complexity of the procedure is \emph{linear} with respect to $i$ and $N$.

If the Frobenius series is convergent, we are able to find $b_1/b_0$ from Eq.~(\ref{two-terms}) and substitute it into
Eq.~(\ref{three-terms})
\begin{equation}\nonumber
\frac{b_1}{b_0}=-\frac{c_{1,1}^{(3)}}{c_{0,1}^{(3)}} = -\frac{c_{2,2}^{(3)}}{c_{1,2}^{(3)}-}
\,\frac{c_{0,2}^{(3)}c_{2,3}^{(3)}}{c_{1,3}^{(3)}-}\,\frac{c_{0,3}^{(3)}c_{2,4}^{(3)}}{c_{1,4}^{(3)}-}\ldots.
\end{equation}
Finally we find
\begin{equation}
0=c_{1,1}^{(3)}-\frac{c_{0,1}^{(3)}c_{2,2}^{(3)}}{c_{1,2}^{(3)}-}
\,\frac{c_{0,2}^{(3)}c_{2,3}^{(3)}}{c_{1,3}^{(3)}-}\ldots\,.
\end{equation}
The latter relation can be inverted $n$ times to give
\begin{eqnarray}\nonumber
&\displaystyle c_{1,n+1}^{(3)}-\frac{c_{2,n}^{(3)}c_{0,n-1}^{(3)}}{c_{1,n-1}^{(3)}-}
\,\frac{c_{2,n-1}^{(3)}c_{0,n-2}^{(3)}}{c_{1,n-2}^{(3)}-}\ldots\,
\frac{c_{2,2}^{(3)}c_{0,1}^{(3)}}{c_{1,1}^{(3)}}=&\\
&\displaystyle \frac{c_{0,n+1}^{(3)}c_{2,n+2}^{(3)}}{c_{1,n+2}^{(3)}-}
\frac{c_{0,n+2}^{(3)}c_{2,n+3}^{(3)}}{c_{1,n+3}^{(3)}-}\ldots\,.&\label{invcf}
\end{eqnarray}
Equation (\ref{invcf}) with an \emph{infinite continued fraction} on the right-hand side can be solved numerically by minimizing the absolute value of the difference between the left- and right-hand sides. The equation has an infinite number of roots (corresponding to the QN spectrum), but for each $n$, the most stable root is different. In general, we have to use the $n$ times inverted equation to find the $n$-th QN mode. The requirement that the continued fraction is convergent itself allows us to truncate its length by some large value, always ensuring that an increase in this value does not change the final results within a desired precision \cite{Leaver:1985ax}.

\subsection{Nollert improvement}

It turns out, that the convergence of the infinite continued fraction becomes worse if the imaginary part of $\omega$ increases with respect to the real part. This means that in order to calculate higher overtones, we must increase the depth of the continued fraction, which dramatically increases the time of computation. The convergence is poor also if $r_0$ in (\ref{Frobenius}) is not a singular point. This fixing of $r_0$ is necessary to move all the singular points for higher-dimensional Schwarzschild black holes outside the unit circle $|z| < 1$.

The problem of slow convergence was circumvented in \cite{PhysRevD.47.5253} for the three-term recurrence relation and generalized for higher $N$ in \cite{Zhidenko:2006rs}. We consider the relation
\begin{equation}\label{NollertR}
-\frac{b_n}{b_{n-1}}=R_n=\frac{c_{2,n+1}^{(3)}}{c_{1,n+1}^{(3)}-}
\frac{c_{0,n+1}^{(3)}c_{2,n+2}^{(3)}}{c_{1,n+2}^{(3)}-}\ldots\,,
\end{equation}
which can be expanded for large $n$ as
\begin{equation}\label{NollertExp}
R_n(\omega)=C_0(\omega)+\frac{C_1(\omega)}{\sqrt{n}}+\frac{C_2(\omega)}{n}+\ldots\,.
\end{equation}

In order to find the coefficients $C_j$ of (\ref{NollertExp}) we divide Eq.~(\ref{rrelation}) by $b_{i-N+1}$ and use the definition $R_n=-b_n/b_{n-1}$. Then one finds the equation with respect to $R_n$:
\begin{equation}\label{NollertHD}
\sum_{j=0}^{N-1}(-1)^j\,c_{j,i}^{(N)}(\omega)\,\prod_{k=0}^{N-2-j}R_{i-k}=0.
\end{equation}
For large $n$ we have $c_{j,n}^{(N)}(\omega)\propto n^2$, thus, substituting the expansion (\ref{NollertExp}) into
(\ref{NollertHD}), we find
\begin{equation}\label{C0eq}
\lim_{n\rightarrow\infty}\frac{1}{n^2}\sum_{j=0}^{N-1}(-1)^j\,c_{j,n}^{(N)}(\omega)\,C_0^{N-1-j}(\omega)=0\,.
\end{equation}

In the general case Eq.~(\ref{C0eq}) has $N-1$ roots (in fact, there are repeated roots among them). If the series has a unit radius of convergence, one of the roots is \textit{always} $C_0=-1$ (it is also a repeated root) (\ref{Frobenius}). Other roots appear due to the existence of additional singular points of Eq.~(\ref{radial}). Thus, we choose $C_0=-1$.

After fixing $C_0=-1$ one can find an equation with respect to $C_1^2$. In order to fix the sign of $C_1$ we can use the convergence of the series (\ref{Frobenius}) at $z=1$. Therefore,
$$\lim_{n\rightarrow\infty}b_n=0,\ \qquad\hbox{i.~e.}\ \nexists N:~\forall n>N, \quad |b_n|>|b_{n-1}|.$$
Since for large $n$
$$\frac{b_n}{b_{n-1}}\sim -R_n\sim -C_0-\frac{C_1}{\sqrt{n}}=1-\frac{C_1}{\sqrt{n}}\,,$$
we find out that the real part of $C_1$ \emph{cannot be negative}.

Once the sign of $C_1$ is fixed, the other coefficients in (\ref{NollertExp}) can be found from Eq.~(\ref{NollertHD}) step by step without encountering indeterminations.

As one calculated the coefficients $C_j$, the expansion (\ref{NollertExp}) could be used as an initial approximation for the ``remaining'' infinite continued fraction. In order to ensure the convergence of (\ref{NollertExp}) for a given value of $\omega$, one has to start from the found approximation deeply enough inside the continued fraction (\ref{invcf}). The expansion gives a good approximation for $R_n$. Therefore, the required depth is less than it would be if we had started from some arbitrary value.

\subsection{Continuation of the Frobenius series through midpoints}\label{sec.midpoints}

Consider now the case when one \emph{cannot fix} the parameter $r_0$ in (\ref{Frobenius}) in such a way that all the singularities, except $r=r_+$ and $r=r_\infty$, can be removed outside the unit circle $|z|<1$. In other words, there is at least one singularity for which $|z|<1$. This singularity implies a smaller radius of convergence for the series (\ref{Frobenius_series}). In order to check if the function $u(z)$ is convergent at $z=1$ we must analytically continue the series by constructing the expansions of $u(z)$ iteratively at some midpoints \cite{Rostworowski:2006bp}.

Namely, we equate the series expansion
\begin{equation}\label{Frobenius_series-midpoint}
u(z)=\sum_{n=0}^\infty b_n z^n = \sum_{n=0}^\infty \tilde{b}_n(z-z_0)^n,
\end{equation}
where $z=z_0$ is a midpoint inside the radius of convergence of (\ref{Frobenius_series}).

The coefficients $\tilde{b}_n$ satisfy the $N$-term recurrence relation, which could be reduced to the three-term relation
\begin{equation}\label{three-terms-midpoint}
\tilde{c}_{0,i}^{(3)}\,\tilde{b}_i+\tilde{c}_{1,i}^{(3)}\,\tilde{b}_{i-1}+\tilde{c}_{2,i}^{(3)}\,\tilde{b}_{i-2}=0, \quad
i>1.
\end{equation}
In order to find $\tilde{b}_1/\tilde{b}_2$, we must use the condition at the event horizon by taking account of (\ref{Frobenius_series-midpoint}),
\begin{equation}
\tilde{b}_0=\sum_{n=0}^\infty b_n z_0^n, \qquad \tilde{b}_1=\sum_{n=1}^\infty n b_n z_0^{n-1}.
\end{equation}
We find the coefficients $b_n$  from Eqs.~(\ref{recurrence-relation}) and substitute them into Eq.~(\ref{three-terms-midpoint}). If $z=1$ is the closest singular point to $z=z_0$, we obtain the equation with respect to $\omega$ as
\begin{equation}\label{continued-fraction-midpoint}
\frac{\tilde{b}_1}{\tilde{b}_0} = -\frac{\tilde{c}_{2,2}^{(3)}}{\tilde{c}_{1,2}^{(3)}-}
\,\frac{\tilde{c}_{0,2}^{(3)}\tilde{c}_{2,3}^{(3)}}{\tilde{c}_{1,3}^{(3)}-}\,\frac{\tilde{c}_{0,3}^{(3)}\tilde{c}_{2,4}^{(3)}}{\tilde{c}_{1,4}^{(3)}-}\ldots
\end{equation}
Otherwise one has to repeat the procedure, constructing the series (\ref{Frobenius_series-midpoint}) for the next midpoints $z_1,\,z_2,\,z_3,\,\ldots$, until the cosmological horizon (or spatial infinity) appears to be inside the radius of convergence.

If the convergence of the continued fraction on the right-hand side of Eq.~(\ref{continued-fraction-midpoint}) is slow, one can use the Nollert improvement. Since the radius of convergence of the Frobenius series is now less than 1 ($R<1$), we must choose $C_0=-R^{-1}$ in (\ref{NollertExp}).

\subsection{Generalization of the Frobenius series}

The series expansion (\ref{Frobenius_series}) does not need to be necessarily some power of a rational function of $r$. For some cases, a more convenient choice is an expansion in terms of another full set of functions in the appropriate Hilbert space. Here we considered, as an illustrative and quite general example, charged scalar field perturbations of Kerr-Newman-de Sitter black holes that are described by the line element
$$ ds^2 = \rho^2
\left(\frac{dr^2}{\Delta_r}+\frac{d\theta^2}{\Delta_\theta}\right)
+
\frac{\Delta_\theta \sin^2\theta}{(1+\alpha)^2 \rho^2}
[adt-(r^2+a^2)d\varphi]^2$$
\begin{equation}
-\frac{\Delta_r}{(1+\alpha)^2 \rho^2}(dt-a\sin^2\theta
d\varphi)^2,
\end{equation}
where
\begin{equation}
\begin{array}{cc}
\multicolumn{2}{c}{{\displaystyle
\Delta_r=(r^2+a^2)\left(1-\frac{\alpha}{a^2}r^2\right)-2Mr+Q^2,}} \\
\Delta_\theta=1+\alpha\cos^2\theta, &
{\displaystyle \alpha=\frac{\Lambda a^2}{3}}, \\
\end{array}
\end{equation}
$M$ is the black hole mass, $Q$ is the charge, $a$ is the rotation parameter, and $\Lambda$ is the cosmological constant.

After the separation of variables, the angular part of the equation of motion for the massless charged scalar field (\ref{em:scalar}) can be reduced to the following form \cite{Suzuki:1998vy}:
\begin{widetext}
\begin{eqnarray}
\left(\frac{d}{dx}(1 + \alpha x^2)(1 - x^2)\frac{d}{dx}+\lambda-s(1-\alpha)+\frac{(1+\alpha)^2}{\alpha}\xi^2-2\alpha x^2 + \right.\nonumber\\ \label{angular-Kerr}\frac{1+\alpha}{1+\alpha x^2}\left(2s(\alpha m - (1+\alpha)\xi)x-\frac{(1+\alpha)^2}{\alpha}\xi^2-2m(1+\alpha)\xi+s^2\right)-\\\nonumber
\left.-\frac{(1+\alpha)^2m^2}{(1+\alpha x^2)(1-x^2)}-\frac{(1+\alpha)(s^2+2smx)}{1-x^2}\right)S(x)=0,
\end{eqnarray}
where $\lambda$ is the separation constant, $\xi=a\omega$, $x=cos\theta$, $s$ is the field spin, and $m$ is the projection of the angular momentum of the field onto the axis of the black hole rotation. Here $s$ and $m$ are (half)integers and $0\leq s\leq 2$.

The appropriate series for the function $S$ is \cite{Suzuki:1998vy}
\begin{equation}\label{angular-Kerr-Frobenius}
S(z)=z^{A_1}(z-1)^{A_2}(z-z_s)^{A_3}(z-z_\infty)\sum_{n=0}^\infty b_n u_n(z),
\end{equation}
where
$$z=\frac{\sqrt{\alpha}-\imo}{2}\frac{x+1}{x\sqrt{\alpha}-\imo}, \quad z_s = -\frac{\imo(1+\imo\sqrt{\alpha})^2}{4\sqrt{\alpha}},
\quad z_\infty = -\frac{\imo(1+\imo\sqrt{\alpha})}{2\sqrt{\alpha}},$$
$$A_1=\frac{|m-s|}{2}, \qquad A_2=\frac{|m+s|}{2}, \quad
A_3 =
\pm\frac{\imo}{2}\left(\frac{1+\alpha}{\sqrt{\alpha}}\xi-\sqrt{\alpha}\xi
-\imo s\right).$$
The expansion is done in terms of the Jacobi polynomials
$$ u_n(z)=F(-n,n+\bar{\omega};\gamma;z)=(-1)^n\frac{\Gamma(2n+\bar{\omega})n!}{\Gamma(n+\gamma)}P_n^{(\bar{\omega}-\gamma,\gamma-1)}(2z-1),$$
where $\bar{\omega} = 2A_1+2A_2+1$ and $\gamma = 2A_1+1$.

The coefficients $b_n$ in (\ref{angular-Kerr-Frobenius}) satisfy the three-term recurrence relation (\ref{three-terms} and
\ref{two-terms}) with
\begin{eqnarray}
c^{(3)}_{0,n} &=&
\pm \frac{i}{\sqrt{\alpha}}\xi \
\frac{n(n+A_1+A_2 \mp s )(n+2A_2)}
{2(2n+2A_1+2A_2+1)(n+A_1+A_2)}, \\
c^{(3)}_{1,n} &=&
\frac{i}{\sqrt{\alpha}} \left\{
\pm \xi \frac{J_n}
{2(n+A_1+A_2)(n+A_1+A_2-1)} \right.
 + \frac{(n-1)(n+2A_1+2A_2)}{4} \\
&& \left. -\frac{1}{4}\left[
\lambda-2A_1 A_2 -A_1 -A_2 +2\big(m+s \mp (2A_1 +1)\big)\xi
-\frac{m^2-s^2}{2}-s \right] \right\}, \nonumber \\
c^{(3)}_{2,n} &=&
\mp\frac{i}{\sqrt{\alpha}}\xi \
\frac{(n-1+A_1+A_2\pm s)(n-1+2A_1)(n-1+2A_1+2A_2)}
{2(2n+2A_1+2A_2-3)(n-1+A_1+A_2)},
\end{eqnarray}
where
$$J_n = (n-1)(n+2A_1+2A_2)(A_1-A_2) +(A_1+A_2 \pm s+1)\left(\ (n-1)(n+2A_1+2A_2)+(2A_1+1)(A_1+A_2) \ \right).$$
\end{widetext}

Since the series (\ref{angular-Kerr-Frobenius}) must be convergent at $z=1$, we can solve Eq.~(\ref{invcf}) numerically in order to find the separation constant as a function of frequency.

In the Reissner-Nordstr\"om-de Sitter limit ($a\rightarrow0$) Eq.~(\ref{invcf}) is reduced to $c^{(3)}_{1,n}=0$. In this case the value of $\lambda$ does not depend on $\omega$
$$\lambda = (\ell-s+1)(\ell+s), \qquad \ell = n+A_1+A_2\geq max(|m|,|s|).$$

\subsection{Example I: Frobenius series for radial part of the charged scalar field in
the Kerr-Newman-de Sitter background}\label{sec:KerrNewmandS}

The radial part of the massless (charged) field equation of motion is of the form (\ref{radial}) and reads
\cite{Suzuki:1998vy}
\begin{eqnarray}\nonumber
\Bigg\{\Delta_r^{-s}\frac{d}{dr}\Delta_r^{s+1}\frac{d}{dr}
\!+\!\frac{1}{\Delta_r}\left(K^2\! -\!\imo s K \frac{d\Delta_r} {dr}
\right)
\!+\!4\imo s(1+\alpha)\omega r
\\\nonumber-\frac{2\alpha}{a^2}(s+1)(2s+1) r^2\!+\!2s(1-\alpha)\!-\!2 \imo s q
Q\!-\lambda\Bigg\} R(r) = 0,
\label{eqn:Rr}
\end{eqnarray}
where $K = [\omega(r^2+a^2)- am](1+\alpha)- q Q r$, and $q$ is the field charge.

The appropriate Frobenius series is found to be
\begin{eqnarray} R(r)=\left(\frac{r-r_+}{r-r_-}\right)^{-s-2\imo
K(r_+)/\Delta_r'(r_+)} \times\nonumber\\
\label{Frobenius-KNdD}
e^{\imo
B(r)}r^{-2s-1}u\left(\frac{r-r_+}{r-r_-}\frac{r_\infty-r_-}{r_\infty-r_+}\right)
\end{eqnarray}
Note that in order to obtain the recurrence relation for both types of the boundaries (asymptotically flat and de Sitter), we introduced the exponent $\displaystyle e^{\imo B(r)}$ such that $\displaystyle\frac{dB(r)}{dr}=\frac{K}{\Delta_r}$. This exponent describes outgoing waves at horizons and spatial infinity. Thus, we have to compensate the outgoing wave at the event horizon. That is why the factor of $2$ appears in the power of the first multiplier in (\ref{Frobenius-KNdD}). The parameter $r_0$ is fixed to be the inner horizon $r_-$ in order to move all the singularities outside the unit circle and, at the same time, to provide the best convergence of the infinite continued fraction (\ref{invcf}).

In the Reissner-Nordstr\"om-de Sitter limit ($a=0$) for the uncharged field ($q=0$) we obtain $K=\omega r^2$ and
\begin{equation}\label{Frobenius-RNdD}
R(r)=\left(\frac{r-r_+}{r-r_-}\right)^{-s-\imo
\omega/\kappa}\!\!\!\!\!\!\!e^{\imo \omega r_\star}r^{-2s-1}u\left(\frac{r-r_+}{r-r_-}\frac{r_\infty-r_-}{r_\infty-r_+}\right).
\end{equation}
The tortoise coordinate is defined as $\displaystyle dr_\star=\frac{r^2 dr}{\Delta_r}$ and $\displaystyle\kappa=\frac{\Delta_r'(r_+)}{2r_+^2}$ is the surface gravity on the event
horizon.

\subsection{Example II: Frobenius series for the massive scalar field in the
higher-dimensional Reissner-Nordstr\"om-de Sitter background}

The $D$-dimensional Reissner-Nordstr\"om-de Sitter black hole is described by the metric
\begin{equation}
ds^2=-f(r)dt^2+\frac{dr^2}{f(r)}+r^2d\Omega_{D-2},
\end{equation}
where $d\Omega_{D-2}$ is a metric of a $(D-2)$-dimensional sphere,
$$
f(r)=1-\frac{2M}{r^{D-3}}+\frac{Q^2}{r^{2D-6}}-\frac{2\Lambda
r^2}{(D-1)(D-2)}.$$

After separation of the angular and time variables the radial part of the massive scalar field equation of motion $(\Box-\mu^2)\Psi=0$ is reduced to the wavelike equation

\begin{equation}
\label{potential-tensortype}
\left(\frac{d^2}{dr_\star^2}+\omega^2-f(r) V(r)\right) r^{\frac{D-2}{2}}R(r) = 0,
\end{equation}
where the effective potential is
$$ V(r)\!=\!\mu^2\!+\!\frac{\ell(\ell\!+\!D\!-\!3)}{r^2}+\!\frac{f'(r)(D\!-\!2)}{2r}+\!\frac{f(r)(D\!-\!2)(D\!-\!4)}{4r^2},$$
where $\ell$ parametrizes the angular separation constant.

The Frobenius series for this case is
$$
R(r) =
\left(\frac{r-r_+}{r-R}\right)^{-\frac{\imo\omega}{\kappa}}e^{\imo
A(r)}r^{-\frac{D-2}{2}}u\left(\frac{r-r_+}{r-R}\frac{r_\infty-R}{r_\infty-r_+}\right),
$$
where $\displaystyle\kappa=\frac{1}{2}f'(r_+)$, $\displaystyle e^{\imo A(r)}$ describes the outgoing wave for spatial infinity and the horizons and satisfies
$$\frac{dA(r)}{dr}=\frac{\sqrt{\omega^2-\mu^2f(r)}}{f(r)}.$$
The sign in the exponent is fixed by the quasinormal boundary condition: The real part of $A(r\rightarrow\infty)$ must be of the same sign as the real part of the eigenfrequency $\omega$. This choice of the sign makes the wave outgoing at spatial infinity.

For the massless field ($\mu=0$) this exponent is $e^{\imo\omega r_\star}$ \cite{Konoplya:2007jv} and the Frobenius series reads
$$
R(r) =
\left(\frac{r-r_+}{r-R}\right)^{-\frac{\imo\omega}{\kappa}}e^{\imo
\omega
r_\star}r^{-\frac{D-2}{2}}u\left(\frac{r-r_+}{r-R}\frac{r_\infty-R}{r_\infty-r_+}\right).
$$
Since for $D=4$ we can choose $R=r_-$, we come to (\ref{Frobenius-RNdD}) ($s=0$).

If $\Lambda>0$, we can observe the same asymptotic behavior of the exponent at the cosmological horizon as for the case of
$\mu=0$:
$$e^{\imo A(r)}\sim e^{\imo\omega r_\star}, \qquad r\rightarrow r_\infty<\infty.$$

For the asymptotically flat case the asymptotic behavior is different in four- and in higher-dimensional space-times. If $D\geq5$, $f(r)=1+o(r^{-1})$ and the Frobenius series can be determined as
\cite{Zhidenko:2006rs}
\begin{equation}\label{Frobenius-msDRN}
R(r) =
\left(\frac{r-r_+}{r-R}\right)^{-\frac{\imo\omega}{2\kappa}}e^{\imo
r\sqrt{\omega^2-\mu^2}}r^{-\frac{D-2}{2}}u\left(\frac{r-r_+}{r-R}\right).
\end{equation}
For $D=4$ the term of order $\sim r^{-1}$ in $f(r)$ leads to the nontrivial contribution \cite{Ohashi:2004wr, Konoplya:2004wg}
\begin{eqnarray}\label{Frobenius-msRN}
&&R(r) = \left(\frac{r-r_+}{r-R}\right)^{-\frac{\imo\omega}{2\kappa}} e^{\imo r\sqrt{\omega^2-\mu^2}} \times\\\nonumber
&&r^{2\imo M\sqrt{\omega^2-\mu^2}+\imo M\mu^2/\sqrt{\omega^2-\mu^2}}r^{-\frac{D-2}{2}}u\left(\frac{r-r_+}{r-R}\right).
\end{eqnarray}

The same approach could be applied for the Maxwell field and the gravitational perturbations, because the radial parts of their wave equations can be reduced to the form (\ref{radial}) \cite{Kodama:2003kk,Kodama:2003jz,Ishibashi:2003ap}.

\subsection{Horowitz-Hubeny method}

In order to find quasinormal modes in the asymptotically anti-de Sitter space-times we normally need to impose Dirichlet boundary conditions at the spatial infinity. Thus, we find the appropriate expansion for the function $R(r)$ in (\ref{radial}) without consideration of the singularity point at the infinity. This method was proposed in \cite{Horowitz:1999jd}. Namely, we define
\begin{equation}\label{newfunction1}
R = z^{-\imo\omega/(2\kappa)}\psi(z),
\end{equation}
where $\kappa$ is the surface gravity at the event horizon, $\displaystyle z=\frac{r-r_+}{r-r_-}$. Substituting (\ref{newfunction1}) into (\ref{radial}) one can rewrite the radial equation in the form
\begin{equation}
s(z)\psi''(z)+\frac{t(z)}{z}\psi'(z)+\frac{u(z)}{z^2}\psi(z)=0,
\end{equation}
$$s(z)=\sum_{n=0}^{N_s}s_nz^n, \quad t(z)=\sum_{n=0}^{N_t}t_nz^n, \quad u(z)=\sum_{n=1}^{N_u}u_nz^n,$$
with respect to the wave function $\displaystyle\psi(z)=\sum_{n=0}^\infty a_n z^n$, which is regular at the event horizon $z=0$. The Dirichlet boundary condition $\psi(z=1)=0$ implies
\begin{equation}\label{HH-equation}
\sum_{n=0}^\infty a_n=0.
\end{equation}
The coefficients $a_n$ can be found through the recurrence relation
\begin{equation}\label{an}
a_n=\frac{\sum_{k=0}^{n-1}a_k(k(k-1)s_{n-k}+kt_{n-k}+u_{n-k})}{n(n-1)s_0+nt_0},
\end{equation}
starting from an arbitrary $a_0$. Substituting (\ref{an}) into Eq.~(\ref{HH-equation}), we find the equation with respect to eigenvalue $\omega$. Since the sum (\ref{HH-equation}) is convergent, one can truncate  the summation at some large $n$ and minimize its absolute value with respect to $\omega$. Doing so we increase the number of terms in the sum until the value of $\omega$ does not change within required precision.

Note that for Eq.~(\ref{HH-equation}) to be convergent, $r_-$ has to be a singular point, and all the other singularities, except $r=r_+$ and $r=r_\infty$, must lie outside the unit circle $|z|>1$. If both of these conditions are impossible to satisfy simultaneously, we must use the continued fraction method with the appropriate fixing of the behavior of $R$ at spatial infinity. The Horowitz-Hubeny method converges slower than the Frobenius one.

\subsection{Limit of high damping: Monodromy method}

The limits of high damping for QN spectra of asymptotically flat or de Sitter and asymptotically anti-de Sitter black holes have qualitatively different behavior. For asymptotically flat or the de Sitter case
\begin{equation}\label{Dsorflat}
|\Im{\omega}| \gg |\Re{\omega}|, \quad n \rightarrow \infty
\end{equation}
while for asymptotically AdS black holes
\begin{equation}\label{AdS}
|\Im{\omega}| \sim |\Re{\omega}|, \quad n \rightarrow \infty.
\end{equation}
The monodromy technique starts by considering the wave equation (\ref{wave1}) in the complex plane, so that instead of the real tortoise variable $r_{*}$ one uses the complex variable $x$. Then, Eq.~(\ref{wave1}) can be written in the form
\begin{equation}\label{omega-x}
\frac{d^{2} \Psi(r)}{d (\omega x)^2} + (1- V(r) \omega^{-2}) \Psi(r) = 0.
\end{equation}
In the limit $|\omega| \rightarrow \infty$ the term proportional to $\omega^{-2}$ in Eq.~(\ref{omega-x}) can be discarded, so that the general solution can be written as a superposition of plane waves
\begin{equation}\label{planewave}
\Psi(\omega x) = A e^{+ \imo \omega x} + B  e^{- \imo \omega x}, \quad  |\omega| \rightarrow \infty.
\end{equation}
Then one needs to find the Stokes line as contours in the complex $\omega x$ plane. The \textit{Stokes lines} are defined as lines satisfying the equation $\Im{\omega x} = 0$. If we apply Eq.~(\ref{choise-omega}) and $x = x_{Re} - \imo x_{Im}$, then
\begin{eqnarray}
&\omega x = \omega_{Re} x_{Re} - \omega_{Im} x_{Im} - \imo (\omega_{Im} x_{Re} + \omega_{Re} x_{Im}) =&\nonumber\\
&\omega_{Re} x_{Re} - \omega_{Im} x_{Im}, \quad Im (\omega x) = 0&
\end{eqnarray}
along the Stokes lines.
Thus, along the Stokes line, the plane wave solution (\ref{planewave}) oscillates without decay (or growth). The next step is to use the basic numerical data for quasinormal modes with large overtones, such as Eq.~(\ref{Dsorflat}) or (\ref{AdS}) in order to translate the condition for the Stokes lines in the $\omega x$-plane to the condition in the complex $x$-plane. For instance, for asymptotically flat or de Sitter black holes the inequality (\ref{Dsorflat}) immediately gives $\Re{\omega} = 0$, which is the \textit{anti-Stokes line} in the complex $x$-plane. Then one needs to determine the closed contour in the complex $x$-plane and calculate the monodromy, which will give an analytic expression for the quasinormal frequency $\omega$. Although the approach is well known from complex analysis, in the present form it was used for the problem of highly damped quasinormal modes only recently \cite{Motl:2002hd,Motl:2003cd} for asymptotically flat cases and developed in \cite{Cardoso:2004up,Natario:2004jd,Ghosh:2005aq} for various asymptotically dS and AdS black holes. Particular elements of the monodromy method, the Stokes lines, were used for QNMs in \cite{Froman}.

\section{Quasinormal modes: summary of results}

\subsection{More on the context in which quasinormal modes are studied}

The quasinormal spectrum of a stable black hole is an infinite set of complex frequencies which describes damped oscillations of the amplitude \cite{Bachelot:1993dp}. It is clear that the oscillation with the smallest damping rate is dominant at late time, while oscillations with higher damping rate are exponentially suppressed. In practice, by studying the signal at the stage when quasinormal ringing is observed, we are able to extract a few dominant modes which have the smallest imaginary parts.

The quasinormal ringing was first described by Vishveshwara when considering the scattered packet of gravitational waves in the
background of the Schwarzschild black hole \cite{Vishveshwara:1970zz}. The term \emph{quasinormal frequencies} was introduced by Press in 1971 \cite{Press:1971}. In the same year the lower quasinormal modes were calculated by studying test particles falling into the Schwarzschild black hole \cite{Davis:1971gg}. Later, quasinormal modes were calculated for Kerr black holes \cite{Detweiler:1977}. The quasinormal spectra of the coupled gravitational and electromagnetic perturbations of the Reissner-Nordstr\"om black holes were first studied in \cite{RNQNMs}.

From that time until now quasinormal modes have been extensively studied within various contexts, as follows:
\begin{enumerate}
\item \textbf{Ringing of astrophysical black holes}: In this context the dominant frequencies of the quasinormal spectrum were studied for astrophysically motivated black holes in four dimensional asymptotically flat and de Sitter space-times (see Sec.~\ref{sec:QNMdS}).
\item \textbf{Analysis of stability and quasinormal modes of mini black holes}: This is an important issue in various extra dimensional scenarios, such as ADD \cite{ADD} and the Randall-Sundrum \cite{Randall-Sundrum} models. In this context dominant quasinormal frequencies of higher-dimensional black holes, black strings and branes, Kaluza-Klein black holes and other black objects have been extensively studied (see Sec.~\ref{sec:QNMHD}).
\item \textbf{In the context of the AdS/CFT correspondence}: Here the quasinormal modes of large AdS black holes and of $Dp$-branes are studied (see Sec.~\ref{sec:AdSCFT}).
\item \textbf{Loop quantum gravity interpretation}: It was suggested that the asymptotic real part of  quasinormal modes of black holes can be interpreted within loop quantum gravity, which allows us to fix the Barbero-Immirzi parameter, the free parameter of the theory (see Sec.~\ref{sec:lqg}).
\item \textbf{Lower-dimensional black holes}: These were studied mainly because of their simplicity, so that in many cases it is possible to find exact solutions for the quasinormal spectra. For instance, it was shown that the quasinormal modes of the $(2+1)$-dimensional AdS black hole exactly coincide with the poles of the retarded Green's function for the two-dimensional CFT (see Sec.~\ref{sec:AdSCFTvoc}).
\end{enumerate}

\subsection{Isospectrality}
Consider two wavelike equations (\ref{wave1}) with the effective potentials $V^+$ and $V^-$ such that
\begin{equation}
V^\pm(r_\star)=W^2(r_\star)\pm\frac{dW(r_\star)}{dr_\star}+\beta,
\end{equation}
where $W(r_\star)$ is some finite function, $\beta$ is a constant, and $r_\star$ is the tortoise coordinate. If $\Psi^+$ is an eigenfunction of the wavelike equation with the potential $V^+$, then the eigenfunction for the potential $V^-$ is given by
\begin{equation}
\Psi^-(r_\star)\propto\left(W(r_\star)-\frac{d}{dr_\star}\right)\Psi^+(r_\star),
\end{equation}
which corresponds to the same eigenvalue $\omega$. Therefore, the quasinormal spectrum is the same for the potentials $V^+$ and $V^-$ \cite{isospectrality}.

If one takes
\begin{eqnarray}\nonumber
W&=&\frac{2M}{r^2}-\frac{3+2c}{3r}+\frac{3c+2c^2}{3(3M+cr)}-\frac{c^2+c}{3M},\\\nonumber
\beta&=&-\frac{c^2(c+1)^2}{9M^2},\qquad c =
\frac{\ell(\ell+1)}{2} - 1,
\\\nonumber r_\star&=&\frac{dr}{f(r)},\qquad f(r) = 1 - \frac{2M}{r},
\end{eqnarray}
then $V^+$ and $V^-$ are the effective potentials for gravitational perturbations of axial and polar types respectively for the Schwarzschild black hole \cite{Chandrabook},
\begin{eqnarray}\nonumber
V^+ &=& f(r)\left(\frac{\ell(\ell+1)}{r^2} - \frac{6M}{r^3}\right),
\\\nonumber
V^- &=& \frac{2f(r)}{r^3}\frac{9M^3 + 3c^2Mr^2 + c^2(1 +
c)r^3 + 9M^2cr}{(3M + cr)^2}.
\end{eqnarray}
This means that the quasinormal spectra of gravitational perturbations of axial and polar types coincide. The same symmetry is preserved in a more general case of the Reissner-Nordstr\"om-de Sitter black hole, where both types of gravitational perturbations are coupled to electromagnetic perturbations. If, for instance, one considers a scalar field which is coupled only to the polar perturbations, the isospectrality is broken.

The isospectrality between axial and polar perturbations does not exist for higher than four dimensional black holes, when, in addition to the latter two types of perturbations, the other dynamical type appears. This new type of gravitational perturbation transforms similar to a tensor with respect to coordinate transformations on a $(D-2)$-sphere. It is interesting to note that the tensor-type gravitational perturbations are described by the same effective potential as a test scalar field in the black hole background.

\subsection{Quasinormal modes of $D=4$ black holes}\label{sec:QNMdS}

\begin{figure}
\resizebox{\linewidth}{!}{\includegraphics*{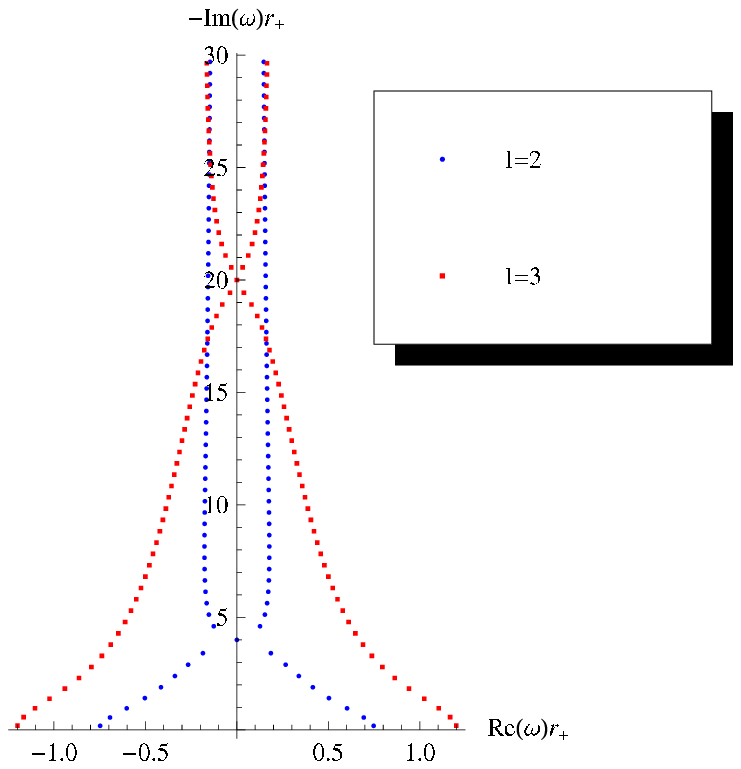}}
\caption{The first 60 quasinormal modes for the gravitational perturbations of the Schwarzschild black hole; the numerical values of 1000 lower quasinormal modes are available from http://qnms.way.to.}
\label{Schwarzschildspectrum}
\end{figure}

The quasinormal spectra of four-dimensional black holes were studied starting from the 1970s in the context of possible observations of ringing from astrophysical black holes (see, e.~g., \cite{Chandrabook}). Figure~\ref{Schwarzschildspectrum} shows the lowest gravitational quasinormal modes of Schwarzschild black holes, calculated accurately with the help of the continued fraction method (see Sec.~\ref{sec:Frobenius}). The imaginary part of the higher overtones grows, while the real part decreases until it becomes almost zero at some moderately large overtone number $n$ (for $\ell=2$, $n=9$, for $\ell=3$, $n=41$). The corresponding purely damped mode is approximately equal to the so-called ``algebraically special mode'', which is given by \cite{algebraicalmode}
\begin{equation}\label{algebraicalmodevalue}
\omega M = -\imo(\ell - 1)\ell(\ell + 1)(\ell + 2)/12.
\end{equation}
For modes which are higher than the algebraically special one, the real part starts growing and approaches its asymptotic value, which can be found exactly (see Sec.~\ref{sec:exactqnms}). Since the damping rate of the algebraically special mode (\ref{algebraicalmodevalue}) grows quickly with $\ell$, the asymptotic regime is achieved at high overtones. For instance, for $\ell=6$ the asymptotic regime of high damping is achieved at $n\sim10^5$ \cite{PhysRevD.47.5253}.

Real astrophysical black holes are rotating. In addition, one can neglect their electric charge so that the Kerr metric is the most astrophysically motivated exact solution of the Einstein equations. Thus, gravitational quasinormal modes of Kerr black holes are of primary interest for observations of gravitational waves. Accurate calculations of quasinormal modes for Kerr black holes were performed by Leaver \cite{Leaver:1985ax}. The basic properties of the spectrum for small rotation were found from the eikonal limit within the slow-rotation approximation. Using the P\"oschl-Teller formula (\ref{PTformula}), one can find that \cite{Ferrari:1984zz}
\begin{eqnarray}
\omega&\approx&\frac{1}{3\sqrt{3}M}\left(\pm\left(\ell+\frac{1}{2}\right)+\frac{2am}{3\sqrt{3}M}-\left(n+\frac{1}{2}\right)\imo\right),\nonumber\\&& \ell\gg|m|\gg1, \qquad a\ll M,
\end{eqnarray}
where $m$ is the azimuthal number and $a$ is the rotation parameter of the black hole.

Rotation gives us the following new properties of the quasinormal spectrum:
\begin{itemize}
\item The real part of the quasinormal mode grows with $am$.
\item The quasinormal modes explicitly depend on the azimuthal number $m$ and there is a symmetry $\omega(m)\rightarrow-\omega^*(-m)$. This allows us to consider only modes with the positive real part.
\end{itemize}

Table~\ref{table:lowermodes} summarized the publications in which the quasinormal ringing of four-dimensional black holes was studied with the help of various numerical methods. The reader is referred to the original works (see Table~\ref{table:lowermodes}) for more details and numerical data. In Sec.~\ref{sec:bhparams} we discussed dependence of the quasinormal modes on the black hole parameters.

\jamode{\begin{table*}}{\begin{table}}
\caption{Publications where the lower modes of various four-dimensional black holes were calculated by different methods: integration of the equations of motion, the WKB formula, and the method of continued fraction.}\label{table:lowermodes}
\tablefont\begin{tabular}{|l|c|c|c|}
\hline
QNMs&Integration&WKB&\jamode{Continued fractions}{Cont. f.}\\
\hline
Schwarzschild&\cite{Davis:1971gg}&\cite{Iyer:1986nq}&\cite{Leaver:1985ax} \\
Reissner-Nordstr\"om&\cite{RNQNMs}&\cite{Kokkotas:1988fm}&\cite{Leaver:1990zz}\\
Reissner-Nordstr\"om (extreme)&&&\cite{Onozawa:1995vu}\\
Kerr&\cite{Detweiler:1977,Detweiler:1980gk}&\cite{Seidel:1989bp}&\cite{Leaver:1985ax}\\
Kerr-Newman&&\cite{Kokkotas:1993ef}&\cite{Berti:2005eb}\\
Reissner-Nordstr\"om+dilaton&&\cite{Ferrari:2000ep,Konoplya:2001ji}&\\
Topological black holes&\cite{Wang:2001tk}&&\\
Stringy black holes&&\cite{Li:2001ct}&\\
Schwarzschild-de Sitter&\cite{Mellor:1989ac}&\cite{Otsuki:1991}&\cite{Moss:2001ga}\\
Reissner-Nordstr\"om-de Sitter&\cite{Molina:2003dc}&\cite{Molina:2003dc}&\\
Kerr-de Sitter&&&\cite{Yoshida:2010zz}\\
Garfinkle-Horowitz-Strominger&&\cite{Shu:2004fj}&\\
\hline
\end{tabular}
\jamode{\end{table*}}{\end{table}}

Lower modes of the Dirac field were found by the WKB approach for Schwarzschild \cite{Cho:2003qe} and Schwarzschild-de Sitter backgrounds \cite{Zhidenko:2003wq} and by using the approximation with the P\"oschl-Teller potential (see Sec.~\ref{sec:PoschTeller}) for Kerr-Newman-de Sitter black holes \cite{Chang:2005ki}. Later the spectrum of the Dirac field was calculated using the accurate Frobenius method for Schwarzschild \cite{Jing:2005dt}, Reissner-Nordstr\"om-de Sitter \cite{Jing:2003wq}, Kerr-Newman \cite{Jing:2005pk} and Kerr-Newman-de Sitter black holes \cite{Konoplya:2007zx}. The quasinormal spectrum of the scalar and Dirac fields around the Born-Infeld black hole was studied in \cite{Fernando:BI}. Dominant gravitational frequencies of black holes in the scalar-tensor gravity were calculated in \cite{Lasky:2010bd}.

Real astrophysical black holes are not isolated, but surrounded by some matter. Theoreticians frequently call such black holes with an ``environment'' as ``dirty'' black holes. These were studied using a perturbative method in \cite{Leung:1999}. Higher modes of the spherically symmetric dirty black holes were studied in \cite{Medved:2003}. Quasinormal modes of black holes surrounded by quintessence were studied in \cite{Chen:2005qh,Ma:2006by,ZhangGui,Varghese:2008ky}, and modes of the phantom scalar field in \cite{Chen:2008xb}.

Recently, perturbations of more exotic objects such as wormholes, white holes, or naked singularities have been investigated. It was found that the quasinormal spectrum of Schwarzschild black holes differs from wormholes \cite{Konoplya:2005et} and white holes \cite{Bishop:2009ba}. Therefore, these objects, if they exist, might be detected through observations of their quasinormal ringing.

\subsection{Quasinormal modes of mini black holes}\label{sec:QNMHD}

Quasinormal spectra of black holes attracted considerable interest in the following extra dimensional models:
\begin{enumerate}
\item The \textbf{large extra dimensions} scenario allows for the size of extra dimensions to be of a macroscopic order \cite{ADD}. When the size of the black hole is much smaller than the size of the extra dimensions, the black hole can be considered as effectively living in a $D$-dimensional world and, thereby, approximated by a solution of higher-dimensional Einstein equations. The simplest example of such a solution is the Tangherlini metric \cite{Tangherlini:1963bw}, which is a generalization of the Schwarzschild metric for $D>4$.
\item \textbf{Randall-Sundrum models} \cite{Randall-Sundrum} assumed that our world is a brane in higher-dimensional anti-de Sitter warped space-time. The AdS space-time implies quick decay of the fields outside the brane. The warp factor, which is the parameter of the theory, can be set up in order to obtain large size (of the order of $TeV$) or, if one wishes, small size of the extra dimensions. In the latter case the gravitational perturbations contain Kaluza-Klein modes, which share many properties of massive fields. For instance, the quasiresonances were observed in the spectrum of black strings \cite{Seahra:2004fg}. The ringing of the Randall-Sundrum braneworld was studied in \cite{Seahra:2005wk,Clarkson:2005mg}, where quasibound states of the gravitational perturbations between the AdS boundary and the brane were observed.
\item \textbf{Brane-localized fields}: When the size of extra dimensions is larger than the size of the black hole, one can consider the model, for which the standard model particles are restricted to live on a $(3+1)$-brane, while gravitons propagate also in the bulk. When considering evolution of the standard model fields in the background of a mini black hole, one can think that the mini black hole effectively behaves similar to a higher-dimensional one projected onto the brane \cite{Kanti:2004nr}.
\end{enumerate}

Quasinormal modes of a scalar field in the background of a $D$-dimensional black hole were estimated in \cite{Cardoso:2002pa} using the third-order WKB approach and more accurate values were obtained in \cite{Konoplya:2003ii} with the sixth-order WKB formula (see Sec.~\ref{sec:WKBmethod}). The scalar field in the background of the five-dimensional rotating black hole was studied in \cite{Ida:2002zk}.

Quasinormal modes of gravitational perturbations of the Tangherlini black hole were calculated in \cite{Konoplya:2003dd} using the higher-order WKB formula. Later, five-dimensional \cite{Cardoso:2003qd} and higher-dimensional  \cite{Rostworowski:2006bp} Tangherlini black hole perturbations were studied using the Frobenius method. Finally, quasinormal modes of all types of perturbations for higher-dimensional Reissner-Nordstr\"om-de Sitter black holes were calculated in \cite{Konoplya:2007jv,KonoplyaPRL}. Quasinormal modes of the Dirac field in the background of higher-dimensional black holes were found in \cite{LopezOrtega:2007sr}.

The quasinormal spectrum of a scalar field in the background of ultraspinning black holes was studied in six- \cite{Morisawa:2004fs} and higher-dimensional space-times \cite{Cardoso:2004cj}. In $D\geq7$, quasinormal modes of tensor type of gravitational perturbations were studied for simply rotating Myers-Perry black holes \cite{Kodama:2009bf} and their asymptotically AdS generalizations \cite{Kodama:2009rq}. It was found that the rotating AdS black holes suffer from the superradiant instability (see Sec.~\ref{sec:superradiance}).

For mini black holes the quantum corrections are significant \cite{Rychkov:2004sf}. These corrections are suggested by the string theory.
Its slope expansion yields higher-order curvature corrections to the Einstein action, which make the gravity be mathematically
non-contradictory in higher dimensions. The first order correction to Einstein gravity is proportional to the square of the curvature and is a topological invariant in four-dimensional space-time \cite{Zwiebach:1985uq}. This correction, called the \emph{Gauss-Bonnet term}, implies a nontrivial correction to the higher-dimensional backgrounds \cite{Boulware:1985wk}. The quasinormal spectrum of the scalar field was studied for the neutral \cite{Iyer:1989rd} and  charged  \cite{Konoplya:2004xx} Gauss-Bonnet black holes, as well as for their asymptotically de Sitter and anti de Sitter generalizations \cite{Abdalla:2005hu}.

The quasinormal modes of tensor- and vector-type gravitational perturbations of Gauss-Bonnet black holes were calculated in \cite{Chakrabarti:2006ei} using the third-order WKB formula. The quasinormal modes of gravitational perturbations of all types and the development of an instability in time domain were found in \cite{GBstability-Konoplya}.

The ringing of the brane-localized standard model fields was studied in the background of Schwarzschild, Schwarzschild-(anti) de Sitter \cite{Kanti:2005xa}, Kerr \cite{Berti:2003yr,Kanti:2006ua}, and Gauss-Bonnet black holes \cite{Zhidenko:2008fp}. The standard model fields were also considered in the scenario of the three-brane with finite tension in the six-dimensional bulk \cite{Chen:2007jz,alBinni:2007gk} and on the two-brane in the four-dimensional version of the Randall-Sundrum model \cite{Nozawa:2008wf}.

In higher-dimensional space-time one can consider more complicated black hole configurations, which were studied for the particular case of five-dimensional space-time. When a gauge field is added, a five-dimensional black hole may be embedded in a rotating universe, which is called the G\"odel universe. The quasinormal modes of the scalar field in the background of such a black hole in the G\"odel universe were studied in \cite{Konoplya:2005sy}. Another family of solutions in the five-dimensional space-time is the squashed Kaluza-Klein black holes, which asymptotically look similar to four-dimensional black holes with the Kaluza-Klein modes due to the compactified extra dimension. The quasinormal spectrum of such black holes allows us to probe the extra dimensions even at small energies, for which the Kaluza-Klein modes are not yet excited. The quasinormal modes of scalar field and gravitational perturbations of squashed Kaluza-Klein black holes were calculated in \cite{Ishihara:2008re} and confirmed by independent calculations of  \cite{He:2008im}. Then, an analysis was continued to the squashed Kaluza-Klein black holes in the G\"odel universe \cite{He:2008kq}.

\subsection{Quasinormal modes of AdS black holes}

In asymptotically AdS backgrounds there is no outgoing wave at spatial infinity. The AdS boundary provides a confinement for perturbations, which are, thereby, localized relatively closely to the black hole. The quasinormal ringing governs evolution of perturbations at all times, i.~e., there is no late-time tail stage \cite{Horowitz:1999jd,Wang:2000gsa} as in the asymptotically flat and de Sitter space-time (see Sec.~\ref{sec:latetimetails}). The high overtone regime is usually already reached  at moderate overtones $n$, and one can find that the real and imaginary parts of asymptotically high overtones have equidistant spacing.

Within the AdS/CFT correspondence the quasinormal modes of black holes in the anti-de Sitter space-time have a direct interpretation in the dual conformal field theory (see Sec.~\ref{sec:AdSCFT}). The quasinormal modes of the conformally invariant scalar field in the Schwarzschild-AdS background were first studied in \cite{ChanMann} prior to the interest in the ADS/CFT correspondence. In the context of the AdS/CFT correspondence, quasinormal modes of a test scalar field were calculated in \cite{Horowitz:1999jd}. Later, quasinormal modes of the electromagnetic field and the gravitational perturbations of the Schwarzschild-AdS black hole were found in \cite{Cardoso:2001bb}. Quasinormal modes of toroidal, cylindrical and planar black holes in anti-de Sitter space-times were calculated in \cite{Cardoso:2001vs}. Quasinormal ringing of the Reissner-Nordstr\"om-AdS background was studied for scalar field \cite{Wang:2000gsa}, gravitational and electromagnetic perturbations \cite{Berti:2003ud}.

When the radius of the black hole is much smaller than the AdS radius, the effect of the black hole is small and the quasinormal modes approach the normal modes of the pure AdS space-time \cite{Zhu:2001vi,Konoplya:2002zu,Konoplya:2002ky}. For large black holes the real and imaginary parts of quasinormal modes grow proportionally to the black hole radius. The only exception is the fundamental vector and scalar modes \cite{Berti:2003ud,Cardoso:2003cj,Konoplya:2003dd}, which are hydrodynamic modes (see Sec.~\ref{sec:hydrodynamic}) \cite{Friess:2006kw}.

Quasinormal modes of the Dirac field were found for the Schwarzschild-AdS \cite{Giammatteo:2004wp} and Reissner-Nordstr\"om-AdS \cite{Jing:2005uy} black holes. In addition, quasinormal ringing was studied for the four-dimensional Kerr-AdS black hole \cite{Giammatteo:2005vu} and planar AdS black holes in four and higher dimensions \cite{Miranda:2005qx,Miranda:2008vb,Morgan:2009pn}. An alternative, perturbative approach for finding quasinormal modes was developed in \cite{Siopsis:2007wn,Alsup:2008fr} for asymptotically flat and AdS black holes.

Using an analytic technique based on the complex coordinate WKB method, the so-called ``highly real'' quasinormal modes were reported to appear in the spectrum of Schwarzschild-anti de Sitter black holes. The real part of these quasinormal frequencies asymptotically approached infinity while the damping rate remained finite \cite{Daghigh:2008jz,Daghigh:2009fy}. Since for large black holes the imaginary part of highly real quasinormal modes is less than those which are established as the fundamental modes, the highly real quasinormal modes must be dominant in the spectrum. Nevertheless, these highly real modes were not observed through time-domain integrations nor any alternative calculations.

\subsection{Exact solutions and analytical expressions}\label{sec:exactqnms}
Here we summarize some analytical formulas for quasinormal modes.
\begin{itemize}
\item \textbf{Large multipole number}: Quasinormal modes can be obtained by substituting the corresponding effective potentials into the first-order WKB formula and taking the limit of large $\ell$.

The large multipole limit for the Schwarzschild black hole was derived in \cite{Mashhoon2} and generalized for the Schwarzschild-de Sitter black hole in \cite{Zhidenko:2003wq} as
\begin{equation}\label{SdSLML}
\omega=\frac{\sqrt{1-9\Lambda M^2}}{3\sqrt{3}M}\left(\ell+\frac{1}{2}-\left(n+\frac{1}{2}\right)\imo\right)+{\cal O}\left(\frac{1}{\ell}\right),
\end{equation}
where $n$ is the overtone number.

This formula is valid also for scalar ($\ell=0,1,2,\ldots$), electromagnetic ($\ell=1,2,3,\ldots$) and Dirac ($\ell=1/2,3/2,5/2,\ldots$) fields. Despite the fact we considered large $\ell$ expansions, the formula~(\ref{SdSLML}) also gives good estimations for the dominant frequencies with small $\ell$. For instance, for the $\ell=3$, $n=0$ gravitational perturbation of the Schwarzschild black hole ($\Lambda=0$), one has $\omega = (0.599-0.093\imo)M^{-1}$, while formula (\ref{SdSLML}) gives $\omega = (0.674-0.096\imo)M^{-1}$.

For higher-dimensional black holes the large multipole expansion reads \cite{Konoplya:2003ii}
\begin{eqnarray}\omega&=&\frac{1}{4}\left(\frac{2}{(D-1)M}\right)^{\frac{1}{D-3}}\sqrt{\frac{D-3}{D-1}}\times
\\\nonumber&&\left(2\ell+D-3-\frac{2n+1}{\sqrt{D-3}}\imo\right)+{\cal O}\left(\frac{1}{\ell}\right).
\end{eqnarray}

This was generalized to the multihorizon black holes in \cite{Vanzo:2004fy}.
\item \textbf{Asymptotically high overtones}: For Schwarzschild black holes, asymptotically high overtones satisfy the relation \cite{Motl:2002hd}
\begin{equation}\label{asymptoteq}
e^{8\pi\omega M}=-1-2cos(\pi s), \qquad \Im{\omega}<0,
\end{equation}
where $s=0,1,2$ for the test scalar, Maxwell, and gravitational fields, respectively. The solution to Eq.~(\ref{asymptoteq}) is
\begin{equation}
2M\omega=\left\{
           \begin{array}{ll}
             \displaystyle\frac{\ln(3)}{4\pi}-\frac{n-1/2}{2}\imo, & \hbox{scalar, gravitational;} \\&\\
             \displaystyle-\frac{n}{2}\imo, & \hbox{Maxwell.}
           \end{array}
         \right.
\end{equation}

The asymptotic behavior of quasinormal modes of the Dirac field is the same as for the Maxwell field: the real part approaches zero, while the spacing of the imaginary part tends to $1/4M$ \cite{Jing:2005dt}.

Formula (\ref{asymptoteq}) was generalized for arbitrary single-horizon black holes as \cite{Tamaki:2004ux}
\begin{equation}
e^{T_H\omega}=-1-2cos(\pi s), \qquad \Im{\omega}<0,
\end{equation}
where $T_H$ is the Hawking temperature.

\jamode{\begin{table*}}{\begin{table}}
\caption{Publications where the higher modes of various black holes were found.}\label{table:highermodes}
\tablefont\begin{tabular}{|l|l|}
\hline
Black hole&Asymptotic formula\\
\hline
Schwarzschild&\cite{PhysRevD.47.5253} (numerical), \cite{Motl:2002hd}\\
Dirac field in Schwarzschild&\cite{Jing:2005dt}\\
Tangherlini&\cite{Motl:2003cd,Birmingham:2003rf}, \cite{Cardoso:2003vt} (numerical)\\
Reissner-Nordstr\"om&\cite{Motl:2003cd} ($D=4$), \cite{Natario:2004jd} ($D>4$)\\
Kerr& \cite{Berti:2004um} (numerical), \cite{Musiri:2003ed,Hod:2005ha}\\
Kerr-Newman&\cite{Keshet:2007nv}\\
Scalar field in Myers-Perry&\cite{Kao:2008sv}\\
Single-horizon (e.~g., dilatonic)&\cite{Tamaki:2004ux}\\
Spherically symmetric&\cite{Das:2004db} (integer spin),\cite{Cho:2005yc}\\
Garfinkle-Horowitz-Strominger&\cite{Chen:2004zr} (scalar), \cite{Chen:2005rm} (Dirac)\\
Gibbons-Maeda dilaton&\cite{Chen:2005pv}\\
Gauss-Bonnet&\cite{Chakrabarti:2005cm}\\
Chern-Simons&\cite{Gonzalez:2010vv}\\
Schwarzschild-de Sitter&\cite{Konoplya:2004uk} (numerical), \cite{Cardoso:2004up}\\
Tangherlini-de Sitter& \cite{Natario:2004jd}\\
Reissner-Nordstr\"om-(A)dS&\cite{Wang:2004bv} (numerical), \cite{Natario:2004jd} ($D\geq4$)\\
EM field in SdS ($D>4$)&\cite{LopezOrtega:2006vn}\\
Extremally charged RN-dS&\cite{Daghigh:2007xj}\\
Large SAdS ($D=4$)&\cite{Cardoso:2004up} \\
Large SAdS ($D=5$)&\cite{Starinets:2002br} (numerical), \cite{Fidkowski:2003nf,Musiri:2003rs}\\
All SAdS&\cite{Natario:2004jd}\\
Massive scalar field&\cite{Konoplya:2004wg} ($D=4$), \cite{Zhidenko:2006rs} ($D>4$)\\
\hline
\end{tabular}
\jamode{\end{table*}}{\end{table}}

The analog of formula (\ref{asymptoteq}) was derived also for charged black holes, for black holes in the asymptotically de Sitter space-time, and for higher-dimensional black holes (see Table~\ref{table:highermodes}). Yet, these equations cannot always be solved algebraically with respect to $\omega$ (see topical review \cite{Natario:2004jd}).

\item \textbf{Near extreme black holes in the de Sitter space-time}: When the black hole horizon approaches the cosmological horizon, the effective potential approaches the P\"oschl-Teller potential (see Sec.~\ref{sec:PoschTeller}) and the quasinormal spectrum can be found analytically \cite{Cardoso:2003sw,Molina:2003ff}. For the four-dimensional Schwarzschild-de Sitter black hole the spectrum has the form
\begin{equation}\label{near-extreme}
\frac{\omega}{\kappa}=\sqrt{(\ell+1-s)(\ell+s)-\frac{1}{4}}-\left(n+\frac{1}{2}\right)\imo,
\end{equation}
where $s=0, 1, 2$ for the test scalar, Maxwell, and gravitational fields, respectively, $\ell=s,s+1,\ldots$ is the multipole number, and $\kappa$ is the surface gravity at the event horizon. The above formula was later confirmed with the help of the accurate Frobenius method \cite{Yoshida:2003zz}.

\item \textbf{Normal modes in the pure anti de Sitter space-time} (without a black hole): The effective potential in this case is an infinite potential well, where bound states exist. There is nothing that could absorb the energy of these states. This is why the quasinormal spectrum consists of normal modes only, i.~e., the imaginary parts of the frequencies are zero. The quasinormal spectrum was first found in \cite{pureAdSqnms} for a massless scalar field. For  gravitational perturbations of four-dimensional AdS space-time the spectrum was calculated in \cite{Cardoso:2003cj} and generalized for an arbitrary number of space-time dimensions in \cite{Natario:2004jd} as
$$\omega R=2n+D+\ell-j,\qquad n\in {\mathbb N},$$
where $R$ is the anti-de Sitter radius, $D$ is the number of space-time dimensions, $\ell$ is the multipole number, $j=1$ for the scalar field and the gravitational perturbations of tensor type, $j=2$ and $j=3$ for vector and scalar types of gravitational perturbations.

\item \textbf{Quasinormal modes in the de Sitter universe}: Perturbations in a de Sitter universe decay as energy can pass through the cosmological horizon. The quasinormal spectrum differs in spaces with odd and even $D$ and can be described by different formulas. The exact solution for odd $D$ was found in \cite{Natario:2004jd} and does not depend on $D$
$$\omega r_\infty=-\imo(2n+\ell+j-1),\qquad n\in {\mathbb N}.$$
Here, $r_\infty$ is the cosmological horizon.
For even $D$ the spectrum has two branches \cite{LopezOrtega:2006my}
$$\omega r_\infty=-\imo(2n+\ell+j-1),\qquad n\in {\mathbb N}$$
and
$$\omega r_\infty=-\imo(2n+D+\ell-j),\qquad n\in {\mathbb N}.$$
It is interesting to note that the second branch of quasinormal modes coincides with the normal modes of AdS space-time if we substitute an imaginary value of the radius of the cosmological horizon.

\item \textbf{Quasinormal modes of the Schwarzschild and Kerr black holes}: The recent development of the theory of Heun equations allows us to find exact solutions for the wave equations in four-dimensional space-time. The Regge-Wheeler equation (\ref{SBH-scalar}), which describes perturbations of the Schwarzschild black hole, can be reduced to the confluent Heun equation and the quasinormal spectrum can be expressed as a solution of nonalgebraic equations with Heun functions. This equation can be solved numerically showing an excellent agreement with the known results \cite{Fiziev:2005ki}. The pair of equations which describes perturbations of the Kerr black hole for any $s$ have been recently solved in \cite{Fiziev:2009ud} as well.

Despite the fact that the quasinormal spectrum for such four-dimensional black holes can be expressed as a solution of a set of equations, the calculation of the quasinormal spectrum remains a nonelementary procedure because of dealing with a much less studied class of special functions.

\item \textbf{BTZ black hole} is the $(2+1)$-dimensional solution of the Einstein equations in asymptotically anti-de Sitter space-time \cite{BTZ}. In $2+1$ space-time there are no gravitational degrees of freedom, although there are dynamical perturbations of test fields. The equations of motion for the test scalar and electromagnetic fields are the same and can be solved analytically in terms of hypergeometric functions. The quasinormal spectrum has the form \cite{Cardoso:2001hn}
\begin{equation}
\omega R = \pm m-2\imo \sqrt{M}(n+1),
\end{equation}
where $R$ is the AdS radius, $M$ is the black hole mass, and $m$ is the azimuthal number.

Since in $2+1$ gravity the Riemann tensor is completely determined by the matter source, the only radiative correction to the geometry comes from quantum excitations of the matter fields, and the perturbative expansion receives no corrections from graviton loops. Thus, it is possible to find the one-loop quantum correction to the BTZ black hole metric and the corresponding correction to the quasinormal spectrum, which is given by \cite{Konoplya:2004ik}
$$\tilde\omega^2=\omega^2-\frac{4m^2+M}{(M)^{3/2}}\frac{l_pF(M)}{R^3},$$
where $\omega$ is the frequency of the BTZ black hole, $l_p$ is the Planck length, and $F(M)$ is a nonlinear function which exponentially approaches zero for large $M$ \cite{Steif:1993zv}.

For the rotating BTZ black hole the exact expression for the quasinormal spectrum of a massive scalar field was also found \cite{Birmingham:2001hc}
\begin{equation}\label{BTZrotation}
\omega R=\pm m-2\imo \frac{r_+\mp r_-}{R}\left(n+\frac{1+\sqrt{1+\mu}}{2}\right),
\end{equation}
where $\mu$ is the mass of the scalar field, and $r_-$ and $r_+$ are the inner and outer horizons, respectively.

Note that for the $(2+1)$-dimensional de Sitter rotating black hole there are several families of the quasinormal frequencies, which are given by (\ref{BTZrotation}) with imaginary values for $R$ and $r_-$. Some of those modes have positive imaginary parts, implying that the black hole is unstable \cite{Abdalla:2002rm}.

\item \textbf{Other exact solutions}: The analytical expressions for the quasinormal spectra of topological AdS black holes were found for the massive scalar field \cite{Aros:2002te} and for gravitational perturbations \cite{Birmingham:2006zx}. Also the exact solutions for quasinormal modes of dilatonic black holes were found in \cite{Becar:2007hu,LopezOrtega:2009zx}.
\end{itemize}

\subsection{Quasinormal modes of massive fields}

Consider a massive scalar field in the space-time of an asymptotically flat black hole. From (\ref{Frobenius-msDRN}) one can see that if one requires a purely outgoing wave at spatial infinity, the behavior of the scalar field amplitude at large distance behaves as
$$\Phi\propto e^{\imo r\sqrt{\omega^2-\mu^2}},$$
where $\mu$ is the mass of the field.

This implies that for $\omega_{Im}\approx0$ and $\omega<\mu$ there is no energy transition to infinity. If, for some values of the black hole parameters, the ingoing wave amplitude at the event horizon is much less than the amplitude far from the black hole, there is no leak of energy from the system and one can observe an analog of standing waves. An almost purely real mode appears in the quasinormal spectrum and the oscillations of the particular frequency become long lived. This phenomenon is called \emph{quasiresonance} \cite{Ohashi:2004wr}.

\begin{figure}
\resizebox{\linewidth}{!}{\includegraphics*{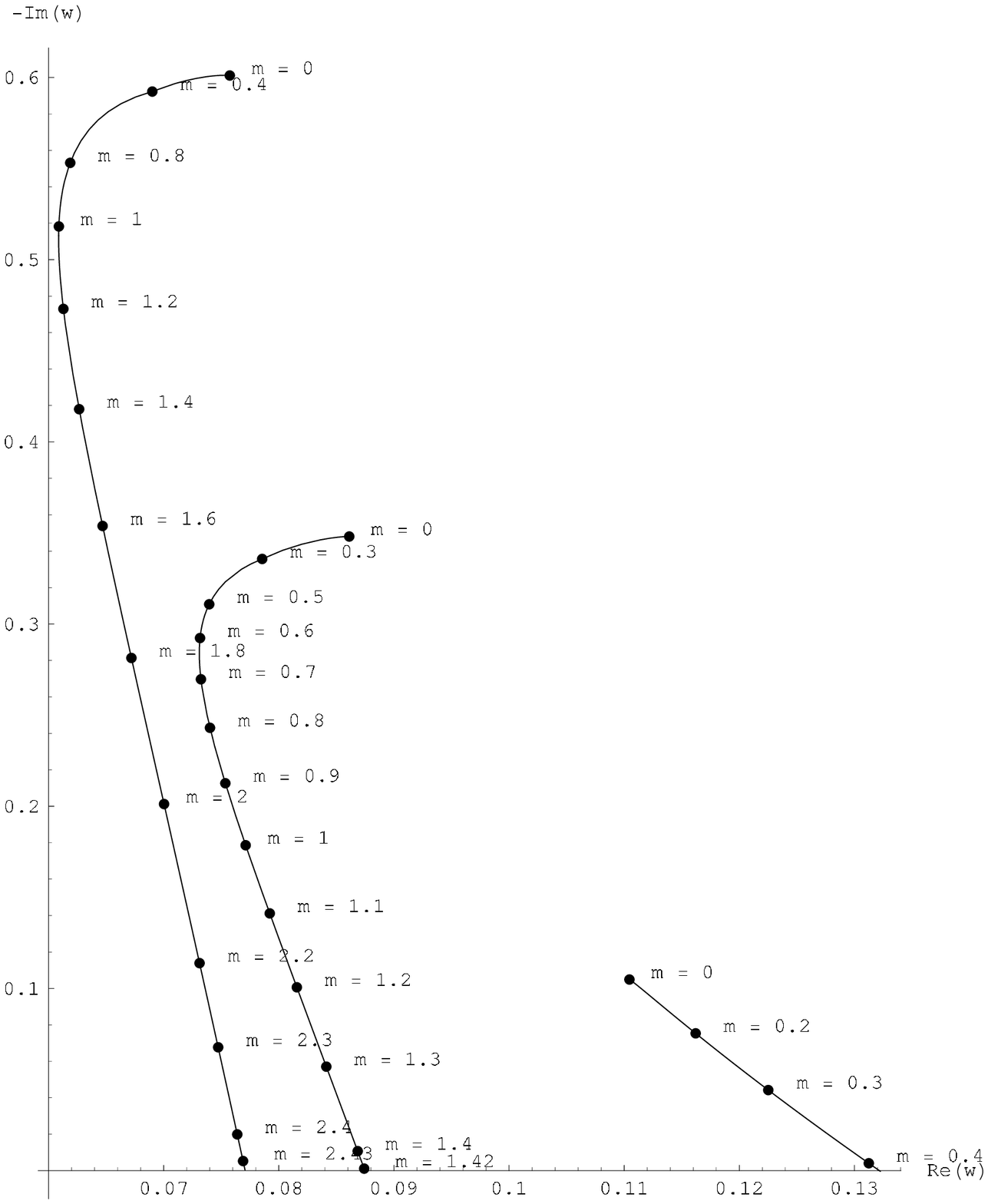}}
\caption{Three dominant quasinormal frequencies of spherically symmetric massive scalar field perturbations in the background of a Schwarzschild black hole ($M=1$) for various scalar field mass $m$.}\label{massivescalar}
\end{figure}

The quasinormal spectra of a massive, neutral scalar field for Schwarzschild and Kerr black holes \cite{Simone:1991wn} and for a massive charged scalar field in the vicinity of a Reissner-Nordstr\"om black hole \cite{Konoplya:2002wt} were studied first within the WKB approximation. It was found that as the field mass increases, the damping rate becomes smaller, which was confirmed by time-domain integration \cite{Xue:2002xa}.
Then, accurate massive modes were obtained in \cite{Konoplya:2004wg} using the convergent Frobenius method (see Sec.~\ref{sec:Frobenius}). It was shown that increasing the mass of the field or the mass of the black hole gives rise to decreasing of the imaginary part of the quasinormal modes until the fundamental mode reaches a \emph{vanishing damping rate}. When some threshold value of the mass is exceeded, the fundamental mode ``disappears'' from the spectrum and the first overtone becomes the fundamental mode (see Fig.~\ref{massivescalar}). Since this mode has a high damping rate, we observed a kind of discontinuity of lifetime values of the oscillations: For some particular values of the field mass the oscillations almost do not decay, while for slightly higher mass we see that the new fundamental mode decays quickly.

The effect of the field mass is crucial for the lower quasinormal modes only. The asymptotic behavior of highly-damped modes does not depend on the field mass \cite{Konoplya:2004wg} and is the same as for the massless field.

Quasiresonance for the Kerr black hole was observed in \cite{Konoplya:2006br}, where the stability of the quasinormal, nonsuperradiant sector of the  spectrum was shown. The spectrum of massive fields in a Kerr background demonstrates a number of phenomena for boundary conditions other than quasinormal, such as amplification of an incident wave (superradiance) and superradiant instability (see Sec.~\ref{sec:superradiance} for details).

For the $D\geq6$ Tangherlini black holes the fundamental mode does not reach quasiresonance, but approaches the field mass asymptotically. Therefore, there is another interesting effect: the appearance of two concurrent long-lived modes with almost the same damping rates. For some values of masses of the field $\mu$ and of the black hole $M$, the superposition of these two oscillations appears \cite{Zhidenko:2006rs}.

The quasinormal spectrum of a massive vector field in the Schwarzschild and Schwarzschild-anti de Sitter backgrounds was studied in \cite{Konoplya:2005hr}. In the Schwarzschild background the fundamental mode of the massive vector field behaves qualitatively similar to the massive scalar one: As the field mass grows, QNMs reach quasiresonance and disappear. The higher overtones become purely imaginary at sufficiently large values of the field mass. The asymptotically high overtones also do not depend on the field mass.

In the asymptotically de Sitter and anti-de Sitter backgrounds, quasiresonances do not exist because the mass term does not change the boundary condition \cite{Zhidenko:2005mv,Konoplya:2004wg}.

Massive fields are short ranged, so their quasinormal ringing will almost certainly not be observed  in near future experiments. Nevertheless, quasinormal modes of massive fields are not only of academic interest: There are numerous of situations where massless fields gain an effective mass. This concerns a self-interacting massless scalar field \cite{Hod:1998ra}, when one takes account of interaction with a black hole, black holes in models with small extra dimensions \cite{Randall-Sundrum} and black holes immersed in a magnetic field \cite{Konoplya:2007yy}.

\subsection{Dependence of the fundamental frequency on black hole parameters}\label{sec:bhparams}
\begin{enumerate}
\item \textbf{Black hole mass}: It is possible to fix one scaling symmetry in the wavelike equation, which allows us to choose the black hole mass $M=1$. In order to convert the frequencies calculated in geometrical units into $kHz$, one should multiply $\omega$ by $2\pi(5.142kHz)M_\odot/M$. For example, for the fundamental mode of the Schwarzschild black hole perturbation $\ell=2$, $$\omega M\approx0.3737-0.0890\imo.$$ Then, one can find that for a black hole of $10$ solar masses the oscillation frequency is $$\nu\approx(1.2074-0.2875\imo)kHz.$$

    The same dependence on the black hole mass ($\omega \sim M^{-1}$) was shown for evaporating black holes \cite{Xue:2003vs}, where the dynamics of the dominant quasinormal frequency was studied when the black hole mass changes. As the black hole mass grows, the real and imaginary parts of the frequencies decrease. The same was observed for the Vaidya background in $4$ \cite{Shao:2004ws,Abdalla:2006vb} and $D$-dimensions \cite{Abdalla:2007hg}. For the evaporating three-dimensional AdS black hole the real and imaginary parts of the quasinormal modes decrease during the black hole evaporation \cite{Shen:2003zm}.

\item \textbf{Electric charge}: For small values of the electric charge $\Re{\omega}$ and the damping rate grow as the charge increases. At some point close to the extremal charge $\Re{\omega}$ reaches its maximum value and then decreases. It has been observed that this point approximately coincides with the point of phase transition, when the heat capacity of the Reissner-Nordstr\"om black hole becomes positive \cite{Jing:2008an}. The damping rate also reaches its maximum after the critical point. However, most probably this is just a numerical coincidence because it takes place only for a particular multipole number.

When we consider charged fields in the background of a charged black hole, we must take into account electromagnetic interactions between the charge of the field and the charge of the black hole. In most cases, as the product of the charges grows, the real part of the quasinormal frequency grows, while the damping rate decreases. This does not happen if the black hole charge is close to its extremal value. At the near extremal charge the behavior of the quasinormal frequencies becomes more complicated: $\omega$ as a function of the black hole charge describes spiral-like curves in the complex plane \cite{Andersson:1996xw}.
\item \textbf{Rotation parameter}: If the angular momentum of the particular component of perturbation ($m>0$ see Sec.~\ref{sec:KerrNewmandS}) coincides with the direction of the black hole rotation, the energy is extracted from the black hole. This causes decreasing of the damping rate and, for some black holes, leads to an instability (see Sec.~\ref{sec:HDinstability}). When the direction of the angular momentum of a perturbation component is opposite to the rotation of the black hole, the perturbation looses additional energy when interacting with the black hole. That is why they usually decay quicker than components with $m>0$. The actual oscillation frequency grows with $m$, which can be seen, e.~g., for the rotating BTZ black hole (see Eq.~(\ref{BTZrotation})), whose QNMs can be obtained as an exact solution \cite{Birmingham:2001hc}. For the Kerr black holes we observe basically the same behavior \cite{Seidel:1989bp}.
\item \textbf{Cosmological constant}: A positive cosmological constant suppresses both the oscillation frequency and the damping rate \cite{Otsuki:1991} for all black holes and all types of perturbations \cite{Zhidenko:2003wq,Konoplya:2007jv,Konoplya:2007zx}. As the black hole approaches its extremal size, i.~e., the event horizon approaches the cosmological horizon, the quasinormal frequencies tend to zero, being proportional to the surface gravity at the event horizon (see Eq.~(\ref{near-extreme})).
\item \textbf{Extra dimensions}: The radius of the Tangherlini black hole horizon $r_+$ is related with its mass $M$ as
    $$r_+^{D-3}=\frac{(D-2)M{\cal A}_{D-2}}{4\pi},$$
where ${\cal A}_{D-2}$ is the area of a unit $(D-2)$-sphere.
Therefore, if one compares scattering properties around black holes living in worlds of different number of space-time dimensions $D$, it is convenient to measure all the quantities in units of the black hole radius.

For a black hole in the scenario with large extra dimensions, the increasing of $D$ leads to increasing of both the real frequency and the damping rate. The damping rate grows quicker than the oscillation frequency and thus, the higher-dimensional black hole is a worse oscillator than the lower-dimensional one. For the brane-localized fields, the real part of the frequency decreases, while the damping rate increases.

\item \textbf{External magnetic field}: For the Ernst black hole, which is a black hole immersed in an asymptotically uniform magnetic field, it was shown that the first order correction from the external magnetic field upon the motion of the scalar field has the form of an effective mass $\mu_{eff}^2 \sim m^2B^2$, where $m$ is the azimuthal number and $B$ is the strength of the magnetic field \cite{Konoplya:2007yy}.
    For a massive charged scalar field in the Kerr-Newman background there are two main effects on the quasinormal spectrum: the Zeeman shift of the particle energy in the magnetic field and the difference of values of an electromagnetic potential between the horizon and infinity, i.e. the Faraday induction \cite{Kokkotas-Konoplya-Zhidenko}.
\item The \textbf{Gauss-Bonnet term} diminishes the damping rate of the fundamental mode, implying a black hole of the same radius is a better oscillator \cite{GBstability-Konoplya,Zhidenko:2008fp}.
\end{enumerate}

\section{Quantization of the black hole area}\label{sec:lqg}

Even though we do not have a fully noncontradictory theory of quantum gravity at hand, there are numerous attempts to study black holes using the general principles of quantum theory. There are indications that the quasinormal spectrum of a black hole possibly allows us to find the law of a black hole area quantization. According to Bohr's correspondence principle, in the regime of  large quantum numbers, i.~e., when the quantum effects are small, one must reproduce classical conservative (adiabatic) quantities. For stationary black holes, one such quantity is its mass, which must have a discrete spectrum in the quantum theory.

It was suggested in \cite{Bekenstein:1995ju} that the highly damped oscillation frequencies of a black hole are associated with the semiclassical limit of the black hole mass transition, while their damping rates correspond to the relaxation time. The relaxation time for asymptotically high overtones goes to zero, which is compatible with the semiclassical approximation.

Thus, one can write the spacing for the Schwarzschild black hole mass as
\begin{equation}
\Delta M=\hbar\Delta\omega,
\end{equation}
and, consequently, for the black hole area as
\begin{equation}
\Delta A=\Delta(16\pi M^2)\simeq32\pi M\Delta M=32\pi\hbar M\Delta\omega.
\end{equation}

S. Hod suggested \cite{Hod:1998vk} taking the asymptotical limit of the real part of the quasinormal frequency as $\Delta\omega$:
$$\Delta\omega=\Re{\omega_\infty}=\frac{\ln3}{8\pi M}.$$

However, some difficulties were found from this conjecture:
\begin{enumerate}
\item The value of the $\Re{\omega_\infty}$ is non-vanishing for gravitational perturbations, while, e.~g., for the vector and Dirac perturbations it is zero (see Sec.~\ref{sec:exactqnms}).
\item The limits of zero rotation and zero charge do not commute with the large overtone limit \cite{Berti:2003zu,Berti:2003jh}. For an infinitesimal rotation the limit of the real part and, consequently, the area quantum becomes arbitrarily small.
\item Such a value of $\Delta\omega$ corresponds to the transitions from the ground state of the black hole to a state with large $n$ \cite{Maggiore:2007nq}.
\end{enumerate}

All these difficulties can be removed if we consider a transition $n\rightarrow (n-1)\gg1$ and associate the energy with the eigenfrequency of some damped oscillator. This eigenfrequency does not coincide with the real part of the quasinormal frequency \cite{Maggiore:2007nq}.

Indeed, consider the damped harmonic oscillator which can be described by the wave equation
$$\ddot{\xi}+2\gamma\dot{\xi}+\omega_0^2\xi=0.$$
The resonant frequencies of the oscillator are the roots of the characteristic equation
$$\omega^2+2\imo\gamma\omega-\omega_0^2=0.$$
The solutions to this equation are
$$\omega=\pm\sqrt{\omega_0^2-\gamma^2}-\imo\gamma.$$
Thus, $\omega_0^2=\Re{\omega}^2+\Im{\omega}^2$.

The $n\rightarrow (n-1)$ transition corresponds to the frequency spacing
$$\Delta\omega_0=\Delta\sqrt{\Re{\omega}^2+\Im{\omega}^2}\simeq\Delta\Im{\omega}=\frac{1}{4M}.$$

Then, the area quantum is
\begin{equation}\label{areaquantum}
\Delta A=8\pi\hbar=8\pi l_p^2,
\end{equation}
where $l_p$ is the Plank length.
Although the result obtained was initially demonstrated for the Schwarzschild black hole,
it was found in a similar way that for the Reissner-Nordstr\"om and Kerr solutions the area quantum is the same \cite{Vagenas:2008yi,Medved:2008iq}.
Thus, the expression for the area quantum (\ref{areaquantum}) seems to not depend much on the details of the black hole geometry.
The quantization of the black hole area in three and five dimensions was considered in \cite{Wei:2009yj}.

Within loop quantum gravity the basis of the Hilbert space is given by spin networks. The surface acquires the area due to spin edges puncturing \cite{Rovelli:1994ge}
\begin{equation}\label{LQGarea}
A=8\pi l_p^2\gamma\sum_{i}\sqrt{j_i(j_i+1)},
\end{equation}
where $\gamma$ is the Barbero-Immirzi parameter \cite{Barbero:1994ap,Immirzi:1996dr}, and $j_i$ is a (half)integer spin.

It was suggested, that the asymptotic behavior of the quasinormal modes allows us to fix $\gamma$ by taking \cite{Dreyer:2002vy}
$$\Delta A = A_{min},$$
where the minimal surface is given by
$$A_{min}=8\pi l_p^2\gamma\sqrt{j_{min}(j_{min}+1)}.$$
Here $j_{min}$ is the lowest possible spin. Then, from (\ref{areaquantum}), we find
$$\gamma\sqrt{j_{min}(j_{min}+1)}=1.$$
However, in order to reproduce the Bekenstein entropy, we need an additional assumption: Either that $j=1/2$ edges puncturing
the horizon is suppressed due to some unknown dynamics \cite{Corichi:2002ty} or an introduction of supersymmetry \cite{Ling:2003iz}.

As was pointed out in \cite{Khriplovich:2004fd}, the semiclassical limit requires consideration of large $j$, for which the area quantum is
$$\Delta A=8\pi l_p^2\gamma\Delta\sqrt{j(j+1)}\simeq 4\pi l_p^2\gamma, \qquad j\gg1.$$
This fixes $\gamma=2$.

The black hole entropy is given by \cite{Maggiore:2007nq},
\begin{equation}
S=\frac{A}{4l_p^2}=2\pi\frac{A}{\Delta A}+{\cal O}(1).
\end{equation}

Finally, we note that all the above arguments are rather more speculative than mathematically strict. The correlation among the quasinormal modes, area quantization, and the Barbero-Immirzi parameter, if it exists, is probably much more subtle than one could naively expect.

\section{Observation of gravitational waves}
\subsection{Gravitational waves from black holes}
Black holes can be perturbed due to different processes, such as formation of a black after collapse of supernovae, black holes, and/or neutron stars mergers, or an infall of matter into a supermassive black hole  \cite{Echeverria,Finn,Anninos,Dreyer}. One of the most frequent phenomenon is the perturbation produced by accretion of matter. Unfortunately, the estimations of the gravitational-wave amplitudes show that these waves are too weak to be detected by gravitational-wave antennas within the present sensitivity. However, within the hyperaccretion scenario gravitational waves are potentially detectable for highly spinning black holes ($a\simeq0.98$) by the next-generation detectors LIGO II \cite{ArayaGochez:2003vb}, LISA \cite{Hughes:2007xm,Jadhav:2008dw}, and ET \cite{ET}.

The estimations for gravitational waves from black holes produced by stellar collapse show that it may be possible to detect gravitational waves from the collapse of stars whose masses are at least $300 M_\odot$ \cite{Fryer:2001zw}. These heavy primordial stars have not been observed; however, currently there is existing evidence of such intermediate-mass black holes, and we can expect that their ringing is observed by the next generation of gravitational-wave antennas \cite{Miller:2003sc}.

In fact, the only realistic scenario for the detection of gravitational waves from astrophysical black holes with the help of the current antennas is observations of collisions of black holes and/or neutron stars. Unfortunately, the estimated binary merger event rate is also small for the present sensitivity (see Sec.~\ref{sec:nstars}). Nevertheless, the recent estimations show that the black hole-black hole inspirals are more likely sources than systems of neutron stars and can be detected in the optimistic scenario even with the help of the current detectors such as LIGO or VIRGO \cite{Belczynski:2010tb}.

The ringing of supermassive black holes in the centers of galaxies \cite{Melia:2007vt} can probably be detected by the future space-based detector LISA. There are many preliminary estimates for event rates of supermassive black hole mergers, in which the number depends on the model and varies from several to thousands of events per year \cite{Berti:2006ew,Berti:2005ys}. LISA should also detect gravitational waves from intermediate-mass black hole mergers, but the estimates of such events are even more uncertain: There are predictions that we will be able to observe several events per year \cite{Miller:2004va}.

Since there are a number of works and reviews that discuss the possibility of observations of gravitational waves from black holes, we do not discuss this topic in more detail here, but instead refer the reader to a recent review \cite{Berti:2009kk}.

\subsection{Ringing of neutron stars}\label{sec:nstars}
Black holes are not the only source of detectable gravitational waves. Another source of gravitational waves are neutron stars, which have been extensively studied during the past four decades. Efforts in this direction are motivated by advances in the first generation of large-scale interferometric gravitational-wave detectors (LIGO, GEO600 and Virgo), which have reached the original design sensitivity \cite{Abbott:2007kv}. During the next five years, the second generation of gravitational-wave detectors is expected to reach their level of sensitivity, which is one order better than the first generation. Gravitational-wave observations are expected to be complementary to the information obtained from electromagnetic observations \cite{Owen:2009tj}. This could probably constrain various theoretical models of neutron stars \cite{Gaertig:2010kc,Andersson:2004bi}.

The most promising sources of gravitational waves among various types of stars are as follows:
\begin{itemize}
\item \textbf{Binary evolution}: Binary systems, i.~e., systems of two stars orbiting around their common center of mass, radiate due to rotation and, at the last stage, the orbits shrink causing an increase in the amplitude of the gravitational radiation. The amplitude of the signal can be calibrated by the two masses. From the observed signal one can extract the spin rates of the objects and the distance to the source \cite{Sathyaprakash:2009xs}. The horizon distance for the current antennas is about $30Mpc$, which gives the estimated frequency of the binary merging of one event per 25-400 years. Since the second generation detectors are expected to provide a horizon distance of about $300Mpc$, which corresponds to several events per year, one can expect that gravitational waves from binary merging will be detected by the advanced LIGO and Virgo detectors \cite{Isenberg:2008kz}.
\item \textbf{Collapse of a supernovae core} is described by complicated dynamics, which requires solving  Einstein's equations together with equations of state for the matter and taking account of neutrino interactions. Recent numerical simulations of core collapse and the evolution of the collapsed remnant indicate that gravitational waves can be detected by modern gravitational antennas \cite{New:2002ew,Ott:2008wt}.
\item \textbf{Rotating stars} with nonaxially symmetric deformation of the star's crust, core, or in an external magnetic field are expected to lose their rotational energy, which is emitted as gravitational waves \cite{Ott:2008wt,Andersson:2010}. Radio pulsars are promising candidates for the observation of gravitational waves because electromagnetic observations can be used to target searches of the gravitational signals. According to a recent simulation, the deformation of the star's crust can be large enough to produce the gravitational-wave signal which can be detected by advanced LIGO and Virgo detectors \cite{Horowitz:2009ya}.
\item \textbf{Oscillations of stars}: Neutron stars have different oscillating modes, which can produce gravitational waves. These modes could allow us to probe internal processes of the stars. Numerical simulations, based on  realistic equations of state, provide a set of empirical relations between the frequencies of the oscillations and star parameters. Some of these modes can be detected by the new generation of the gravitational-wave detectors. If such gravitational waves are detected, we will be able to deduce a star's mass and radius with relative error of less than 1\% \cite{Andersson:1997rn,Gaertig:2010kc}.
\end{itemize}

Because of their high density, neutron stars have to be excellent sources of gravitational waves. The detection of their ringing, in particular with the help of the next generation of gravitational-wave detectors such as the Einstein Telescope, will give us the possibility to probe the physics of their interior (see topical review \cite{Andersson:2009yt}).

\section{Late-time tails}\label{sec:latetimetails}

\subsection{Generic approach}

As mentioned, as a manifestation of the noncompleteness of the set of quasinormal modes, at sufficiently late time the quasinormal modes are suppressed by the exponential or power-law tails. The generic approach to analysis of the late-time tails, which we relate here, was suggested in \cite{CLSY}. We start from the wave equation (\ref{wave1}) without implying the stationary ansatz $\Psi \sim e^{i \omega t}$, i.~e.,
\begin{equation}
-\frac{d^2\Psi}{dr_*^2} + \frac{d^2\Psi}{dt^2} + V(r_{*}) \Psi = 0
\end{equation}
We define the pair of the following operators:
\begin{eqnarray}
D &=& -\frac{d^2}{dr_*^2} + \frac{d^2}{dt^2} + V(r_{*}),\\
\widetilde{D}(\omega) &=& -\frac{d^2}{dr_*^2} - \omega^2 + V(r_{*}).
\end{eqnarray}

The wave function $\Psi(r_{*}, t)$ can be written for $t>0$ as the integral over the spatial coordinate
\begin{eqnarray}
\Psi(r_{*}, t) = \int d r_{*}' G(r_{*}, r_{*}'; t) \partial_{t}  \Psi(r_{*}', 0) + \nonumber\\
\int d r_{*}' \partial_{t} G(r_{*}, r_{*}'; t)  \Psi(r_{*}', 0),
\end{eqnarray}
where $G(r_{*}, r_{*}'; t)$ is the retarded Green's function
\begin{equation}
D G(r_{*}, r_{*}'; t) = \delta (t) \delta (r_{*} - r_{*}').
\end{equation}
The Fourier transform reads
\begin{equation}
\widetilde{D}(\omega) \widetilde{G}(r_{*}, r_{*}'; \omega) = \delta (r_{*} -
r_{*}').
\end{equation}
The boundary conditions for black holes and stars differ, though the late-time behavior does not because it depends only upon the asymptotics of the effective potential at spatial infinity. Therefore, we think of the radial tortoise coordinate $r_{*}$ as the one that changes from zero to infinity $0 \leq r_{*} < \infty$ (half-line problem) for some nonsingular systems, such as relativistic stars, or as the one that changes from the event horizon to infinity  $-\infty < r_{*} < \infty$ (full-line problem) for black holes. The boundary condition at infinity for both cases is the same: pure outgoing waves at infinity, while at the left boundary the boundary conditions read
\begin{equation}
\widetilde{\Psi}(r_{*}, \omega) \propto e^{- i \omega r_{*}}, \quad r_{*} \rightarrow - \infty
\end{equation}
for black holes, and
\begin{equation}
\widetilde{\Psi}(r_{*}, \omega) \rightarrow 0, \quad r_{*} \rightarrow 0
\end{equation}
for some nonsingular objects. For simplicity we consider potentials which vanish at both boundaries.

We define the two functions $f(\omega, r_{*})$ and $g(\omega, r_{*})$ which are solutions to the equation
$$\tilde{D}(\omega) f(\omega, r_{*}) = \tilde{D}(\omega) g(\omega, r_{*}) =
0,$$
where $f$ satisfies the left boundary condition (at the horizon or in the origin) and $g$ satisfies the right boundary condition (at spatial infinity). The normalization condition reads
$$\lim_{r_{*}\rightarrow \infty} [e^{-i \omega r_{*}} g(\omega, r_{*})] = 1; $$
and for the half-line problem
$$f(\omega, r_{*}=0) = 0, \qquad f^\prime (\omega, r_{*}=0) = 1,$$
while for the full-line problem
$$\lim_{r_{*} \rightarrow - \infty } [e^{i \omega r_{*}} f(\omega, r_{*})] = 1.$$
The Wronskian has the form
$$ W(\omega) = W(g,f) = g \partial_{r_{*}} f  - f \partial_{r_{*}} g$$
and is independent of $r_{*}$. Then, one can find that
\begin{equation}
\tilde{G}(r_{*}, r_{*}';\omega )= \begin{cases}
{\ds f(\omega, r_{*}) g(\omega, r_{*}') / \ds W(\omega )} \ , &  r_{*} < r_{*}' \cr \cr
{\ds f(\omega, r_{*}') g(\omega, r_{*}) / \ds W(\omega )} \ , &  r_{*}' < r_{*}
\end{cases}
\ \  \ \ .
\end{equation}

Now one needs to integrate $f$ from the left and $g$ from the right, until some common point.

\begin{figure}
\resizebox{\linewidth}{!}{\includegraphics*{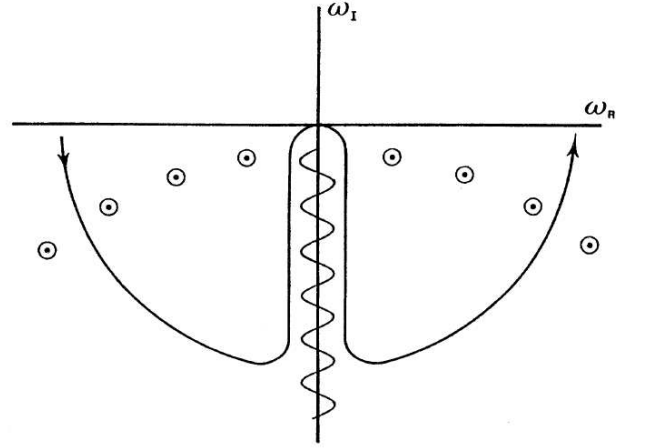}}
\caption{Singularity structure of $\tilde{G}(r_{*}, r_{*}'; \omega)$ in the lower half $\omega$ plane, and the contributions of the corresponding residues to the Green's function $G(r_{*}, r_{*}'; t)$. The figure is taken from \cite{CLSY}.}\label{poles}
\end{figure}

We consider the inverse transformation
\begin{equation}
\tilde{G}(r_{*}, r_{*}'; \omega) = \intop_{0}^{\infty} dt G(r_{*}, r_{*}'; t) e^{i \omega t},
\end{equation}
and for $t > 0$ we close the contour in the complex $\omega$-plane by a semicircle of radius $C$ in the lower halfplane, taking the limit $C \rightarrow  \infty $. Then, one must take into consideration poles of $\tilde{G}$ for $\Im{\omega} < 0$ (Fig.~\ref{poles}). Further, one can distinguish three different contributions to $G$.

\begin{enumerate}
\item For a wide class of potentials, for instance, if $V(r_{*} \rightarrow \infty) \sim r_{*}^{-\alpha} (\ln r_{*})^{\beta}$, $\beta = 0, 1$, the function $g(\omega, r_{*})$ has singularities on the $-Im \omega$ axis and has, thereby, the form of a branch cut along the imaginary axis. This branch cut contribution to the Green's function is called the \textit{tail contribution}.

\item At some complex $\omega = \omega_j$, the Wronskian $W(\omega)$ has zeros and, therefore, the functions $f$ and $g$ are linearly dependent, i.~e., one can find a solution that satisfies both left and right boundary conditions, i.~e., $ \omega_j$ are quasinormal modes. This contribution is naturally called the \textit{quasinormal modes contribution}. This contribution decays exponentially as $t$ increases, so that for the asymptotically late-time behavior $t \rightarrow \infty$, the quasinormal modes contribution is negligible for most cases (although not for AdS space-times, see Sec.~\ref{sec:defQNMs}).

\item The contribution from the semicircle $|\omega| = C$, $C \rightarrow \infty$ corresponds to large $\omega$ or short time. This contribution vanishes after some time and therefore is called the \textit{prompt contribution}.
\end{enumerate}

Now we are in a position to summarize the results obtained for asymptotically late-time behavior of various black holes.

\subsection{Late-time tails of massive and massless fields}

The first study of late-time tails was performed by R. Price \cite{Price} who analyzed the asymptotic behavior of the linearized wave equation for the Schwarzschild black hole and showed that perturbations of the massless scalar and gravitational fields decay according to the power law
\begin{equation}
|\Psi| \sim t^{-(2 \ell + 3)},
\end{equation}
where $\ell$ is the multipole number. Soon after, Bi\`{c}\'{a}k found \cite{Bicak} that the Reissner-Nordstr\"{o}m tails for the scalar field are different for nonextremal and extremal charge,
\begin{equation}
|\Psi| \sim \left\{%
\begin{array}{ll}
    t^{-(2l+2)}, & \hbox{if $Q<M$;} \\
    t^{-(l+2)}, & \hbox{if $Q=M$.} \\
\end{array}%
\right.
\end{equation}
The asymptotic tails for the Kerr space-time were shown to be the same as for the Schwarzschild one. We note that the above results hold when the problem allows for ``good'' initial conditions of compact support.

Unlike asymptotically flat solutions, de Sitter black holes have qualitatively different late-time behavior \cite{Brady:1997,Brady:1999}. First, the asymptotic tails are not power law but exponential, and second, the $\ell =0$ mode does not go to zero, as takes place for the Schwarzschild case, but asymptote to some constant
\begin{equation}
|\Psi| \sim e^{-\ell k_c t} \quad if \quad \ell = 1, 2, \ldots,
\end{equation}
\begin{equation}
|\Psi| = |\Psi_{0}| + |\Psi_{1}| e^{-2 k_c t}, \quad \mbox{if}
\quad
\ell = 0,
\end{equation}
where $k_c$ is the surface gravity on the cosmological horizon.

The late-time behavior of massive fields is drastically different from massless ones. The late-time tails are not purely decaying but intensively oscillating for massive fields (see, for instance, Fig.~\ref{instability}). One of the reasons is that for massive fields tails appear already for Minkowski space-time, which serves as a dispersive medium for massive scalar field. Using the exact Green's functions it was shown \cite{Burko1} that the late-time tails of the scalar field in flat space-time are
\begin{equation}
|\Psi_{flat}| \sim t^{-\ell - \frac{3}{2}} \sin (\mu t),
\end{equation}
where $\mu$ is the inverse Compton wavelength. The Minkowski space-time tail shows itself in the black hole tails
at the so-called \emph{intermediately late time}, i.~e., when
$$1 \ll t/M < (\mu M)^{-3}, $$
for massive scalar and Dirac fields.
For the Proca field, the intermediate late-time tail depends on the polarization of the field and can be either $t^{-(\ell + 1/2)} \sin (\mu t)$, $t^{-(\ell + 3/2)} \sin (\mu t)$, or $t^{-(\ell + 5/2)} \sin (\mu t)$ \cite{Konoplya-Molina-Zhidenko}.

In the presence of a black hole the massive scalar field decays at asymptotically late time (which now can be defined as $t/M > (\mu M)^{-3} $) as \cite{Koyama123}
\begin{equation}\label{5over6}
|\Psi| \sim t^{-\frac{5}{6}} \sin (\mu t), t \rightarrow \infty,
\end{equation}
independently of the angular number $\ell$. This asymptotic decay is quite universal not only because of independence on $\ell$ but also because it takes place for a quite general class of black holes, which includes Schwarzschild \cite{Koyama123}, Kerr \cite{Burko1} or dilaton \cite{Rogatko} solutions as well as for other massive fields (Dirac \cite{Jing-dirac-tail} and Proca \cite{Konoplya-Molina-Zhidenko}). One may expect, therefore, that this behavior (\ref{5over6}) is universal for all massive fields and does not depend much  on the details of the black hole horizon geometry \cite{Konoplya-Molina-Zhidenko}. Although this supposition sounds too strong and was not checked until recently.

Finally, it should be noted that the Price decay law $|\Psi| \sim t^{-(2 \ell + 3)}$ holds for some time-independent backgrounds, while for time-dependent solutions a different law $|\Psi| \sim t^{-(2 \ell + 2)}$ was recently found \cite{time-dependent-tail}.

\subsection{Are asymptotic tails pure nonlinear phenomena?}

In the linear approximation, the study of asymptotic tails in higher than four dimensions was initiated by Cardoso et al. in \cite{evenDtail}. There it was stated that the late-time decay law reads
\begin{equation}
|\Psi_{even}| \sim t^{- (2 \ell + 3 D -8)}
\end{equation}
for an even number of space-time dimensions $D>4$ and is
\begin{equation}
|\Psi_{odd}| \sim t^{- (2 \ell + D - 2)}
\end{equation}
for odd $D$ \cite{oddDtail}. Now the case $D=4$ looks quite special. Further, it was shown that in odd dimensions the late-time tails are independent on the details of the black hole geometry and exist already in the Minkowski space-time even for massless fields. Therefore, it did not come as a surprise that odd-dimensional Gauss-Bonnet black holes have the same decay law at asymptotically late time \cite{Abdalla:2005hu}.

Recently, however, the above results, obtained in the linear approximation for $D>4$ black holes and probably for some $D=4$ solutions, have been challenged in \cite{nonlineartails1} by consideration of the fully nonlinear perturbations of the $D$-dimensional Schwarzschild black hole. In \cite{nonlineartails1} it is stated that the nonlinear approach gives slower decay at asymptotically late time for all $D$, except $D=4$, so that the asymptotic tails seem to be essentially a nonlinear phenomenon. This implies that the $D=4$ case appears as an exception when both linear and nonlinear analyses lead to the same result. The possible nonlinear effect on asymptotic tails was also reported in \cite{nonlineartails2}, although, nonlinear analysis of black hole tails is still poor and needs further effort in order to prove or disclaim the results obtained in the linear approximation.

\section{Stability}\label{sec:stability}
\subsection{Gravitational (in)stability: General aspects}

Gravitational stability of any static or stationary solution describing a compact object which is expected to exist in nature is a principal issue of the physical adequateness of the solution. Objects which are unstable under small perturbations will be inevitably destroyed by them and, thereby, simply cannot exist. A compact object can be a black hole or brane, a star or a wormhole, or even more exotic objects such as a black ring, black saturn, or a naked singularity. Any of them can be accepted as properly described by its metric only after the metrics is proved to be stable.

Since dynamical perturbations of a gravitational system can usually be described by a single wavelike equation (\ref{wave1}) (or a set of such equations), one can analyze stability against linearized perturbations in the following way. Under existence of the well-posed initial value problem, the complete spectrum of quasinormal modes, we judge about stability: if all quasinormal modes are decaying in time, the solution is evidently stable. If there is at least one growing mode, the space-time is unstable with the instability growth rate, which is proportional to the imaginary part of the growing QNM.

If the effective potential $V_{eff}$ in the wave equation (\ref{wave1}) is positive definite, the differential operator
\begin{equation}
A= - \frac{\partial^{2}}{\partial r_{*}^2} + V_{eff}
\end{equation}
is a positive self-adjoint operator in the Hilbert space of square integrable functions $L^{2}(r_{*}, d r_{*})$.
Then, there are no negative (growing) mode solutions that are normalizable, i.~e., for a well-behaved initial data (smooth data of compact support), all solutions are bounded all of the time.

If the effective potential is not positive definite everywhere outside the horizon, a special trick, called the \emph{Friedrichs extension}, can be used, which allows us to prove stability (if there is one) in some cases. If the effective potential $V_{eff}$ is bounded from below, the operator $A$ is a symmetric semibounded operator for smooth functions with compact support in $r_{*}$ Then, $A$ can be extended to a semibounded self-adjoint operator in such a way that the lower bound of the spectrum of the extension is the same as that of $A$. Such an extension is unique for asymptotically flat space-times, because the range of $r_{*}$ is complete $(-\infty, +\infty)$, and thereby the time evolution of the wave function $\Psi$ is uniquely determined by the chosen initial data. Finally, if one can find the positive definite extended operator, this means that the space-time is stable.

For asymptotically anti-de Sitter space-times, the range of $r_{*}$, which is $(-\infty, r_{*}^{0})$, where $r_{*}^{0}$ is some constant, is not complete and the Schwarzschild wedge is not globally hyperbolic. Nevertheless, upon imposing the Dirichlet boundary conditions (\ref{BC3}), one can use the Friedrichs extension as well \cite{Kodama:2003jz,Ishibashi:2003ap,Kodama:2003kk}. Unfortunately in the majority of interesting cases, such as higher-dimensional Schwarzschild-AdS, Schwarzschild-dS, Reissner-Nordstr\"om black holes and various other generalizations, the Friedrichs extension method (which is also called the $S$-deformation method) does not work well, because it is difficult to find an appropriate ansatz for the function which performs the extension of the operator. Therefore \emph{a practical tool for testing stability in all those cases is numerical investigation of the quasinormal spectra } \cite{Konoplya:2007jv,KonoplyaPRL}.

\subsection{Gravitational stability in four dimensions}

The first work on the black hole stability (by Regge and Wheeler \cite{Regge:1957td}) proved stability of the $(3+1)$-dimensional Schwarzschild black holes against axial type of perturbations (though polar perturbations were considered there as well but with a mistake which was
corrected later by Zerilli \cite{Zerilli}). Since that time, the Schwarzschild solution gained status of a physically adequate model for neutral isolated nonrotating black holes.

The next step was an analysis of perturbations and proof of stability of Reissner-Nordstr\"om black holes performed, almost at the same
time and independently, by N. Sibgatulin and G. Alexeev \cite{Alexeev1} and by Moncrief \cite{Moncrief:1974}. In \cite{Alexeev1} in addition to the decoupling of variable and deduction of the master wave equations, the gravitational scattering around Reissner-Nordstr\"om black holes was considered that included construction of the Green's function for the wave equations.

The Kerr black hole was shown to be stable against linear perturbations by Teukolsky and Press by using the first-order WKB approach \cite{Press:1973zz,Teukolsky:1974yv}. A mathematical proof of stability was performed much later by Whiting \cite{Whiting:1988vc}.

The $(3+1)$ Schwarzschild-de Sitter and Reissner-Nordstr\"om de Sitter black holes were proven to be stable as well \cite{Mellor:1989ac}, and the stability of the Schwarzschild-anti-de Sitter solution was shown in \cite{Cardoso:2001bb}. The stability of Kerr-de Sitter black holes was shown recently in \cite{Suzuki:1999nn}, and of Kerr-AdS black holes in \cite{Giammatteo:2005vu}.

Finally, the $D=4$ string theory inspired black holes, such as dilaton black holes, Born-Infeld black holes, and Gauss-Bonnet coupled to dilaton black holes, were shown to be gravitationally stable as well \cite{Fernando:2004pc,Kanti:1997br,Holzhey:1991bx,Ferrari:2000ep,Torii:1998,Pani:2009wy}. Although the stability analysis of the Born-Infeld black holes \cite{Fernando:2004pc} is still incomplete and includes only axial perturbations, leaving therefore a chance of instability in the polar modes.

\jamode{\begin{table*}}{\begin{table}}
\caption{Stability of four-dimensional black holes. All of the four-dimensional black holes considered here are stable, except Kerr-Newman black holes and its string theory generalization (dilaton-axion black holes) for which the variables cannot be decoupled in an easy way.}
\label{tableone}
\tablefont\begin{tabular}{|c|c|}
  \hline
 Black hole solution (parameters) & Publication \\
   \hline
  Schwarzschild ($M$) & \cite{Regge:1957td} \\
   \hline
  Reissner-Nordstr\"om ($M$, $Q$) & \cite{Moncrief:1974,Alexeev1} \\
    \hline
  Schwarzschild-dS (M, $\Lambda >0$)& \cite{Mellor:1989ac} \\
      \hline
  Schwarzschild-AdS (M, $\Lambda <0$)& \cite{Cardoso:2001bb} \\
    \hline
  Reissner-Nordstr\"om-dS (M, Q, $\Lambda$)  & \cite{Mellor:1989ac} \\
      \hline
  Kerr (M, J) & \cite{Press:1973zz,Teukolsky:1974yv} \\
    \hline
  Kerr-dS (M, J, $\Lambda >0$) & \cite{Suzuki:1999nn} \\
    \hline
  Kerr-AdS (M, J, $\Lambda<0$)& \cite{Giammatteo:2005vu} \\
    \hline
  Kerr-Newmann (M, J, Q)& ? \\
    \hline
  Kerr-Newman-A(dS) (M, J, Q, $\Lambda$)& ? \\
    \hline
  Dilaton (M, Q, $\phi$) & \cite{Kanti:1997br,Holzhey:1991bx,Ferrari:2000ep} \\
    \hline
  Dilaton-axion  (M, Q, J, $\phi$, $\psi$) & ? \\
    \hline
  Dilaton-GB (M, $\phi$) & \cite{Torii:1998,Pani:2009wy} \\
    \hline
  Born-Infeld (M, Q) & axial \cite{Fernando:2004pc} \\
       \hline
  Black universes (M, $\phi$) & \cite{Broonikov-Kon-Zhid} \\
  \hline
  BHs in the Chern-Simons theory (M, $\beta$) & \cite{Cardoso-CS} \\
  \hline
\end{tabular}
\jamode{\end{table*}}{\end{table}}

All of the above mentioned works are summarized in Table~\ref{tableone}. We concluded, therefore, that all of the considered here four-dimensional black holes, tested for stability, proved to be stable. Although Kerr-Newman black holes, and their string theory generalizations, which include axion and dilaton fields, are still not tested for stability because the perturbation equations do not allow for an easy decoupling of the angular variables.

\subsection{Gravitational instabilities in higher than four dimensions: Gregory-Laflamme and non-Gregory-Laflamme instabilities}\label{sec:HDinstability}

We have seen in the previous subsection that four-dimensional black holes are usually stable against gravitational perturbations. The situation in higher dimensions is much richer, where one has various instabilities.  A wide class of $D\geq4$ objects, black strings and their various generalizations, such as
black branes, suffer from a general type of  gravitational instability, called the \emph{Gregory-Laflamme instability} \cite{Gregory:1993vy}. The essence of this phenomenon can be easily understood when considering linear perturbations of black strings \cite{Gregory:1994bj}.

Unlike Kaluza-Klein black holes, the black string metric is a solution to the Einstein equations in five- or higher-dimensional gravity which has a factorized form consisting of the Tangherlini black hole and an extra flat compact dimension. According to the brane-world scenarios, if the matter localized on the brane undergoes gravitational collapse, a black hole with the horizon extended to the transverse extra direction will form. This object looks similar to a black hole on the brane, but is, in fact, a black string in the full $D$-dimensional theory. The generic black string metric has the following form:
\begin{equation}
ds^2 = g_{ab} dy^{a} dy^{b} + d z^2,
\end{equation}
where $g_{ab}$ describes the black hole behavior on the brane, while the ``extra'' dimension is in the $z$-direction. The z-direction is periodically identified by the relation $z = z+ 2 \pi R$.

The perturbation of the Einstein equations for black strings can be reduced to scalar, tensor, and vector types of perturbations in the same way as was done for higher-dimensional black holes (see Sec.~\ref{sec:equations}). An analysis of H.~Kudoh \cite{Kudoh} showed that vector and tensor gravitational perturbations are stable, as well as $\ell =1, 2, ...$ scalar perturbations. The only unstable type of  perturbations is $\ell =0$ scalar gravitational type, also called an \emph{s-wave}. This s-wave shows the Gregory-Laflamme instability at long wavelengths in the $z$-direction.

The zero mode ($k_z = 0$) of the scalar gravitational perturbation with $\ell= 0$ corresponds to a shift of the mass of higher-dimensional Schwarzschild black holes, and, therefore, this mode is not dynamical due to Birkhoff's theorem. However, the Kaluza-Klein mode with $k_z \neq 0$ and $\ell = 0$ is essentially different from the gravitational perturbations of Schwarzschild black holes, because it cannot be interpreted as an infinitesimal change of mass of the black string, i.~e., it corresponds to a dynamical perturbation. Thus, from the viewpoint of an effective theory on the $z = const$ plane, the Gregory-Laflamme instability is apparently related to the inapplicability of Birkhoff's theorem. An interesting membrane illustration of the Gregory-Laflamme instability was suggested in \cite{CardosoPRL-GL}.

\begin{figure}
\resizebox{\linewidth}{!}{\includegraphics*{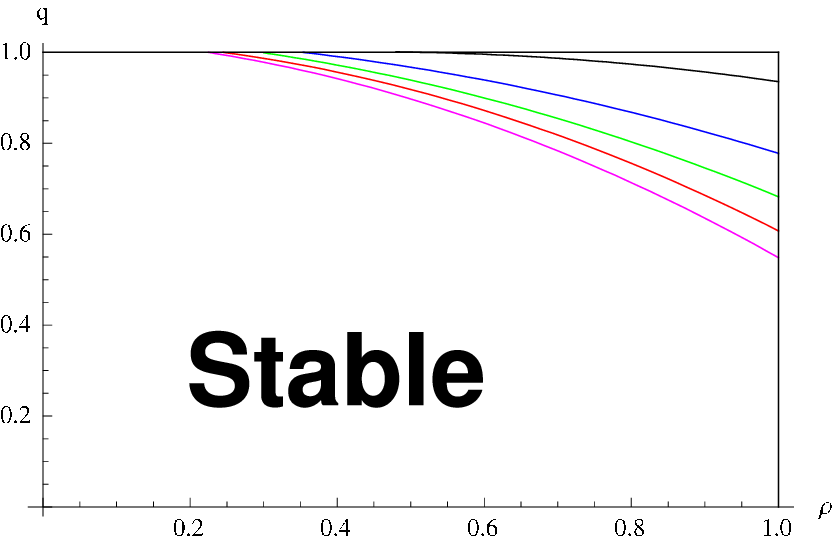}}
\caption{The parametric region of instability in the right upper corner of the square in the $\rho-q$ ``coordinates'' for $D=7$ (top, black), $D=8$ (blue), $D=9$ (green), $D=10$ (red), $D=11$ (bottom, magenta). The units $r_+ =1$ are used; $\rho=r_+/r_c=1/r_c<1$, $r_c$ is the cosmological horizon. The charge can be normalized by its extremal quantity $q=Q/Q_{ext}<1$.}
\label{RNdSparameter}
\end{figure}

\begin{figure}
\resizebox{\linewidth}{!}{\includegraphics*{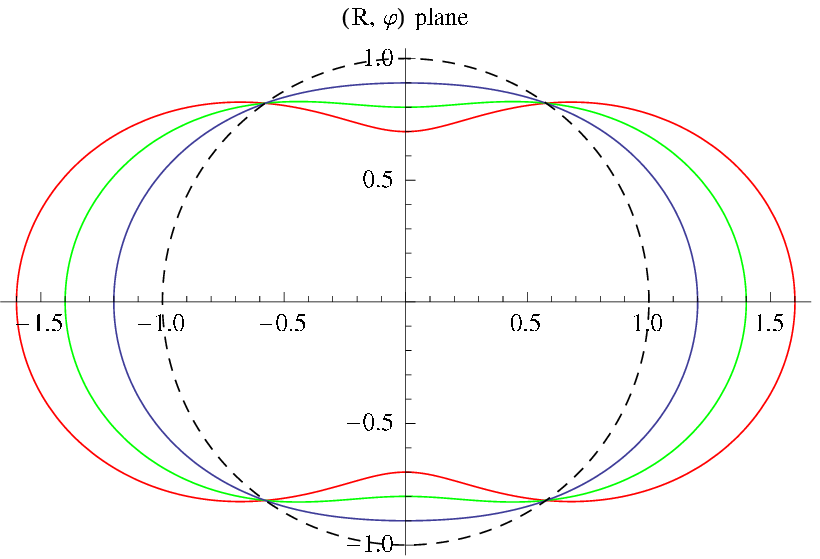}}
\caption{The equatorial plane of the black hole horizon hypersurfaces at the edge of stability. The dashed line corresponds to the unperturbed BH of unit horizon. The blue, green and red lines correspond to the perturbed BHs (for different values of $\cal A$) after decay of all the dynamical modes.}
\label{RNdSdeformed}
\end{figure}

A qualitatively different type of instability, which occurs not at the s-wave but at the first multipoles, takes place for some higher-dimensional black holes. For example, the $D$-dimensional Reissner-Nordstr\"om-de Sitter black hole, described by the metric
\begin{equation}\label{metricRNdS}
ds^2=f(r)dt^2- f^{-1}(r) dr^2-r^2d\sigma_{D-2}^2,
\end{equation}
where $d\sigma_{D-2}^2$ is the line element of a unit $(D-2)$-sphere,
$$ f(r)=1-\frac{2M}{r^{D-3}}+\frac{Q^2}{r^{2(D-3)}}-\frac{2\Lambda
r^2}{(D-1)(D-2)}, $$
is unstable for large values of the charge and cosmological term \cite{KonoplyaPRL}. This instability, unlike the Gregory-Laflamme one, occurs at the $\ell =2$ multipole, and thereby is not related to the inapplicability of the Birkhoff's theorem. The parametric region of instability can be seen in Fig.~\ref{RNdSparameter}, while the form of the slightly deformed black hole at the threshold of instability is shown in Fig.~\ref{RNdSdeformed}.

In order to distinguish the latter instability from the Gregory-Laflamme one, we call this instability a \emph{non-Gregory-Laflamme instability}. In addition to black string, black brane, and Reissner-Nordstr\"om-de Sitter instabilities, there is an instability of higher-dimensional black holes in the Einstein-Gauss-Bonnet theory \cite{GBstability1,GBstability2,GBstability3,GBstability-Konoplya}. The summary of (in)stability analysis in higher dimensions is given in Table~\ref{tabletwo}.

\begin{table*}
\caption{(In)stability of higher-dimensional black holes}\label{tabletwo}
\tablefont\begin{tabular}{|c|c|}
  \hline
  Black hole solution (parameters) & Publication \\
   \hline
  Schwarzschild (M) & Stable for all $D$ \cite{Kodama:2003jz,Ishibashi:2003ap} \\
   \hline
  Reissner-Nordstr\"om (M, Q) & Stable for $D =5,6,\ldots,11$ and nonextremal charge \jamode{\\&}{} \cite{KonoplyaPRL} \\
    \hline
  Schwarzschild-dS (M, $\Lambda$),  (M, $\Lambda>0$) & Stable for $D =5,6,\ldots,11$ \cite{Konoplya:2007jv} \\
    \hline
  Schwarzschild-AdS (M, $\Lambda$) (M, $\Lambda<0$) & Stable in EM theory for $D =5,6,\ldots,11$ \cite{Konoplya:2008rq} \\
    \hline
  Reissner-Nordstr\"om-dS (M, $\Lambda$) (M, Q, $\Lambda>0$) & Unstable for $D =7,8,\ldots,11$ \cite{KonoplyaPRL} \\
    \hline
  Reissner-Nordstr\"om-AdS (M, $\Lambda$) (M, Q, $\Lambda<0$) & Stable in EM theory  \cite{Konoplya:2008rq} \jamode{\\&}{} and unstable in supergravity \cite{Gubser:2000mm} \\
    \hline
  Gauss-Bonnet (M, $\alpha$) & Unstable for moderate and large $\alpha$ \jamode{\cite{GBstability1,GBstability2}\\&\cite{GBstability3,GBstability-Konoplya}}{\cite{GBstability1,GBstability2,GBstability3,GBstability-Konoplya}} \\
    \hline
 Myers-Perry and its generalizations (M, J) & ? Only particular types of perturbations \jamode{\\&}{} \cite{Murata:2008yx,Kunduri:2006qa,Kodama:2007sf,Kodama:2009rq,Kodama:2009bf} \\
    \hline
  Dilaton (M, Q, $\phi$) & ? \\
    \hline
  Dilaton-axion  (M, Q, J, $\phi$, $\psi$) & ? \\
    \hline
  Dilaton-Gauss-Bonnet (M, $\phi$, $\alpha$) & ? \\
    \hline
\end{tabular}
\end{table*}

The case of the Reissner-Nordstr\"om-AdS solution (given by the metric (\ref{metricRNdS}) with a negative cosmological constant) deserves special comments. Solution (\ref{metricRNdS}) satisfies not only the ordinary Einstein-Maxwell equations, but also the ${\cal N}=8$ supergravity equations which include, among a number of fields, the dilaton. The stability properties of the above solution naturally depend upon the theory in which the solution is considered. In \cite{Gubser:2000mm} it was shown that in the ${\cal N}=8$ supergravity, large Reissner-Nordstr\"om-AdS black holes are unstable for large values of charge, while in the Einstein-Maxwell theory these black holes are stable \cite{Konoplya:2008rq}. If one remembers that in the ${\cal N}=8$ supergravity a dilaton allows for a dynamical $\ell =0$ (i.~e., pure spherically symmetric) mode, then it is evident that the instability is of the Gregory-Laflamme type. This type of instability is certainly impossible in the pure Einstein-Maxwell theory, where the Reissner-Nordstr\"om-AdS black hole has a nondynamical s-wave mode \cite{Konoplya:2008rq}.

Although the stabilities of nonrotating solutions are studied relatively well, the stability of rotating black holes and branes in higher dimensions is almost a black spot. The most appealing problem is certainly stability of higher-dimensional Myers-Perry black holes and of their generalizations. There are two qualitatively different cases here: If all of the angular momenta are equal $a_1 = a_2 = \cdots = a_N$, then, as in the four-dimensional case, the momentum of the black hole is bounded. The perturbation equations in this case allow for a particular separation of variables \cite{Murata:2008yx,Kunduri:2006qa,Murata:2007gv}. Thus, in the $D =5$ case with $a_1=a_2$ the separation of variables is possible for the so-called zero mode, i.~e., for lowest eigenvalues of the momentum \cite{Murata:2007gv}. Tensor gravitational perturbations allow for the separation of variables for any $D$ and all the eigenvalues in this case.

\begin{figure}
\resizebox{\linewidth}{!}{\includegraphics*{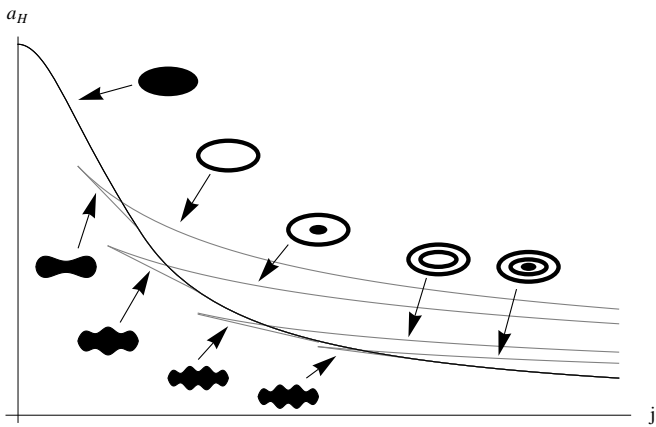}}
\caption{The qualitative phase diagram for the black objects in $D\geq 6$, proposed in \cite{Emparan:2010sx}. The horizontal and vertical axes correspond, respectively, to the spin and area of a black object. If thermal equilibrium is not imposed, multirings are possible in the upper region of the diagram.}\label{hidphases}
\end{figure}

The second, less symmetric case, when at least one of the angular momenta is different from the others, is much more interesting. Then there is no bound for the rotation parameter $a$ (for $D>5$) and an infinitely high angular speed of rotation is possible for the Myers-Perry solution, at least formally. Intuitively we expect that such a highly rotating black hole \emph{must be unstable} when the centrifugal forces exceed the gravitational attraction (see Fig.~\ref{hidphases}). However, until recently, no instability was proved for such black holes, because of the impossibility of the separation of variable in the perturbation equations. In the linearized theory some particular results were obtained for tensor gravitational perturbations \cite{Kodama:2007sf,Kodama:2009bf,Kodama:2009rq}, where no instability was found even for asymptotically high rotation. This apparently means that the instability is in the scalar gravitational modes, which are responsible for deformation of the horizon. Recently, a static limit of perturbations was considered \cite{Dias:2009iu,Dias:2010gk} and the onset of the bifurcation point toward a new solution was shown for some critical value of the angular momentum. Finally, the instability was detected by numerical simulations in the fully nonlinear theory for large angular momenta \cite{Shibata:2009ad}.

\subsection{Two types of developing of gravitational instabilities in time domain}

The development of instability in the time domain is different for static, spherically symmetric nonrotating black holes and rotating black holes. The spherically symmetric black holes have damped QNMs in the form of damped oscillations, i.~e., with nonvanishing real and imaginary parts of $\omega$, while growing, unstable modes \emph{must be pure imaginary}, i.e., nonoscillatory. Indeed, we multiply Eq.~(\ref{wave1}) by the complex conjugated function $\Psi^\star$ and assume that the dependence on time is $\Psi(t,r_\star)=e^{-\imo\omega t}\Psi(r_\star)$. The integral of the obtained equation reads
$$I=\intop_{-\infty}^\infty \left(\Psi^\star(r_\star)\frac{d^2\Psi(r_\star)}{dr_\star^2}+\omega^2|\Psi(r_\star)|^2 -V|\Psi(r_\star)|^2\right)dr_\star.$$
Integration of the first term by parts gives
\begin{eqnarray}\nonumber &&I=
\Psi^\star(r_\star)\frac{d\Psi(r_\star)}{dr_\star}\Biggr|_{-\infty}^\infty + \\\nonumber&&+\intop_{-\infty}^\infty \left(\omega^2|\Psi(r_\star)|^2 -V|\Psi(r_\star)|^2 - \left|\frac{d\Psi(r_\star)}{dr_\star}\right|^2\right)dr_\star = 0.
\end{eqnarray}
Taking account of the boundary conditions (\ref{Frobenius}), one can find the imaginary part of the integral
\begin{eqnarray}\nonumber \Im{I}&=&\Re{\omega}|\Psi(\infty)|^2+\Re{\omega}|\Psi(-\infty)|^2+
\\\nonumber&&+2\Re{\omega}\Im{\omega}\intop_{-\infty}^\infty|\Psi(r_\star)|^2dr_\star=0.
\end{eqnarray}
The nonzero real part of the quasinormal frequency implies that the imaginary part is negative. Therefore, the unstable modes ($\Im{\omega}>0$) must have zero real part. In other words, for static spherically symmetric black holes \emph{unstable modes cannot be oscillating}.

\begin{figure}
\resizebox{\linewidth}{!}{\includegraphics*{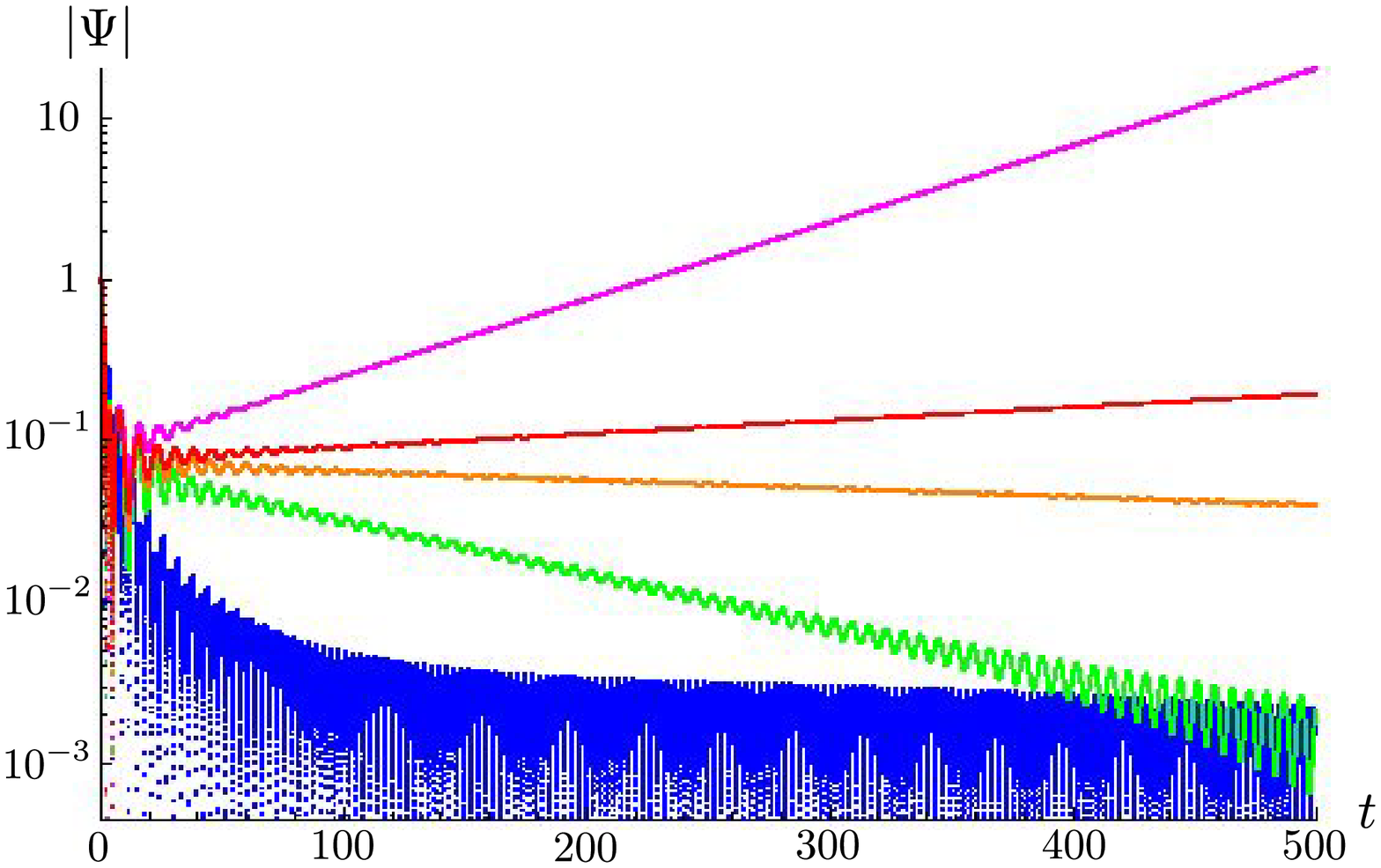}}
\caption{Time-domain profiles of black string perturbations for $n=1$ $k=0.84$ (magenta, top), $k=0.87$ (red), $k=0.88$ (orange), $k=0.9$ (green), $k=1.1$ (blue, bottom). We can see two concurrent modes: for large $k$ the oscillating one dominates , while near the critical value of $k$ the dominant mode does not oscillate (looks like exponential tail), and for unstable values of $k$ the dominant mode grows. The plot is logarithmic, so that straight lines correspond to an exponential decay.}\label{instability}
\end{figure}

\begin{figure}
\resizebox{\linewidth}{!}{\includegraphics*{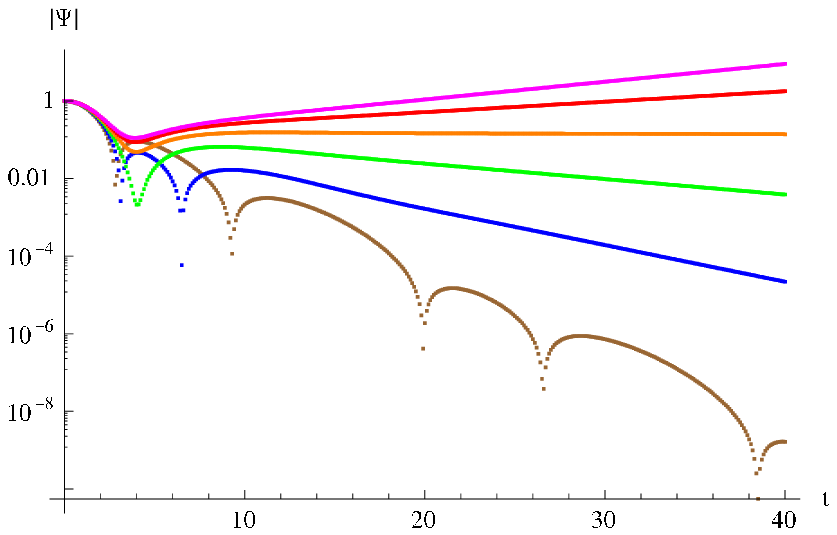}}
\caption{Time-domain profile of near extremal RN BH perturbation ($D=11$, $\rho=0.8$), q=0.4 (brown), q=0.5 (blue), q=0.6 (green), q=0.7 (orange), q=0.8 (red), and q=0.9 (magenta). The smaller $q$ is, the slower the growth of the profile is.}\label{RNdS.profiles.D=11.rho=0.8}
\end{figure}

In addition to the nonoscillatory character of the unstable modes, there is another distinction of the evolution of instabilities in the time domain. The first type of the time-domain instability develops immediately after the initial outburst. This type of instability is shown in Fig.~\ref{instability} for the black string \cite{KonoplyaBlackString} and in Fig.~\ref{RNdS.profiles.D=11.rho=0.8} for the Reissner-Nordstr\"om black holes \cite{KonoplyaPRL}.

\begin{figure}
\resizebox{\linewidth}{!}{\includegraphics*{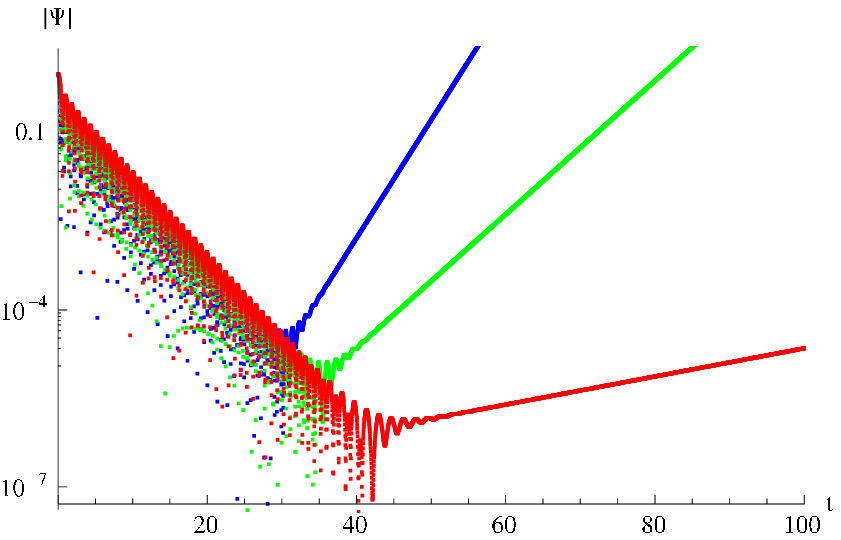}}
\caption{The picture of instability, developing at large multipole numbers: $D=6$, $\ell =8$ (red, bottom line), $\ell = 12$ (green), $\ell=16$ (blue, top line), $\alpha =1.4$. Tensor type of gravitational perturbations.}\label{GBtimedomain}
\end{figure}

The second type of time-domain evolution is much more exotic: The instability develops after a rather long period of damped quasinormal oscillations. This type of instability takes place for black holes in the Einstein-Gauss-Bonnet theory (see Fig \ref{GBtimedomain}) \cite{GBstability1,GBstability-Konoplya}. The instability occurs at higher multipoles $\ell$, while the first few lowest multipoles are stable \cite{GBstability-Konoplya}. We believe that the stability of lowest multipoles gives rise to a prolonged period of damped oscillations and the ``delayed'' appearance of the instability in this case.

\subsection{Correlation between gravitational and thermodynamic instabilities}

\begin{figure}
\resizebox{\linewidth}{!}{\includegraphics*{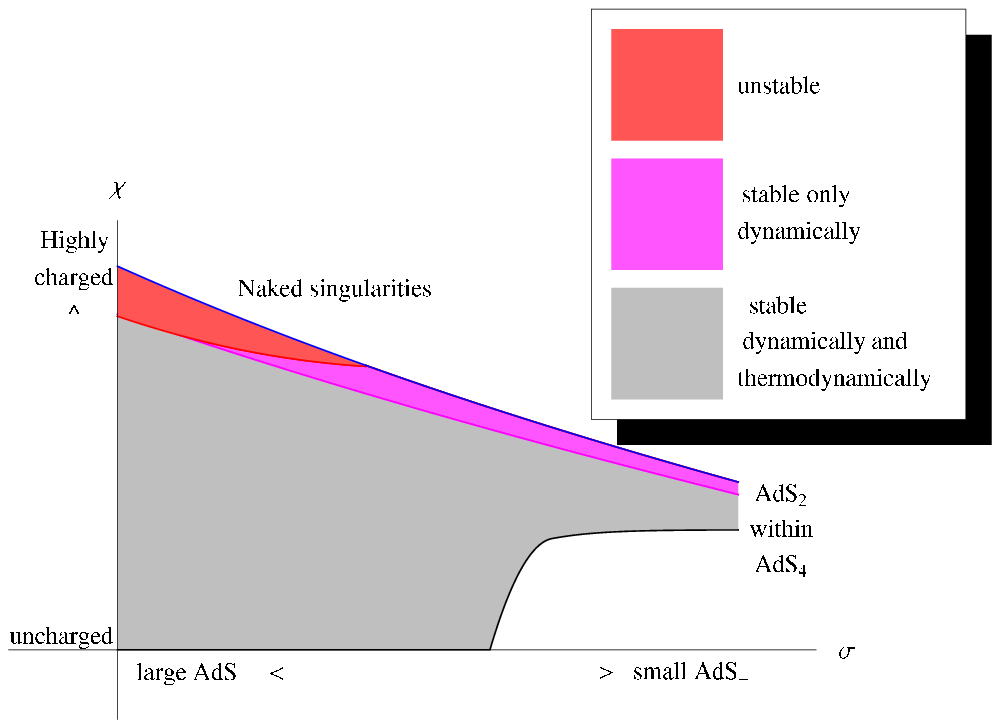}}
\caption{Instability region for RNAdS black holes in ${\cal N}=8$ supergravity. The plot was proposed in \cite{Gubser:2000mm}.}\label{GubserMitra}
\end{figure}

According to Gubser and Mitra \cite{Gubser:2000mm}, gravitational instability is correlated to the local thermodynamic instability for the RNAdS black holes in ${\cal N}=8$ supergravity. Mathematically the local thermodynamic instability means that the Hessian matrix of the second derivatives of the mass with respect to the entropy (and the conserved charges or angular momenta) has a negative eigenvalue. As we have already learned, this instability is of the Gregory-Laflamme type and, thus, directly connected to the s-wave mode and inapplicability of Birkhoff's theorem. The parametric regions of gravitational and thermodynamic stability do not coincide exactly as shown in Fig.~\ref{GubserMitra} \cite{Gubser:2000mm}; the region of gravitational stability is larger than that of the thermodynamic one and includes the latter completely. In other words, the Reissner-Nordstr\"om-AdS black hole that is thermodynamically stable is also gravitationally stable but not otherwise, the gravitationally stable black hole is not necessarily thermodynamically stable. For large highly charged anti-de Sitter black holes the regions of gravitational and thermodynamic instabilities asymptotically coincide (Fig.~\ref{GubserMitra}). Most probably this happens because in this limit the AdS black hole approaches the black brane regime. For the black brane, Gubser and Mitra claimed that thermodynamical stability is the sufficient and necessary condition for the absence of gravitational instability. They also claimed that, in the limit of large AdS black holes, gravitational and thermodynamic stabilities coincide, while the small discrepancy observed in Fig.~\ref{GubserMitra} is probably a numerical error \cite{Gubser:2000mm}. The physical explanation of the instability consists of an energetically more preferable state of a set of black holes other than of a single black brane: The entropy of an array of black holes is larger than the entropy of the uniform black brane of the same total mass.

\subsection{Superradiant instability}\label{sec:superradiance}

Consider the classical scattering problem for a massless scalar field in the background of the Kerr black hole. The problem can be reduced to the wavelike equation (\ref{wave1}) with the following effective potential
\begin{eqnarray}
V &=& \left(\omega - \frac{a m}{r^2 + a^2}\right)^2 -
\frac{\lambda^2 \Delta_r}{(r^2 + a^2)^2} +\nonumber\\
&&\frac{\Delta_r}{(r^2 + a^2)^{3/2}} \frac{d}{d r}\left(\Delta_r \frac{d}{d r} \frac{1}{\sqrt{r^2 +
a^2}}\right),\\\nonumber
\Delta_r&=&r^2+a^2-2Mr,
\end{eqnarray}
where $\lambda$ is the separation constant \cite{Teukolsky}. The effective potential has the following asymptotic behavior:
\begin{eqnarray}
V \rightarrow \omega^2, &\quad& r \rightarrow \infty,\\
V \rightarrow (\omega - m \Omega_{h})^2,  &\quad& r \rightarrow
r_{+}.
\end{eqnarray}
The asymptotic forms of the solutions near the black hole and at spatial infinity are
\begin{eqnarray}
\label{infty-asym}
\Psi(r_{*}) = e^{-i \omega r_{*}} + \mathcal{R} e^{+ i \omega r_{*}}, &\quad& r \rightarrow \infty,\\
\label{horizon-asym}
\Psi(r_{*}) = \mathcal{T} e^{- i (\omega - m \Omega) r_{*}}, &\quad& r \rightarrow r_{+}.
\end{eqnarray}
Here $\mathcal{R}$ is called the amplitude of the reflected wave or the reflection coefficient, and $\mathcal{T}$ is the transmission coefficient. Since the effective potential is real, the Wronskian of the complex conjugate solutions $\Psi(r_{*})$ and
$\Psi^{*}(r_{*})$ obeys the relation
\begin{equation}\label{Wronskian1}
\imo \frac{d}{dr_{*}} W(\Psi, \Psi^{*}) = \imo \frac{d}{dr_{*}} \left(\Psi \frac{d \Psi^{*}}{dr_{*}} - \Psi^{*} \frac{d \Psi}{dr_{*}} \right) = 0.
\end{equation}
Integrating the above relation from $r_{+}$ until spatial infinity and using the asymptotic form (\ref{infty-asym}) at infinity, we obtain
\begin{equation}\label{Wronskian2}
|\mathcal{R}|^2 = 1 + (\imo/2 \omega) W|_{r_{+}}.
\end{equation}
Using the asymptotic form (\ref{horizon-asym}), we find that
\begin{equation}
|\mathcal{R}|^2 = 1+\left(\frac{m \Omega_h}{\omega}-1\right)|\mathcal{T}|^2.
\end{equation}
If $|\mathcal{R}| > 1$, that is,
\begin{equation}\label{superrad-cond}
\frac{m \Omega}{\omega} > 1,
\end{equation}
the reflected wave has larger amplitude than the incident one. This amplification of the incident wave is called the \emph{superradiance} and was first predicted by Zeldovich \cite{Zeldovich}. The superradiance effect for Kerr black holes was first calculated by Starobinsky \cite{Starobinsky,StarobinskyChurilov}, who showed that the superradiance  is much stronger for the gravitational field than for scalar and electromagnetic fields. The process of superradiant amplification occurs due to extraction of rotational energy from a black hole, and, therefore, it happens only for modes with positive values of azimuthal number $m$ that corresponds to ``corotation'' with a black hole.  The superradiance is absent for fermion fields \cite{Maeda:1976tm,Iyer:1978du}.

The condition (\ref{superrad-cond}) can be generalized to perturbations of fields of other integer spin, as well as to other than rotating black hole systems \cite{Unruh-superradiance}, for instance, to an electrically conductive rotating cylinder or another system with absorption. In the general case the effective potential has a nonvanishing imaginary part
\begin{equation}
V(r_{*}) = U(r_{*}) + \imo \Gamma(r_{*}).
\end{equation}
Then, Eq.~(\ref{Wronskian1}) reads
$$\imo\frac{d}{dr_{*}} W(\Psi, \Psi^{*}) = 2 \Gamma |\Psi(r_{*})|^{2},$$
and Eq.~(\ref{Wronskian2}) gets an extra term
\begin{equation}
|\mathcal{R}|^2 = 1 + (\imo/2 \omega) W|_{r_{+}} - \frac{1}{\omega}
\intop_{r_{*}^{0}}^{\infty} \Gamma(r_{*}) \Psi(r_{*}) d(r_{*}).
\end{equation}
The condition of superradiance is
$$(i/2 \omega) W|_{r_{+}} -
\frac{1}{\omega} \int_{r_{*}^{0}}^{\infty} \Gamma(r_{*})
\Psi(r_{*}) d(r_{*}) > 0.$$
The existence of an ergoregion is not sufficient for superradiance: An appropriate boundary condition is also necessary, which, for rotating black holes, is the requirement of only the ingoing group velocity waves at the event horizon \cite{Unruh-superradiance}.

\begin{figure*}
\resizebox{\linewidth}{!}{\includegraphics*{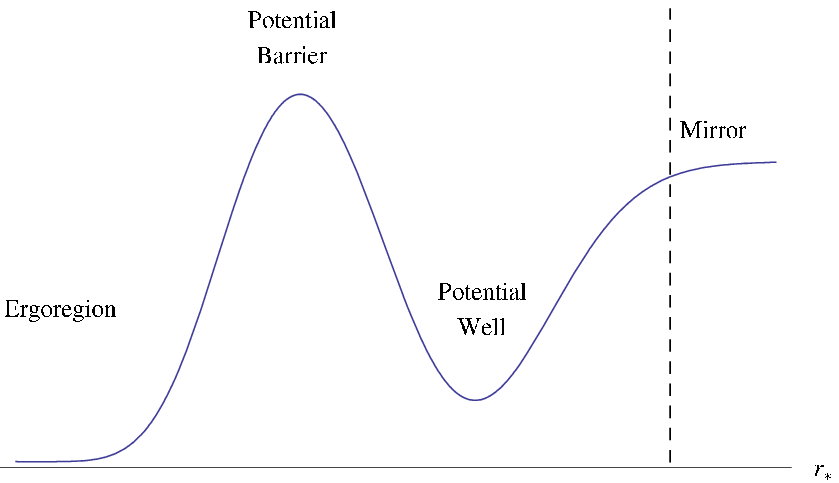}\includegraphics*{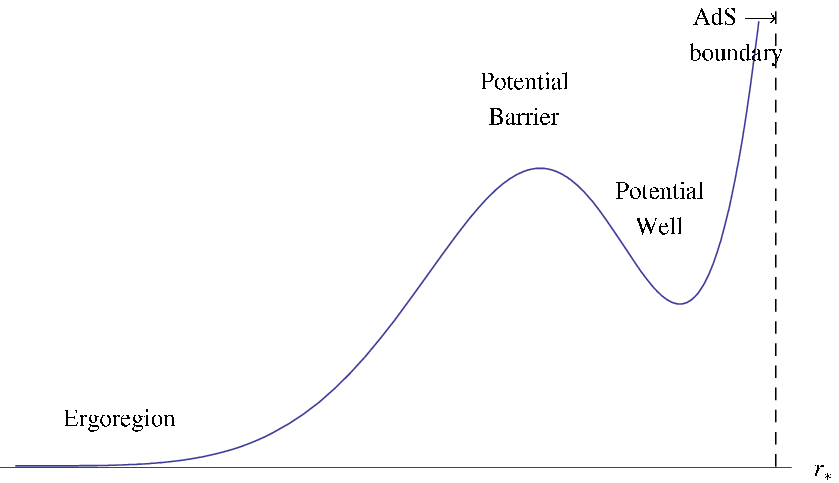}}
\caption{The qualitative behavior of the effective potential for massive fields in the Kerr background (left) and for the massless fields in the Kerr-AdS background (right).}\label{superradiant}
\end{figure*}

When one has the massive scalar or other integer spin field in asymptotically flat or de Sitter space-time or massless field but an AdS boundary at spatial infinity, the superradiance has unstable modes. The essence of this phenomenon can be understood from the plot of the effective potentials for massive scalar field in asymptotically flat black holes and for massless fields in AdS black holes (Fig.~\ref{superradiant}). The effective potentials for both cases, in addition to the local maximum, have an extra local minimum far from the black hole which creates a secondary reflection of the wave reflected from the potential barrier. This secondary reflected wave, when incident on a potential barrier, will be reflected again at the far region. As each reflection from the potential barrier in the superradiant regime increases the amplitude of the wave, the process of reflections will continue with increased energies of waves and, thus, one has an instability. The superradiance for various black holes was considered in \cite{Koga:1994wt,Shiraishi:1992yv,Andersson:1999wj,Winstanley:2001nx,Lepe:2004kv,Jung:2005,Dias:2007nj,Kobayashi:2008xh}.

We now discuss in more detail the instability due to the massive term. This case implies that at least in four space-time dimensions the meta-stable bound states can be formed \cite{BHboundstates} in the valley of the local minimum. These bound states are characterized by the product of the field mass $\mu$ and the black hole mass $M$, i.~e., by the ratio of the characteristic size of a black hole to the Compton wavelength of the particle
\begin{equation}
\mu M = GM \mu / \hbar c \sim r_{+}/\lambda_c.
\end{equation}
Detweiler \cite{Detweiler} first estimated the instability growth rate for the $\ell =m=1$ bound state in the limit $\mu M \ll 1$,
\begin{equation}
\tau \approx 24 (a/M)^{-1} (\mu M)^{-9} (GM/c^{3}),
\end{equation}
where $\tau$ is an $e$-folding time. Then, Zouros and Eardley \cite{Zouros} computed the WKB instability growth rate in the opposite limit $\mu M \gg 1$,
\begin{equation}
\tau \approx 10^{7} e^{1.84 \mu M} (GM/c^{3}).
\end{equation}
Finally Dolan \cite{Dolan} analyzed the instability for the whole range of values of $\mu M$ and found that the maximal instability is for the $\ell = 1$, $m = 1$ state, for $\mu M \lesssim 0.42$, $a = 0.99$ and equals $\tau^{-1} \approx 1.5\cdot 10^{-7} (GM/c^{3})^{-1}$. In \cite{Konoplya-superradiant} it was shown that the presence of the strong magnetic field can enhance the superradiant instability. The superradiant instability of massive particles in the vicinity of black holes appears always as a negligible process: The reason is that the instability growth rate is always small for the standard model particles in comparison to the decay rate of particles or of the Hawking evaporation rate of black holes.

\begin{figure}
\resizebox{\linewidth}{!}{\includegraphics*{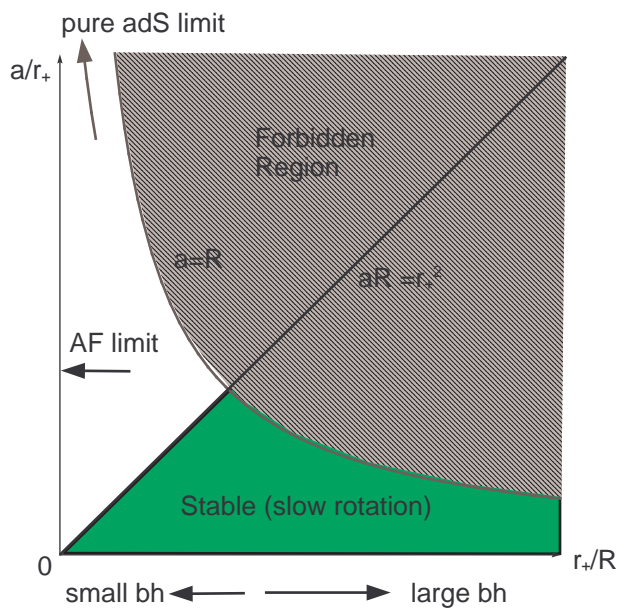}}
\caption{The stable region in the parameter plane for the simply rotating higher-dimensional asymptotically AdS black hole. The figure is taken from \cite{Kodama:2009rq}.}\label{paramregion}
\end{figure}

The superradiant instability due to the presence of the AdS boundary was studied in \cite{Cardoso:2004hs,Cardoso:2004nk} for the scalar field and in \cite{Kodama:2009rq,Kunduri:2006qa} for gravitational perturbations. In \cite{Kodama:2009rq} it was shown that the instability of the tensor gravitational perturbations of simply rotating $D>6$ Myers-Perry black holes is due to superradiant modes only and also has a small growth rate of order $10^{-12} (r_{+})^{-1}$. As the instability occurs for small AdS black holes (one can see the plot of the parametric range of instability in Fig.~\ref{paramregion}), it is certainly negligible and will be suppressed by the violent Hawking evaporation.

Note that at the quantum level superradiant instability of a massive particle simply means that the particle is made to leave a given superradiant state in favor of a ``normal'' (nonsuperradiant) one.

\section{AdS/CFT interpretation of QNMs}\label{sec:AdSCFT}
\subsection{A brief overview of AdS/CFT}

While string theory is often regarded as a theory of quantum gravity and grand unification, it also encompasses a wide range of dualities, including AdS/CFT which relates strongly coupled field theories to weakly coupled gravitational duals \cite{Maldacena:1997re}. As a strong/weak coupling duality, AdS/CFT has been applied to the study of the strongly interacting quark-gluon plasma as well as to condensed matter systems such as high $T_c$ superconductors. Here we give some basic ideas about AdS/CFT, and refer the reader to a review \cite{Aharony:1999ti} for further study.

The key theoretical problem of modern particle physics is the search for an adequate description of interactions of quarks and gluons, i.~e., quantum chromodynamics that would be valid at all energies. For small values of the coupling constant $g$, the Yang-Mills theory with an SU($3$) group of symmetry is a good approximation, where $3$ stands for the three colors of quarks. However, at small energies and larger distances the coupling constant $g$ is growing, which does not allow us to use an expansion in $g$. An approach that could hopefully solve this problem is based on the consideration of the SU($N$) Yang-Mills theory in the limit $N \rightarrow \infty$ \cite{Polyakov:1987ez}, a theory with an ``infinite number of quarks' colors''. In this limit the perturbation theory becomes much simpler so that all nonvanishing diagrams look similar to triangulations of a sphere \cite{tHooft}. It is then natural to expect that in the limit $N \rightarrow \infty$, the Yang-Mills theory can be described within the formalism of two-dimensional strings \cite{tHooft}. The contributions to these triangulations are expansions in terms of $g^2 N$, so that one keeps $g^2 N$ finite in the regime of large $N$.

The so-called conformal-invariant supersymmetric Yang-Mills theories have gained a great success in this way during the past decade \cite{Aharony:1999ti}.
According to string theory, the world surface of a string has an infinite tower of quantum excitations which look similar to particles. A finite number of these quantum states of a string corresponds to massless particles, and the infinite ``rest'' of the states represents massive particles. The mass of a particle is proportional to the string tension, so that at large distances ($r \gg l_{s}$) only massless particles survive. Among massless excitations of closed strings there is one corresponding to the spin-2 field, that is, to a graviton.
Consequently from string theory at large distances, one can deduce the Einstein-Hilbert theory of gravity, and from superstrings -- supergravity.
The full theory which includes gravitation and interacting ``matter'' fields can be described by the system of both closed and open strings and a set of the so-called $D$-branes. The $D$-branes are submanifolds to whose ends open strings are attached, while closed strings can propagate in the bulk as well. Our world is supposed to be such a $D3$-brane (here $3$ is for three spatial dimensions) to which strings are attached and fluctuate in the bulk. $D$-branes must keep some part of the supersymmetry and this stipulates the particular form of the space-time geometry in the vicinity of $D$-branes. The geometry near the brane's throat is the anti-de Sitter one and the anti-de Sitter radius $R$ of the $D3$ brane is
\begin{equation}
R = \ell_s (g^2 N)^{1/4},
\end{equation}
where $\ell_s$ is the string length. If a closed string has energy which is less than the brane curvature, the string cannot overwhelm the brane's gravitational attraction and will leave the brane, staying near its throat. If $g^2 N \gg 1$, one has the regime of a classical superstring in the background of the $D3$-brane.

\subsection{The AdS/CFT vocabulary and interpretation of quasinormal modes}\label{sec:AdSCFTvoc}

Now we are in a position to relate, briefly, the famous AdS/CFT correspondence, the duality between string theory and field theory \cite{Maldacena:1997re}.
On the string theory side we have the following constants: the anti-de Sitter radius $R$, string's length $\ell_s$ and the string coupling $g_s$. On the field theory side one has the 't Hooft coupling $N g^2$. The duality between string and field theory works through the following relations between the coupling constants:
\begin{equation}
4 \pi g_s = g^2, \quad g^2 N =  \frac{R^4}{\ell_{s}^{4}}.
\end{equation}
The main advantage of this mapping is that one can describe the strong coupling regime in field theory (large $g^2 N$) by string theory in the regime $R \gg \ell_s$, that is, by supergravity.  The AdS/CFT vocabulary says that there is a correspondence between some operator $\mathcal{O}$ in the field theory and its dual bulk field $\phi$ in supergravity
\begin{equation}\nonumber
bulk~field~(supergravity)
\sim operator~(field~theory),
\end{equation}

In particular,

\begin{enumerate}
\item
the dilaton field $\phi$  is dual to the $\mathcal{O} = - \mathcal{L}$, where $\mathcal{L}$ is the Lagrangian density,

\item
the gauge field $A_{\mu}^{a}$ is dual to the $R$-charge current $J^{a \mu}$,

\item
the metric is dual to the stress-energy tensor in the sense that
$Z_{4D}[g_{\mu \nu}^{0}] = e^{i S_{cl}[g_{\mu \nu}]}$.

\end{enumerate}

More generally, the AdS/CFT correspondence states that
\begin{equation}\label{ADS-CFT}
Z_{4D}[J] \equiv \int d \phi e^{i S + \int i d^{4} x J \mathcal{O}}= e^{i S_{cl}},
\end{equation}
and the partition function of the field theory $Z[J]$, where $J$ is the source coupled to the operator $\mathcal{O}$.
Differentiating Eq.~(\ref{ADS-CFT}) with respect to $J$ gives us various correlation functions.

It is well known from the spectral theorem that the poles of the retarded Green's function of a wave equation coincide with the normal modes of the wave functions under appropriate boundary conditions. Thus, quasinormal modes of some gravitational background are naturally poles of the correlation functions. At the same time we can see that correlation functions in the field theory are connected through Eq.~(\ref{ADS-CFT}) with derivatives of the classical action on the gravity side. Thus, it does not come as a surprise that the quasinormal modes of an asymptotically AdS gravitational background coincide with the poles of the correlation functions \cite{Kovtun:2005ev}. When the field theory is at zero temperature, the dual gravitational background is described by some regular asymptotically AdS metric. If the quantum field theory is at finite temperature, the gravitational background \textsl{must have a horizon}, and \emph{the Hawking temperature of the horizon corresponds to the  temperature in the dual field theory}.
A supposition about this was done by Horowitz and Hubeny who suggested that the quasinormal modes of the large $D$-dimensional asymptotically AdS black holes are poles of the retarded Green's functions in the dual conformal field theory in $(D-1)$ dimensions \cite{Horowitz:1999jd}. As the calculations of the Green's functions directly on the CFT side in four and higher dimensions are complicated, no proof of this QNM interpretation was found.
Nevertheless, soon after Birmingham, Sachs and Solodukhin \cite{Birmingham:2001pj} showed that for the $(2+1)$-dimensional asymptotically AdS (BTZ) black hole, the quasinormal modes, which are given by the relation \cite{Cardoso:2001hn}
\begin{equation}
\omega = \pm q - 4 \pi T i (n+1),
\end{equation}
exactly coincide with the poles of the two-point retarded Green's function,
$$ G^{R} = A \frac{\omega^2 - q^2}{4 \pi^2} \times $$
\begin{equation}
\left(\psi \left(1- \frac{i(\omega - q)}{4 \pi T}\right) + \psi \left(1- \frac{i(\omega + q)}{4 \pi T}\right)\right),
\end{equation}
in two-dimensional CFT. This was the first proof of the AdS/CFT interpretation of QNMs \cite{Birmingham:2001pj,Son:2002sd}.

Currently, the variety of gravitational backgrounds considered in string theory literature is quite large, because of considerable interest in a number of condensed matter phenomena at strong coupling which can be modeled through the AdS/CFT correspondence. These are interesting problems because there are plenty of condensed matter strong coupling phenomena (which includes, for instance, strongly correlated electrons) which could be engineered in a laboratory. This area of research is now called the AdS/condensed matter theory (AdS/CMT) correspondence and one example of AdS/CMT problems, the holographic superconductor, we discuss later. Meanwhile, we mention here a recent review devoted to the AdS/CMT correspondence  \cite{Hartnoll:2009sz}.

Next we discuss only the most universal properties of the AdS/CFT applications of QNMs which are valid for a wide class of dual gravitational backgrounds, thereby, which describe the hydrodynamic regime of strongly coupled conformal field theories.

\subsection{Universality of the hydrodynamic regime}\label{sec:hydrodynamic}

A great advantage that perturbations of black holes give through the AdS/CFT correspondence is the description of the large distance collective behavior of the quark-gluon plasmas at strong coupling and, more generally, the hydrodynamic regime of the dual field theory.
In the regime of long-wavelength perturbations hydrodynamics can be described by the energy-momentum tensor of a system,
\begin{equation}
T^{\alpha \beta} = (\varepsilon + P) u^{\alpha} u^{\beta} + P g^{\alpha \beta} - \sigma^{\alpha \beta},
\end{equation}
where $\sigma^{\alpha \beta}$ is proportional to derivatives of local temperature $T$ and four-velocity $u^{\alpha}$, is called the \emph{dissipative part} of $T^{\alpha \beta}$. The equation of motion is the conservation law
\begin{equation}
T^{\alpha \beta}_{,\beta} = 0
\end{equation}
and, if there are other conserved currents, an extra conservation law is added
\begin{equation}
j^{\beta}_{,\beta} = 0,
\end{equation}
where
\begin{equation}
j^{\alpha} = \rho u^{\alpha} - {\cal D} (g^{\alpha \beta}+ u^{\alpha} u^{\beta}) \rho.
\end{equation}
In the frame of reference in which the fluid is at rest, this gives the law of diffusion
\begin{equation}\label{Fick}
\overrightarrow{j} = - {\cal D} \nabla \rho.
\end{equation}
Here, ${\cal D}$ is the constant of diffusion.
When considering hydrodynamical processes related to the behavior of a quark-gluon plasma, one can distinguish three types of perturbations of the energy-momentum tensor which can be treated within independent sets of equations (in analogy, for instance, with axial and polar perturbations of black holes). These modes are as follows \cite{Son:2007vk}:

\begin{itemize}
\item The \emph{shear} mode, corresponding to perturbations of the $T^{01}$, $T^{02}$, $T^{31}$, and $T^{32}$ components.
This mode is responsible for translational fluctuations of the ``cross section'' of the system, a flow, also allowing, in the general case, for small rotations.
In the frequency domain, the shear channel has the so-called hydrodynamic mode which is purely imaginary and, in the linear approximation, is proportional to $q^2$ \cite{Policastro:2001yc,Policastro:2002se},
\begin{equation}
\omega \sim - i q^2,
\end{equation}
where $q$ is momentum normalized by the temperature, $q = k/2\pi T$.

\item The \emph{sound} mode corresponding to perturbations of the $T^{00}$, $T^{03}$, and $T^{33}$ components.
In simple systems this mode is responsible for fluctuations in the direction of the flow, thereby generating fluctuations of energy density, i.~e., sound waves. The sound mode usually has the following form \cite{Policastro:2002tn}:
\begin{equation}
\omega = c_s q - i \gamma q^2,
\end{equation}
where $c_s$ is the speed of sound and $\gamma$ is some constant that depends on the energy density, pressure, shear, and bulk viscosities of the system \cite{Son:2007vk}.

\item The \emph{bulk} mode corresponds to perturbations of the $T^{12}$ component. Because of the conformality of the theory, the viscosity of the bulk mode equals zero. Therefore, we shall not discuss this mode in detail.
\end{itemize}

Following \cite{Kovtun:2003wp,Starinets:2008fb,Buchel:2003tz}, we discuss two relatively easy and at the same time universal examples.

Consider a quite general class of gravitational backgrounds given by the metric
\begin{equation}\label{backgroundADS-CFT}
ds^2 = g_{tt} dt^2 + g_{rr} dr^2 + g_{xx} d\textbf{x}^2.
\end{equation}
The metric coefficients can be arbitrary, assuming only the following behavior near the event horizon $r = r_+$:
\begin{equation}
g_{tt} = -\gamma_t (r-r_+), \quad g_{rr} = \frac{\gamma_r}{r-r_+}.
\end{equation}
Here $\gamma_t$ and $\gamma_r$ are constants. As a particular example we consider the process of diffusion of the conserved $R$-charge in the $\mathcal{N} =4$ super Yang-Mills theory.  The field equations for the gauge field dual to the conserved current have the form
\begin{equation}\label{gauge-waveAdS}
\partial_{\alpha} (g_{eff}(r)^{-2} \sqrt{-g} F^{\alpha \beta}) = 0,
\end{equation}
where $g_{eff}(r)$ is the effective gauge coupling. The conserved current $j^{\alpha}$, which is associated with the wave equation (\ref{gauge-waveAdS}), obeys the equation
\begin{equation}
\partial_{\alpha}j^{\alpha} = 0.
\end{equation}
Using the membrane paradigm, Kovtun, Son, and Starinets \cite{Kovtun:2003wp} showed that
\begin{equation}
j_{i} + {\cal D} \partial_{i} j^{0} =0,
\end{equation}
and found the diffusion coefficient to be
\begin{equation}
{\cal D} = \frac{\sqrt{-g}}{ g_{xx} g_{eff}^2 \sqrt{-g_{tt} g_{rr} }}(r_+) \int_{r_+}^{\infty} dr\frac{-g_{tt} g_{rr} g_{eff}^2}{\sqrt{-g}}(r).
\end{equation}
Indeed, the current $j^{\alpha}$ can be written in the form
\begin{equation}
j^{\alpha} = n_{\beta} F^{\alpha \beta} |_{r_h}, \quad j^{r} = 0,
\end{equation}
where $r_h$ is the radius of the stretched horizon in the membrane description. Then, it can be shown that in the radial gauge \cite{Buchel:2003tz}
\begin{equation}
F_{i r} = \sqrt{\gamma_{r}/\gamma_{0}} F_{ti} (r-r_+)^{-1},
\end{equation}
\begin{equation}
F_{ti} \approx - \partial_i A_t, \quad (A_r = 0).
\end{equation}
The wave equation for the $A_t$-component of the gauge potential is
\begin{equation}\label{Ateq}
\partial_r (\sqrt{-g} g^{rr} g^{tt} \partial_r A_t) = 0.
\end{equation}
The solution to the above equation (\ref{Ateq}) which vanishes at infinity has the form
\begin{equation}\label{189}
A_t(r) = C \int_{r}^{\infty} dr' \frac{g_{tt}(r') g_{rr}(r')}{\sqrt{-g(r')}}.
\end{equation}
From the above equation (\ref{189}) it follows that
\begin{equation}\label{190}
\frac{A_t}{F_{tr}}|_{r=r_+} = \frac{\sqrt{-g}}{g_{tt} g_{rr}}(r_+) \int_{r_+}^{\infty} dr \frac{g_{tt} g_{rr}}{\sqrt{-g}}(r).
\end{equation}
Then using the expressions for $j^{0}$ and $j^{i}$, one can find Fick's law (\ref{Fick}).
From the above it follows that on the gravity side of the duality, independent of the details of the background metric, there is always the \emph{hydrodynamic quasinormal mode}
\begin{equation}
\omega = - i {\cal D} k^2.
\end{equation}

Another example concerns the calculation of the ratio of the shear viscosity to the density of entropy $\eta/s$ \cite{Buchel:2003tz}.
Following Buchel and Liu \cite{Buchel:2003tz} and Son and Starinets \cite{Son:2007vk}, we perform the following boost of the metric (\ref{backgroundADS-CFT}):
\begin{equation}
r = r',
\end{equation}
\begin{equation}
t = \frac{t' + v y'}{\sqrt{1-v^2}} \approx t' + v y',
\end{equation}
\begin{equation}
y = \frac{y' + v t'}{\sqrt{1-v^2}} \approx y' + v t',
\end{equation}
\begin{equation}
x_i = x_{i}'.
\end{equation}
Then, the metric (\ref{backgroundADS-CFT}) transforms to the following form:
\begin{eqnarray}
ds^2 = g_{tt} dt^2 + g_{rr} dr'^{2} + g_{xx}(r) \sum_{i=1}^{p}(dx'^{i})^2 +\nonumber\\
+ 2 v (g_{tt} + g_{xx}) dt' dy'.
\end{eqnarray}
The above metric is a $k=0$ perturbation, so that $$A_t = v g^{xx} (g_{tt} + g_{xx})$$ can be considered as a gauge potential, similar to the one in (\ref{189}). The potential vanishes at infinity and Eq.~(\ref{190}) now has the form
\begin{equation}
\frac{A_{t}}{F_{tr}}|_{r \rightarrow r_+} = - \frac{1 + g^{xx} g_{tt}}{\partial_r (g^{xx} g_{tt})} |_{r \rightarrow r+0} = \frac{g_{xx}(r_+)}{\gamma_0}
\end{equation}
Then, the coefficient of diffusion is
\begin{equation}
{\cal D} = \sqrt{\frac{\gamma_r}{\gamma_0}} = \frac{1}{4 \pi T}.
\end{equation}
Using the thermodynamic relations under zero chemical potential
\begin{equation}
{\cal D} = \frac{\eta}{\epsilon + P}, \quad \epsilon + P = T s,
\end{equation}
one finds \cite{Kovtun:2003wp}
\begin{equation}
\frac{\eta}{s} = \frac{1}{4 \pi}.
\end{equation}
Thus, \emph{for all field theories at finite temperature, which can be described by some dual gravitational background, the ratio of
the shear viscosity $\eta$ to the entropy density $s$ is always $\hbar/ 4 \pi k_{b}$ to leading order in the 't Hooft coupling}, where $k_{b}$ is the Boltzmann constant.
The next correction in the series of powers of the inverse 't Hooft coupling is positive \cite{Buchel:2004di,Buchel:2008sh},
\begin{equation}
\frac{\eta}{s} = \frac{1}{4 \pi}\left(1 + \frac{15 \zeta(3)}{(g^2 N)^{3/2}}\right), \quad \zeta(3) \approx 1.202.
\end{equation}
This suggested a conjecture that \emph{for all systems that can be obtained from relativistic quantum field theory under finite temperature and chemical potential, the ratio $\eta/s$ cannot be less than $1/4\pi$}.
There are now various counterexamples to the above conjecture, showing that $\eta/s \ge 1/4\pi$. For example, this happens when considering higher derivative quantum corrections \cite{Myers:2008yi,Buchel:2008vz}.

We now see what value of the viscosity/entropy ratio would produce the weakly coupled theory when naively extrapolated to a strong coupling regime.
Viscosity, being the measure of the diffusion of the momentum, is evidently larger for a larger temperature $T$ and is decreasing when the coupling $\lambda$ is weaker (because the free way of particles is longer which makes an easier transfer of momentum). More accurately, for example, for the
$\lambda \phi^4$ theory, one has \cite{Son:2007vk}
\begin{equation}
\eta \sim \frac{T^3}{\lambda^2}, s \sim T^3.
\end{equation}
Thus,
\begin{equation}
\frac{\eta}{s} \sim \lambda^{-2}.
\end{equation}
For the strong coupling regime $\lambda \sim 1$  and $\eta/s \sim 1$, which is one order more than the AdS/CFT value $1/4\pi$.
Experimental data suggests that $\eta/s$ is small, which means that the quark-gluon plasma is in the strongly coupled regime.
The data extracted from relatively recent experiments with collisions of heavy atomic nuclei in the Relativistic Heavy Ion Collider \cite{Teaney:2003,Shuryak:2003xe} say that if one models the quark-gluon plasma by the Navier-Stokes equation with parameters fixed to fit the experimental data, then \emph{the ratio $\eta/s$ must be indeed about one order less than unity and is close to $1/4\pi$} \cite{Teaney:2003,Shuryak:2003xe}.
This gives optimism to string theorists because the observation of the universal $\eta/s$ ratio in the RHIC or the LHC might be an experimental
confirmation of the AdS/CFT correspondence and direct connection with a real experiment, which was always lacking in string theory \cite{Birmingham:2002ph,Son:2006em,Haack:2008xx,Ritz:2008kh,Myers:2009ij,Cremonini:2009sy,Maeda:2006by,Kats:2007mq,Matsuo:2009yu,Amado:2007pv,Michalogiorgakis:2006jc,Chen:2009hg,Chen:2010ik}.

\subsection{Holographic superconductor}

In this subsection we showed how the WKB formula (\ref{QNM_WKB6}), used for finding QNMs, can be quite unexpectedly used for estimation of the conductivity of superconductors constructed in the spirit of the AdS/CFT correspondence \cite{Gubser-superconducotr1,Horowitz-superconductor3}.

The effect of superconductivity consists of vanishing of the electrical resistivity of some metals (superconductors) below some critical temperature $T_c$. In addition, the magnetic field is expelled from such a superconductor at low temperature (Meissner effect), so that the superconductor is also a perfect diamagnetic. The phenomenological theory of superconductivity by Bardeen, Cooper and Schrieffer (BCS) \cite{BCS} suggested that pairs of electrons with opposite spin can form a bound state by interacting with phonons. This bound state is a boson, called the Cooper pair, and below the critical temperature these pairs condense, thereby inducing superconductivity. However, the BCS theory works well for superconductors in the weak coupling regime, i.~e., for those superconductors whose critical temperature $T_c$ is quite low. Superconductors for which $T_c$ is high are expected to involve strongly correlated electrons. For a description of this class of superconductors the holography could provide some insight.

On the gravity side of the gauge/gravity duality the superconductivity can be modeled by an asymptotically AdS black hole with scalar hair: A nonzero condensate is ``represented'' by the black hole ``hair''. The field theory temperature is again dual to the Hawking temperature of the black hole so that one needs a black hole that would possess scalar hair at low temperature and would loose it at $T > T_c$.
Gubser showed that a charged black hole with charged scalar field around it satisfies these requirements \cite{Gubser-superconducotr1,Gubser-superconducotr2}.
Thus, the Lagrangian has the form
\begin{equation}\label{supercond-larg}
\mathfrak{L }= R + \frac{6}{L^2} - \frac{1}{4} F^{\mu\nu} F_{\mu\nu}
- |\nabla \psi - i q A \psi |^2 -V( |\psi|), \,
\end{equation}
where $\psi$ is the scalar field, $F_{\mu\nu}$ is the strength tensor of the electromagnetic field, $q$ is the scalar field's charge and $A$ is the vector potential ($F=dA$). The cosmological constant is $-3/L^2$.

According to the gauge/gravity dictionary the conductivity can be found by solving for fluctuations of the Maxwell field. Therefore, further we study fluctuations of the potential $A_x$.

The plane symmetric solution can be written in the following general form:
\begin{equation}\label{metric}
 ds^2=-g(r) e^{-\chi(r)} dt^2+{dr^2\over g(r)}+r^2(dx^2+dy^2),
\end{equation}
\begin{equation}
A=\phi(r)~dt, \quad \psi = \psi(r),
\end{equation}
\begin{equation}\label{207}
A_x'' + \left(\frac{g'}{g} - \frac{\chi'}{2} \right) A_x' + \left(\left(\frac{\omega^2}{g^2} - \frac{\phi'^2}{g} \right) e^{\chi} - \frac{2 q^2
\psi^2}{g} \right) A_x  =  0.
\end{equation}
Using a new radial variable $dz = {e^{\chi/2}\over g} dr$, at large $r$, $dz  \approx dr/r^2$, and we choose the constant of integration so that $ z = -1/r$. The horizon is located at $z=-\infty$. Then, Eq.~(\ref{207}) has the following wavelike form:
\begin{equation}\label{schr}
-A_{x,zz} + V(z) A_x = \omega^2 A_x,
\end{equation}
where the effective potential is \cite{Horowitz-superconductor2,Horowitz-superconductor3}
\begin{equation}\label{potential}
V(z) = g[\phi_{,r}^2  + 2q^2 \psi^2  e^{-\chi}].
\end{equation}

Now, an incoming wave from the right  will be partly transmitted and partly reflected by the potential barrier. The transmitted wave is purely ingoing at the horizon and the reflected wave satisfies the scattering boundary conditions at $z \rightarrow \infty$, and  will obey, at the same time, the Dirichlet boundary condition at $z=0$.
Thus, the scattering boundary conditions for $z>0$ are
\begin{equation}\label{BCs1}
A_x = e^{-i\omega z} + {\cal R} e^{i\omega z}, \quad z \rightarrow +\infty,
\end{equation}
and at the event horizon
\begin{equation}\label{BCs2}
A_x = {\cal T} e^{-\imo\omega z}, \quad z \rightarrow -\infty,
\end{equation}
where ${\cal R}$ and ${\cal T}$ are reflection and transmission coefficients.
Then, one has
\begin{equation}
A_x(0) = 1+{\cal R}, \quad A_{x,z}(0) = -\imo\omega(1-{\cal R}).
\end{equation}
As shown in \cite{Horowitz-superconductor1}
\begin{equation}
A_x = A_x^{(0)} + \frac{A_x^{(1)}}{r}+\ldots\,,
\end{equation}
Within the gauge/gravity vocabulary the limit of the electric field and the first subleading term on the boundary are dual to the field strength and the induced current, respectively,
\begin{equation}
E_x = -A_x^{(0)}, \qquad J_x = A_x^{(1)}.
\end{equation}

Thus, we find the conductivity
\begin{equation}\label{oldcond}
\sigma(\omega) = \frac{J_x}{E_x}= -{\imo\over \omega} {A_x^{(1)}\over A_x^{(0)}}.
\end{equation}
Since $ A_x^{(1)} = - A_{x,z} (0)$, then,
\begin{equation}\label{cond}
\sigma(\omega) =  {1-{\cal R}\over 1+{\cal R}}
\end{equation}
The above boundary conditions (\ref{BCs1}), (\ref{BCs2}) are nothing but the standard scattering boundary conditions for finding the $S$-matrix. The effective potential has the distinctive form of the potential barrier, so that the WKB approach can be applied for finding ${\cal R}$ and $\sigma$ \cite{Konoplya:2009hv}. We note that as the wave energy (or frequency) $\omega$ was real, the first order WKB values for ${\cal R}$ and ${\cal T}$ are real \cite{WKB} and
$
{\cal T}^2 + {\cal R}^2 = 1.
$
Next, one can distinguish two qualitatively different cases: First, when $\omega^2$ is much less than the maximum of the effective potential $V_0$ $\omega^2 \ll V_{0}$, and second when $\omega^2 \simeq V_{0}$. When $\omega^2 \gg V_0$, usually the reflection coefficient ${\cal R}$ decreases too quickly with $\omega$, so that $\sigma$ reaches its maximal value (unity) even at moderate $\omega > V_{0}$.
Therefore, the case $\omega^2 \simeq V_{0}$ also works well for large $\omega$ and one can use the WKB formula (\ref{QNM_WKB6}), for scattering around black holes related in (see Sec.~\ref{sec:WKBmethod} in this review), which gives
\begin{equation}\label{WKB1}
{\cal R} = \left(1 + e^{\displaystyle- 2 i \pi (\nu + (1/2))}\right)^{-1/2}, \quad \omega^2 \simeq V_{0},
\end{equation}
where
\begin{equation}
\nu + \frac{1}{2} = i \frac{(\omega^2 - V_{0})}{\sqrt{-2 V_{0}^{\prime \prime}}} + \Lambda_2 +\Lambda_3.
\end{equation}
Further and extensive literature on this subject considers various generalizations of the above example. This mainly includes the addition of a magnetic field and various corrections, such as higher curvature corrections \cite{Amado:2009ts,Jing:2010zp,Zeng:2009dr,Nishioka:2009zj,Gregory:2009fj,Horowitz-superconductor3}.

\section{Conclusions}

This review should be considered as an introduction into the analysis of quasinormal modes of various black holes which are studied in astrophysics, higher-dimensional gravity, and string theory. Currently three comprehensive reviews on quasinormal modes \cite{Nollert:1999ji,Kokkotas-review,Berti:2009kk}, discuss in detail observational aspects of quasinormal ringing. We concentrated on the issues which were not reviewed or only briefly touched on there: methods of calculations of QNMs, stability of black holes and branes, late-time tails, the holographic superconductor, etc. Therefore, we sketched rather than discussed in detail some questions, referring the reader to more specialized literature.

\begin{acknowledgments}
The work of A. Z. was supported by the \emph{Funda\c{c}\~ao de Amparo \`a Pesquisa do Estado de S\~ao Paulo} (FAPESP), Brazil. At the initial stage of this work R. A. K. was supported by \emph{the Japan Society for the Promotion of Science} (JSPS), Japan, and at the final stage by \emph{the Alexander von Humboldt foundation} (AvH), Germany. At its final stage this work was also partially funded by the Conicyt grant ACT-91: ``Southern Theoretical Physics Laboratory'' (STPLab). \emph{The Centro de Estudios Cientificos} (CECS) is funded by the Chilean Government through the Centersof Excellence Base Financing Program of Conicyt.

The authors thank Elcio Abdalla, Victor Cardoso, Panagiota Kanti, Hideo Kodama, Kostas Kokkotas, Alberto Saa, Jiro Soda, Harvey Reall, and Andrei Starinets for useful discussions. Our special thanks to Paul Lasky for his careful reading of the manuscript and useful comments.
\end{acknowledgments}

\end{document}